\documentclass[%
 aip,
 amsmath,amssymb,
 reprint,%
]{revtex4-1}

\usepackage{graphicx}
\usepackage{dcolumn}
\usepackage{bm}
\usepackage{siunitx}
\usepackage[utf8]{inputenc}
\usepackage[T1]{fontenc}
\usepackage{mathptmx}
\usepackage{etoolbox}
\usepackage{ragged2e}
\usepackage[version=4]{mhchem}
\usepackage{hyperref}

\newcommand{\Ir}{\ce{IrO2} }
\newcommand{\Ru}{\ce{RuO2} }

\usepackage{tabularx}
\usepackage{multirow}
\usepackage[super]{nth}

\usepackage[normalem]{ulem}
\DeclareSIUnit\bar{bar}
\DeclareSIUnit\angstrom{\text {Å}}

\makeatletter
\def\@email#1#2{%
 \endgroup
 \patchcmd{\titleblock@produce}
  {\frontmatter@RRAPformat}
  {\frontmatter@RRAPformat{\produce@RRAP{*#1\href{mailto:#2}{#2}}}\frontmatter@RRAPformat}
  {}{}
}%
\makeatother

\begin{document}

\preprint{AIP/123-QED}


\title[]{Epitaxial RuO$_2$ and IrO$_2$ films by pulsed laser deposition on TiO$_2$(110)}

\author{P. Keßler}
\affiliation{Physikalisches Institut, Universität Würzburg, 97074 Würzburg, Germany}
\affiliation{Würzburg-Dresden Cluster of Excellence ct.qmat, Universität Würzburg, 97074 Würzburg, Germany}

\author{T. Waldsauer}
\affiliation{Physikalisches Institut, Universität Würzburg, 97074 Würzburg, Germany}
\affiliation{Würzburg-Dresden Cluster of Excellence ct.qmat, Universität Würzburg, 97074 Würzburg, Germany}

\author{V. Jovic}
\affiliation{Earth Resources and Materials, Institute of Geological and Nuclear Science, 5010 Lower Hutt, New Zealand}
\affiliation{MacDiarmid Institute for Advanced Materials and Nanotechnology, 6012 Wellington, New Zealand}

\author{M. Kamp}
\affiliation{Physikalisches Institut, Universität Würzburg, 97074 Würzburg, Germany}
\affiliation{Röntgen Center for Complex Material Systems, Universität Würzburg, 97074 Würzburg, Germany}

\author{M. Schmitt}
\affiliation{Physikalisches Institut, Universität Würzburg, 97074 Würzburg, Germany}
\affiliation{Würzburg-Dresden Cluster of Excellence ct.qmat, Universität Würzburg, 97074 Würzburg, Germany}

\author{M. Sing}
\affiliation{Physikalisches Institut, Universität Würzburg, 97074 Würzburg, Germany}
\affiliation{Würzburg-Dresden Cluster of Excellence ct.qmat, Universität Würzburg, 97074 Würzburg, Germany}

\author{R. Claessen}
\affiliation{Physikalisches Institut, Universität Würzburg, 97074 Würzburg, Germany}
\affiliation{Würzburg-Dresden Cluster of Excellence ct.qmat, Universität Würzburg, 97074 Würzburg, Germany}

\author{S. Moser}
\affiliation{Physikalisches Institut, Universität Würzburg, 97074 Würzburg, Germany}
\affiliation{Würzburg-Dresden Cluster of Excellence ct.qmat, Universität Würzburg, 97074 Würzburg, Germany}
\email{simon.moser@uni-wuerzburg.de}

\date{\today}

\begin{abstract}
We present a systematic growth study of epitaxial RuO$_2$(110) and IrO$_2$(110) on TiO$_2$(110) substrates by pulsed laser deposition. We describe the main challenges encountered in the growth process, such as a deteriorating material flux due to laser induced target metallization or the delicate balance of under- vs over-oxidation of the ’stubborn’ Ru and Ir metals. We identify growth temperatures and oxygen partial pressures of \SI{700}{K}, \SI{1e-3}{mbar} for RuO$_2$ and \SI{770}{K}, \SI{5e-4}{mbar} for IrO$_2$ to optimally balance between metal oxidation and particle mobility during nucleation. In contrast to IrO$_2$, RuO$_2$ exhibits layer-by-layer growth up to 5 unit cells if grown at high deposition rates. At low deposition rates, the large lattice mismatch between film and substrate fosters initial 3D island growth and cluster formation. In analogy to reports for RuO$_2$ based on physical vapor deposition,\cite{He2015} we find these islands to eventually merge and growth to continue in a step flow mode, resulting in highly crystalline, flat, stoichiometric films of RuO$_2$(110) (up to 30~nm thickness) and IrO$_2$(110) (up to 13~nm thickness) with well defined line defects. 
\end{abstract}

\maketitle

\section{Introduction}\label{sec: Introduction}


The binary oxides of ruthenium (RuO$_2$) and iridium (IrO$_2$) are functional Dirac semi-metals that have recently attracted considerable interest in both applied and fundamental materials science.\cite{Over2012,Weaver2013,Scarpelli2022,Jang2020} They both are important co-catalysts for the oxygen evolution reaction (OER) in electrocatalytic water splitting, with RuO$_2$ being more active, but IrO$_2$ being more corrosion resistant.\cite{Over2012,Spoeri2017,Weber2022,Naito2021,Hess2023} Their catalytic efficiency depends on crystal size and surface orientation,\cite{Assmann2005a,Stoerzinger2014} on surface order and stoichiometry,\cite{Moser2021,Reiser2023} as well as on the lattice structure of the supporting substrate.\cite{Seki2010}

RuO$_2$ is the only stable solid oxide phase of ruthenium and crystallizes in a non-symmorphic rutile crystal structure ($a = b = \SI{4.48}{\angstrom}$, $ c = \SI{3.11}{\angstrom}$, space group 136: P$4_2$/mnm).\cite{Jain2013} With its Fermi level well positioned within the $t_{2g}$ derived conduction band, it is a good electrical conductor that below 2\;K can be tuned into the superconducting regime by epitaxial strain.\cite{Uchida2020,Ruf2021} The metallic conductivity of RuO$_2$ as well as its favorable thermal and chemical stability are the main reasons for its industrial utility, e.g., as contact material in microelectronic devices \cite{Bai1998} or as electrocatalyst in a variety of oxidation and dehydrogenation reactions.\cite{Over2012,Weber2022} The Fermi surface of RuO$_2$ was characterized by transport and calorimetric methods in the 1970s,\cite{Mattheiss1976} and recently mapped by angle resolved photoemission spectroscopy (ARPES).\cite{Jovic2018,Jovic2019} The non-symmorphic crystal structure of the RuO$_2$ rutile lattice produces a Fermi surface composed of Dirac nodal lines (DNL).\cite{Sun2017,Jovic2018,Jovic2019} Strong nesting of these DNLs might be prone to Fermi surface instabilities,\cite{Ahn2019} a postulated driving force of collinear magnetic ordering in RuO$_2$ that was derived from neutron \cite{Berlijn2017} and resonant x-ray scattering,\cite{Zhu2019, Lovesey2021} but recently challenged by muon spin rotation experiments.\cite{Hiraishi2024,Kessler2024}

\begin{table*}
\setlength\doublerulesep{0.2cm} 
\def\arraystretch{1.2}
\begin{tabularx}{\textwidth}{|p{2.7cm}|p{2cm}|X|}
\hline
  \textbf{Material}  & \textbf{Substrate}  & \textbf{Growth parameters} \\
\cline{1-2}\cline{3-3}
 \textbf{Method} & \textbf{Lattice}  & \textbf{Evaluation of growth process}   \\
 \cline{1-1}
 \textbf{Reference} & \textbf{Mismatch\cite{Jain2013}}  & \\
\hline
\hline
  \textbf{RuO$_2$(110)}  & Ru(0001)  & $T = \SI{650}{K}$, \ce{O2}-exposure: $\SI{2e6}{L}$
  \\
\cline{1-2}\cline{3-3}
 Substrate oxidation & [001]: \SI{13.0}{\percent} & \multirow{2}{\hsize}{\justifying Scanning tunneling microscopy indicates a well ordered RuO$_2$(110) surface with small regions of RuO$_2$(100) growth. Annealing above \SI{500}{K} partly restores a damaged surface.}
 \\
 \cline{1-1}
 Over (2004)\cite{Over2004a} &[1$\bar{\textrm{1}}$0]: \SI{-11.0}{\percent} & \\
\hline
\hline
  \textbf{RuO$_2$(100)}  & Ru($10\overline{1}0$)  & $T = \SI{700}{K}$, \ce{O2}-exposure: $\SI{3e5}{L}$\\
\cline{1-2}\cline{3-3}
 Substrate oxidation &[001]: \SI{13.0}{\percent}  & \multirow{2}{\hsize}{\justifying After oxidation, a predominating RuO$_2$(100) surface is accompanied by regions of RuO$_2$(110), RuO$_2$(101) as well as an catalytically inactive RuO$_2$(100)-c(2x2) phase. }
 \\
 \cline{1-1}
 Over (2004)\cite{Over2004a} &[010]: \SI{4.8}{\percent} & \\
\hline
\hline
  \textbf{RuO$_2$(110)}  & TiO$_2$(110) & $T = \SI{600}{K}$, $p(\ce{O2}) = \SI{1e-6}{mbar}$\\
\cline{1-2}\cline{3-3}
 PVD & [001]: \SI{4.9}{\percent}  & \multirow{2}{\hsize}{\justifying A well ordered film with surface clusters resulting from island merging. The XPS area ratio of the screened and unscreened Ru 3d peaks hints to a non-stoichiometric film composition.
}   \\
 \cline{1-1}
 He (2015) \cite{He2015} & [$1\bar10$]: \SI{-2.6}{\percent}& \\
\hline
\hline
  \textbf{RuO$_2$(110)}  & MgO(001) & $T = \SI{970}{K}$, $p(\ce{O2}) = \SI{9e-4}{mbar} - \SI{1e-1}{mbar}$, $F = \SI{0.45}{J/cm^2}$, $\lambda = \SI{193}{nm}$\\
\cline{1-2}\cline{3-3}
 PLD & [001]: \SI{-1.1}{\percent} & \multirow{2}{\hsize}{\justifying The best crystalline quality is achieved for films grown at \SI{970}{K}. Changing the oxygen pressure influenced the growth rate, but lead to no / small changes of the crystalline quality.}   \\
 \cline{1-1}
 Fang (1997) \cite{Fang1997} & [$1\bar10$]: \SI{0.8}{\percent} & \\
\hline
\hline
  \textbf{RuO$_2$(200)}  & LaAlO$_3$(001) & $T = \SI{770}{K}$, $p(\ce{O2}) = \SI{1e-1}{mbar}$, $F = \SI{3}{J/cm^2}$, $\lambda = \SI{248}{nm}$\\
\cline{1-2}\cline{3-3}
 PLD & [011]: \SI{2.0}{\percent} & \multirow{2}{\hsize}{\justifying AFM suggests a transition from layer-by-layer to island growth at a film thickness of 7 RuO$_2$ unit cells.}   \\
 \cline{1-1}
 Wang (2006) \cite{Wang2006} & n.a. & \\
 \hline
\hline
  \textbf{RuO$_2$(001)}   & TiO$_2$(001) & $T = \SI{970}{K}$, $p(\ce{O2}) = \SI{6.7e-3}{mbar} - \SI{6.7e-2}{mbar}$, $F = \SI{2}{J/cm^2}$, $\lambda = \SI{248}{nm}$\\
\cline{1-2}\cline{3-3}
 PLD & [010]: \SI{-2.6}{\percent}& 
  \multirow{2}{\hsize}{\justifying The RuO$_2$ sheet resistance can be reduced by applying a lower $p(\ce{O2})$ during growth (\SI{6.7e-2}{} - \SI{1.3e-2}{mbar}). AFM reveals island growth for a \SI{10}{nm} thick RuO$_2$ film.}   \\
 \cline{1-1}
 Kim (2019) \cite{Kim2019a} & [100]: \SI{-2.6}{\percent}& \\
\hline

\end{tabularx}
\caption{\label{tab:review_RuO2} Literature review of single crystalline RuO$_2$ thin film synthesis. Summarized are the respective fabrication method, the substrate type and orientation, the growth parameters along with a short summary of the study. The lattice mismatch is calculated based on structural data summarized in Ref. \cite{Jain2013} according to the formula $1 - a_{\text{substrate}}/a_{\text{film}}$, where $a$ is the respective lattice constant. The following symbols are used to describe physical quantities: $T$: substrate temperature; $p(\ce{O2})$: oxygen partial pressure; $F$: deposition laser fluency; $\lambda$: deposition laser wavelength; $L$: dosage ($\SI{1}{L} =\SI{1}{Langmuir} = \SI{1.33e-6}{\milli\bar\s}$).}
\end{table*}

In unison with the crystal field, this collinear order would produce a non-relativistic spin splitting that breaks time reversal symmetry in the band structure of RuO$_2$, and is thus consistent with recent observations of the anomalous Hall effect (AHE),\cite{Feng2022,Tschirner2023} spin transfer torque,\cite{Bose2022a} and giant tunneling magnetoresistance\cite{Shao2021} in RuO$_2$ thin film structures. RuO$_2$ would thus classify as a so called \textit{altermagnet}, i.e., as a compensated collinear magnet of net zero magnetization, whose spin splitting alternates in reciprocal space, setting altermagnets fundamentally apart from ferro- and antiferromagnets.\cite{Smejkal2022,Smejkal2022a} RuO$_2$ thus lends itself as a formidable playground to study the intimate relationship between its structural, electronic and magnetic properties that potentially culminate in altermagnetism. 

IrO$_2$ on the other hand ($a = b = \SI{4.51}{\angstrom}$,
$c = \SI{3.18}{\angstrom}$, space group 136: P$4_2$/mnm),\cite{Jain2013} the 5d sister compound of RuO$_2$, offers itself as useful comparative material to benchmark potential altermagnetism in RuO$_2$. It shares the non-symmorphic rutile crystal structure and thus the spin orbit gapped DNL structure with RuO$_2$,\cite{Sun2017,Nelson2019} and was shown to also exhibit exotic experimental effects such as spin-orbit torque,\cite{Bose2020, Patton2023} electrochromism,\cite{Gottesfeld1978, Patil2005} an inverse spin-Hall effect\cite{Fujiwara2013} and magnetic field dependent charge carrier switching.\cite{Uchida2015} In contrast to RuO$_2$, however, consensus about Pauli paramagnetism in IrO$_2$ remains unchallenged, even though a slight ferromagnetic order seems to be theoretically possible.\cite{Ping2015} 

A meaningful investigation of such physical phenomena requires reproducible samples of excellent crystallinity and well defined stoichiometry.\cite{Martin2024} In this work, we thus optimized the epitaxial synthesis of RuO$_2$ and IrO$_2$ thin films on TiO$_2$(110) substrates by pulsed laser deposition (PLD), and report the individual growth modes, defect structures and trends that we encountered in our extensive in- and ex-situ chemical and structural analysis. Specifically, we demonstrate the PLD film growth of stoichiometric, uniform, closed and thickness controlled epitaxial RuO$_2$(110) (up to 30~nm) and IrO$_2$(110) (up to 13~nm) films on TiO$_2$(110) substrates of high crystalline quality with few well defined line defects and suppressed island formation.

To achieve this goal, we performed a systematic exploration of the multidimensional growth control parameter space, including target and laser flux as well as substrate temperature and oxygen partial pressure. We thoroughly analyzed this growth matrix by a combination of \textit{in situ} and \textit{ex situ} micro-spectroscopy techniques, identifying the individual growth modes along with the optimal parameter set for epitaxial film growth and the most common defect structures. In the following, we detail out the technical challenges for the PLD growth of RuO$_2$ and IrO$_2$, and highlight the most pertinent commonalities and differences encountered in these two material systems.

\section{State-of-the-art \protect\Ru and \protect\Ir epitaxial synthesis}

\begin{table*}
\setlength\doublerulesep{0.2cm} 
\def\arraystretch{1.2}
\begin{tabularx}{\textwidth}{|p{2.7cm}|p{2cm}|X|}
\hline
  \textbf{Material}  & \textbf{Substrate}  & \textbf{Growth parameters} \\
\cline{1-2}\cline{3-3}
 \textbf{Method} & \textbf{Lattice}  & \textbf{Evaluation of growth process}   \\
 \cline{1-1}
 \textbf{Reference} & \textbf{Mismatch\cite{Jain2013}}  & \\
\hline
\hline
 \textbf{IrO$_2$(110) / (100)}  & Ir(111)  & $T = \SI{775}{K} - \SI{875}{K}$, $p(\ce{O2}) = \SI{100}{mbar}$\\
\cline{1-2}\cline{3-3}
 Substrate oxidation & [001]: \SI{14.2}{\percent}  & \multirow{2}{\hsize}{\justifying 
The bulklike IrO$_2$ mainly consists of domains with an (110) and (100) oriented surface. Changing the oxidation parameters leads to the formation of an O-Ir-O trilayer as well as to corundum (Ir$_2$O$_3$).}\\
 \cline{1-1}
 He (2008)\cite{He2008} & [$1\bar10$]: \SI{25.9}{\percent}  & \\
 \hline
 \hline
 \textbf{IrO$_2$(100)}  & Ir(111)  & $T = \SI{600}{K}$, \ce{O}-plasma: $\SI{3.6e5}{L}$\\
\cline{1-2}\cline{3-3}
 Substrate oxidation & [001]: \SI{14.2}{\percent}  & \multirow{2}{\hsize}{\justifying Applying an oxygen plasma enables the selective preparation of a pure IrO$_2$(100) phase without IrO$_2$(110) domains. Corundum can also be obtained for higher plasma exposures.}\\
 \cline{1-1}
 Chung (2012)\cite{Chung2012} & [$0\bar10$]: \SI{-4.8}{\percent}  & \\
\hline
\hline
 \textbf{IrO$_2$(110)}  & TiO$_2$(110) & $T = \SI{570}{K}$, $p(\ce{O2}) = \SI{6.67e-6}{mbar}$ \\
\cline{1-2}\cline{3-3}
 MBE & [$00\bar1$]: \SI{6.8}{\percent}  & \multirow{2}{\hsize}{\justifying Production of high phase purity films by solid-source metal-organic MBE from 3 - 50 monolayer thickness. The ratio between Ir and IrO$_2$ can be engineered by epitaxial strain.}   \\
 \cline{1-1}
 Nair (2023) \cite{Nair2023} & [$1\bar10$]: \SI{-2.1}{\percent} & \\
\hline
\hline
  \textbf{IrO$_2$(110)}  & MgO(001)& $T = \SI{1020}{K}$, $p(\ce{O2}) = \SI{2.67e-1}{mbar}$, $\lambda = \SI{248}{nm}$ \\
\cline{1-2}\cline {3-3}
 PLD & [$00\bar1$]: \SI{1.0}{\percent}   & \multirow{2}{\hsize}{\justifying A BaTiO$_3$ buffer layer is needed to prevent the formation of IrMg intermetallics. XRD indicates a single domain film and AFM shows island growth with a root-mean-square roughness of $\approx$ \SI{6}{nm}.}   \\
 \cline{1-1}
 Stoerzinger (2014) \cite{Stoerzinger2014} & [$1\bar10$]: \SI{1.3}{\percent} & \\
\hline
\hline
  \textbf{IrO$_2$(100)}   & SrTiO$_3$(100) & $T = \SI{670}{K} - \SI{870}{K}$, $p(\ce{O2}) = \SI{5e-2}{mbar} - \SI{1}{mbar}$, $F = \SI{0.5}{J/cm^2} - \SI{0.7}{J/cm^2}$, $\lambda = \SI{248}{nm}$\\
\cline{1-2}\cline{3-3}
 PLD & [202]: \SI{-0.4}{\percent}  & 
  \multirow{2}{\hsize}{\justifying The thin films are insulating due to island growth up to a film thickness of \SI{25}{nm}. Thicker films of \SI{40}{nm} show metallic transport behaviour likely resulting from a continuous but granular structure.}   \\
 \cline{1-1}
 Bhat (2017) \cite{Bhat2017} & n.a.  & \\
 \hline
\hline
  \textbf{IrO$_2$(110)}  & TiO$_2$(110) & $T = \SI{770}{K}$, $\lambda = \SI{248}{nm}$ \\
\cline{1-2}\cline{3-3}
 PLD & [$00\bar1$]: \SI{6.8}{\percent}  & \multirow{2}{\hsize}{\justifying The IrO$_2$(110) films were grown up to a thickness of \SI{92.2}{nm}. Data on the morphology of the sample was not provided in this study.}   \\
 \cline{1-1}
 Arias (2021) \cite{Arias-Egido2021} & [$1\bar10$]: \SI{-2.1}{\percent}  & \\
\hline
\hline
  \textbf{IrO$_2$(110)}  & RuO$_2$(110)  &  $T = \SI{700}{K}$, $p(\ce{O2}) = \SI{3e-7}{mbar}$, $\lambda = \SI{248}{nm}$ \\
\cline{1-2}\cline{3-3}
PVD & [001]: \SI{2.1}{\percent}  & \multirow{2}{\hsize}{\justifying The IrO$_2$ films with a thickness of up to \SI{10}{nm} exhibit a high crystalline and surface quality. They are prepared by an initial nucleation step that is followed by layer-by-layer growth.}   \\
 \cline{1-1}
 Abb (2018) \cite{Abb2018} & [$1\bar10$]: \SI{0.5}{\percent}  & \\
\hline
\end{tabularx}
\caption{\label{tab:review_IrO2} Literature review of single crystalline IrO$_2$ thin film synthesis. Summarized are the respective fabrication method, the substrate type and orientation, the growth parameters along with a short summary of the study. The lattice mismatch is calculated based on structural data summarized in Ref. \cite{Jain2013} according to the formula $1 - a_{\text{substrate}}/a_{\text{film}}$, where $a$ is the respective lattice constant. The following symbols are used to describe physical quantities: $T$: substrate temperature; $p(\ce{O2})$: oxygen partial pressure; $F$: deposition laser fluency; $\lambda$: deposition laser wavelength; $L$: dosage ($\SI{1}{L} =\SI{1}{Langmuir} =\SI{1.33e-6}{\milli\bar\s}$). }
\end{table*}

Fabrication methods of crystalline RuO$_2$ and IrO$_2$ are extensive and outlined in excellent reviews.\cite{Over2012,Scarpelli2022, Jang2020} Here, we are exclusively interested in the synthesis of RuO$_2$ and IrO$_2$ thin films of high bulk and surface crystalline order and provide a detailed summary of literature reports in Tab. \ref{tab:review_RuO2} and Tab. \ref{tab:review_IrO2} for both materials. In addition to this overview, much progress in sample production was carried by (i) the oxidation of Ru(0001), Ru(10$\bar1$0), Ir(111) and Ir(100) single crystal surfaces (RuO$_2$: Refs. \onlinecite{Over2004a,Herd2012,Abb2018,Weber2019}, IrO$_2$: Refs. \onlinecite{He2008,Rai2016,Liang2017a,Chung2012}) as well as a variety of different bottom-up growth procedures such as (ii) molecular beam epitaxy (MBE, RuO$_2$: Refs. \onlinecite{Kim1997,Nunn2021}, IrO$_2$: Refs. \onlinecite{Uchida2015,Nair2023,Kawasaki2018b,Kawasaki2018a,Kawasaki2016,Kuo2019}), (iii) pulsed laser deposition (PLD, RuO$_2$: Refs. \onlinecite{Fang1997,Wang2006,Kim2019a,Iembo1997a}, IrO$_2$: Refs. \onlinecite{Zhang2007,ElKhakani1997,ElKhakani1998,ElKhakani1996,Kim2016b,Hou2017a,Arias-Egido2021,Stoerzinger2014}), (iv) physical vapor deposition (PVD, RuO$_2$, Ref. \onlinecite{He2015}, IrO$_2$: Ref. \onlinecite{Abb2018}), (v) reactive magnetron sputtering (RuO$_2$: Refs. \onlinecite{Wang1996,Huang2001,Meng1999,Paoli2015,Liu2023}, IrO$_2$: Refs. \onlinecite{Klein1995,Klein1995a,Balu1996,Shimizu1997,Liu2008,VanOoyen2009,Cogan2009}), (vi) atomic layer deposition (ALD, RuO$_2$: Refs. \onlinecite{Wang2006,Kim2007,Kwon2007,Park2014,Jung2014}, IrO$_2$: Refs. \onlinecite{Hamalainen2011,Mattinen2016,Kim2008a,Hamalainen2008,Simon2021}), or more recently (vii) thermal laser epitaxy (TLE, RuO$_2$: Ref. \onlinecite{Kim2021}). To our knowledge, high quality single crystalline thin films with a well defined surface orientation to date have solely been reported from methods (i) single crystal oxidation (RuO$_2$: Ref. \onlinecite{Over2004a}, IrO$_2$: Refs. \onlinecite{He2008,Chung2012}), from (ii) MBE (IrO$_2$, Ref. \onlinecite{Nair2023}), (iii) PLD (RuO$_2$: Refs. \onlinecite{Fang1997,Wang2006,Kim2019a}, IrO$_2$: Refs. \onlinecite{Stoerzinger2014,Bhat2017,Arias-Egido2021}), and from (iv) PVD (RuO$_2$, Ref. \onlinecite{He2015}, IrO$_2$: Ref. \onlinecite{Abb2018}). The respective results are reviewed in Tabs. \ref{tab:review_RuO2} and \ref{tab:review_IrO2}.

In method (i), single Ru(0001), Ru(10$\overline{1}$0), Ir(111) or Ir(100) crystal surfaces are prepared by repeated cycles of sputtering and annealing to form a clean, well ordered surface, as well as subsequent thermal flashing in oxygen to remove residual carbon contamination. The oxide layer is then formed in a following oxidation step at elevated temperatures (Tabs. \ref{tab:review_RuO2} and \ref{tab:review_IrO2}), which results in oxide flakes with a well ordered crystal surface.\cite{Over2004a,He2008} While such samples can be well suited for some surface spectroscopy as well as catalytic experiments,\cite{Knapp2007b} they are limited by the low achievable film thickness of only a few nanometers, \cite{Kim2000,He2008,Assmann2005} by the formation of multi-domain structures with different surface orientations due to the substrate symmetry,\cite{Kim2000,Herd2012} as well as by the metallicity of the substrate that prevents meaningful transport experiments in the oxide over-layer.

Alternatively, RuO$_2$ and IrO$_2$ films have been grown on insulating single crystal substrates of varying lattice parameteres and surface orientations. The most usual substrates along with their lattice mismatches to RuO$_2$ and IrO$_2$, the respective sample growth parameters, and a brief summary of the growth process and resulting crystalline quality are summarized in Tab. \ref{tab:review_RuO2} and Tab. \ref{tab:review_IrO2}. The most commonly used substrate material is rutile TiO$_2$. Beyond the commercial availability of high quality polished wafers of variable surface orientation at acceptable cost, its isostructural crystal lattice with regards to RuO$_2$ and IrO$_2$ enables epitaxial growth without the formation of rotational domains. This allows for experiments that depend explicitly on the crystal orientation, such as bulk transport\cite{Bose2020,Feng2022,Tschirner2023} or surface catalytic measurements.\cite{Stoerzinger2014,Reiser2023} While the lattice constants of the materials are comparable, their mismatch to \ce{TiO2} is still significant with 4.9\% and -2.6\% along $[00\overline{1}]$ and $[1\overline{1}0]$ for RuO$_2$,\cite{Kim1997} as well as 6.8\% and -2.1\% along $[00\overline{1}]$ and $[1\overline{1}0]$ for IrO$_2$.\cite{Jain2013} This fosters the formation of defects and disorder in the epitaxial films resulting from strain release. 

To cope with problems arising due to the lattice mismatch induced strain, different strategies have been reported in literature including the growth of material at a reduced deposition rate,\cite{Fang1997,Jia1995} the application of a temperature gradient across the substrate to induce an initial intermixing of substrate and film,\cite{Kim1997} the growth of islands,\cite{Wang2006,Kim2019a} and the coalescence of such initial islands (island merging) followed by a final step flow process.\cite{He2015} Moreover, the lattice mismatch can also be used for strain engineering, e.g., to induce superconductivity in RuO$_2$,\cite{Ruf2021} or to reduce the formation of metallic Ir droplets at the surface of IrO$_2$.\cite{Nair2023}

Comparing the methods that are commonly employed in the synthesis of low dimensional RuO$_2$ and IrO$_2$ samples, we notice a variety of problems that are specific to the individual deposition technique, and mostly relate to the low vapor pressure of Ru, Ir and their oxides in combination with high oxidation potentials. In MBE, e.g., the high source temperatures needed to evaporate the metal in combination with high oxygen partial pressures\cite{Ruf2021,Abb2019} require the use of customized reactive-oxide MBEs to minimize the risk of corroding (oxidizing) machine components. To minimize this effect, atomic oxygen or ozone is typically used to locally increase the oxidation potential right at the substrate.\cite{Ruf2021} Further, the metal vapor pressure can be artificially increased and the source temperature consequently decreased by applying metal organic precursors.\cite{Nair2023} 

While these drawbacks of MBE are probably most optimally solved by the adsorption controlled TLE approach,\cite{Braun2019,Kim2021} they can also be appropriately addressed by the more common PLD method where a strong pulsed laser ablates the target material pulse by pulse. The plasma that forms in this way is ill defined and the deposited material does not arrive uniformly on the substrate surface.\cite{Zheng1993} This has to be counteracted by additional reactive oxygen, and thus pushes PLD far off the adsorption-controlled limit. In PLD of RuO$_2$ and IrO$_2$, in particular, we further find a considerable loss of target oxygen upon ablation, leading to a massive increase in target reflectivity and a consequent drop in deposition rate, which ultimately limits the achievable film thickness. In combination with the well-known coating problem of the deposition laser entry window, this is likely the reason why a wide range of laser energy densities has been reported in the PLD literature for RuO$_2$\cite{Fang1997,Jia1995,Wang2006,Kim2019a} and IrO$_2$\cite{Adiga2022,Serventi2001} growth. This might also be the reason why to date -- at least to our knowledge -- there are no reports on the PLD growth of epitaxial RuO$_2$ and IrO$_2$ thin films with high quality bulk as well as surface order as achieved, e.g., by MBE.\cite{He2015,Nair2023}
 
\section{Experimental details}
The sample growth was conducted in a commercially available PLD setup (TSST B.V.) with a base pressure typically below \SI{5e-9}{mbar}. The substrate temperature was controlled by a laser heating setup ($\lambda$ = \SI{976}{nm}, ComPACT-Evolution DILAS), where the temperature is measured by a two-color pyrometer (IMPAC) directed to the backside of a \SI{1}{mm} thick, sandblasted Inconel sample holder. A capillary, mounted at a distance of \SI{10}{cm} from the substrate, directs N5.5 oxygen at its surface to locally adjust the oxygen partial pressure. The pressure is measured by a Pfeiffer full range gauge mounted on the chamber walls, resulting in an effectively higher pressure at the sample. The targets are placed at a distance of \SI{55}{mm} in front of the substrate and are ablated by an excimer laser ($\lambda=$ \SI{248}{nm}, COMPex Pro 205/KrF). The growth was monitored by a reflection high-energy electron diffraction setup (RHEED, Staib Instruments: CB801420) that was operated at electron energies of \SI{30}{keV}. The PLD chamber is connected \textit{in vacuo} to an x-ray photoelectron spectroscopy (XPS, Scienta Omicron) setup with a monochromatic Al K$\alpha$ X-ray source of \SI{1486.7}{eV}, and to a low-energy electron diffraction system (LEED, Omicron: SPECTALEED) with a LaB$_6$ cathode. \textit{Ex-situ} measurements include atomic force microscopy (AFM, VEECO, tapping mode in air), scanning transmission electron microscopy (STEM, uncorrected FEI Titan 80-300, \SI{300}{kV}, \SI{100}{} - \SI{200}{pA}), scanning electron microscopy (SEM, Zeiss Ultra Plus, 15 kV) including energy-dispersive X-ray spectroscopy (EDX), as well as  X-ray diffraction (XRD) and reflectivity (XRR, BRUKER, Cu K$\alpha$ \SI{8.05}{keV}).

\subsection{\label{sec:Substrate}TiO$_2$(110) substrate preparation}

\begin{figure}[!htbp]
\includegraphics[width = \linewidth]{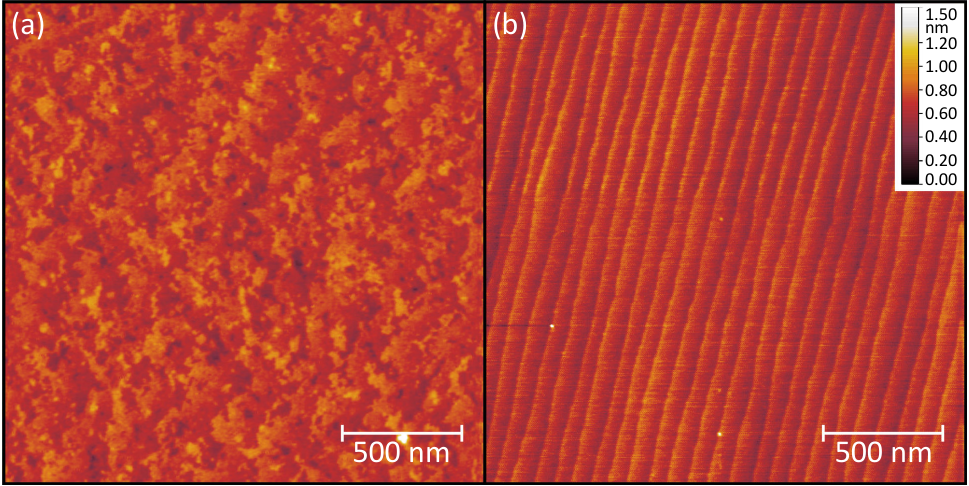} \caption{\label{fig:sub_prep}Atomic force microscope measurements of TiO$_2$(110) substrate: (a) Untreated substrate as delivered. (b) Stepped terrace surface after two step ultrasonic cleaning and tube furnace annealing.}
\end{figure}

For the film growth, we employed (110) terminated TiO$_2$ substrates supplied by Crystal GmbH, Berlin, Germany. The untreated substrates show a surface with randomly sized patches of a few atoms height difference as shown in Fig. \ref{fig:sub_prep} (a). To remove surface contamination, the substrates were initially cleaned for \SI{20}{minutes} in subsequent ultrasonic baths of acetone and isopropyl alcohol. 
Following a protocol developed for TiO$_2$(001),\cite{Wang2016a} we further annealed these substrates in a tube furnace at \SI{820}{\degreeCelsius} under an oxygen flow of \SI{20}{l/h} for \SI{5}{h}. This annealing step routinely yields well-defined TiO$_2$(110) surfaces with a stepped terrace morphology as shown in Fig. \ref{fig:sub_prep} (b). As XPS and EDX on our RuO$_2$ films occasionally revealed calcium contamination that likely resulted from polish residues of our commercial substrates, the  TiO$_2$(110) employed for the IrO$_2$ growth was further treated by a short etching step (t = \SI{5}{min}, ultrasonic bath) in 6:1 buffered hydrofluoric acid. After etching, the substrates were rinsed by high-purity water and the acetone / isopropyl cleaning step was repeated. Substrates that were treated in this way did no longer exhibit the calcium contamination.

\subsection{Target control}

\begin{figure}[!htbp]
\includegraphics[width = \linewidth]{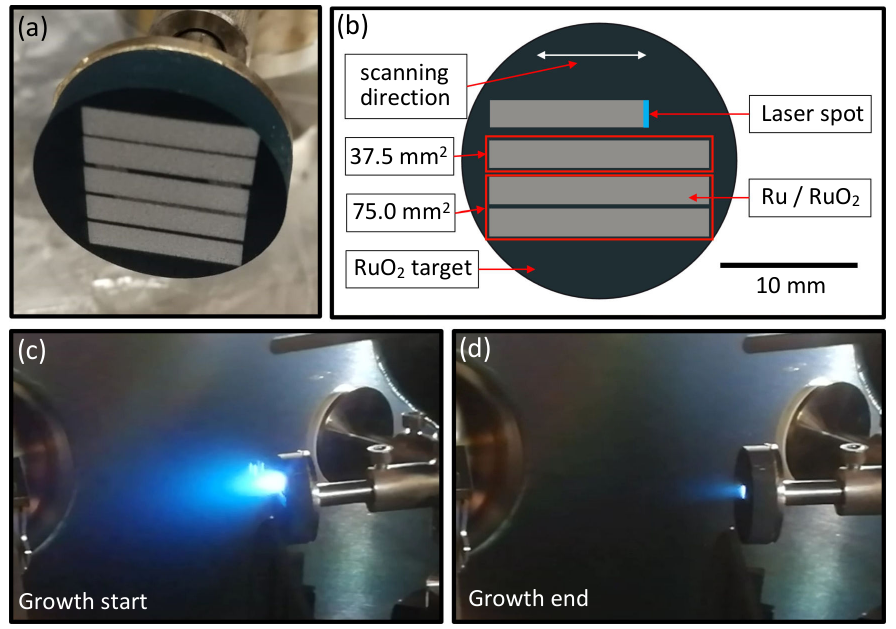}
\caption{\label{fig:Target} (a) RuO$_2$ target metallization due to laser ablation. (b) Definition of the target areas. Size of the plasma plume at (c) the start and (d) the end of the film growth.}
\end{figure}

For the RuO$_2$ film growth, we used a disk-shaped commercial target (TOSHIMA manufacturing Co. Ltd.) from compressed and sintered RuO$_2$ powder. Ablation was achieved by scanning the target relative to the fixed ablation laser spot at a constant scan rate of \SI{1.5}{mm/s}. The scanning area was \SI{37.5}{mm^2} (gray stripes in Figs. \ref{fig:Target} (a) and (b)), determined by the laser spot width of \SI{2.5}{mm}, and the scanning width of \SI{15.0}{mm} that is limited by the target diameter. Repeated scanning of the same target area increases the target reflectivity and leads to a steady reduction of the deposition rate. For a given film growth, we thus typically scanned two fresh, separate target regions, which resulted in a total ablated area of \SI{75.0}{mm^2} (Fig. \ref{fig:Target} (b)).

\begin{figure*}[!htbp]
\includegraphics[width = \linewidth]{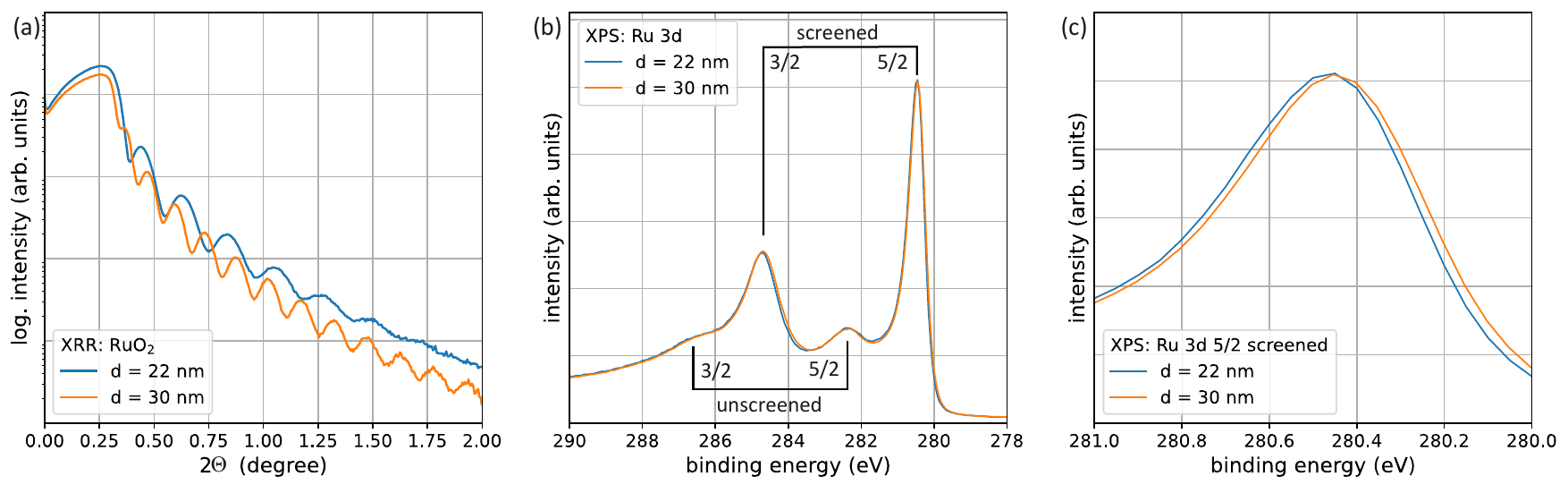}
\caption{\label{fig:coating} (a) XRR measurements of two RuO$_2$ films that were grown consecutively at nominally identical growth parameters. While the first film is \SI{30}{nm} thick, a reduction in the deposition rate due coating of the laser entry window yields only a thickness of \SI{22}{nm} for the second film. (b) XPS measurements of both films, highlighting the unscreened (u) and screened (s) contributions to the Ru 3d 5/2 and 3/2 multiplets, respectively. A closeup in (c) of the screened Ru 3d 5/2 peak highlights the second film to be slightly less metallic than the first one, indicated by a shift to lower binding energies.}
\end{figure*}

The IrO$_2$ target was manufactured in house using a \SI{99.9}{\%} pure trace metal basis IrO$_2$ powder supplied by Sigma-Aldrich. After X-ray powder diffraction to confirm the crystalline structure, the IrO$_2$ powder was ground to obtain a uniform grain size and then compressed for 5 minutes by a force of \SI{50}{kN}. In a subsequent annealing step at \SI{870}{\degreeCelsius}, the target was sintered for \SI{12}{h} under an oxygen flow of \SI{25}{l/h} to increase its solidity. The ablated target area was limited to \SI{12}{mm^2} due to cracks in the target disk. However, we recommend a larger scanning area -- similar to the one used for RuO$_2$ growth -- due to a rapid loss of oxygen and consequent metallization of the target surface that drastically increases its reflectivity.

Every RuO$_2$ and IrO$_2$ film reported in this study was grown from a freshly prepared target surface. Once the entire target surface was consumed, it was revived by sanding off the reflective surface layer \textit{ex vacuo}.

\begin{figure*}[!htbp]
\includegraphics[width = \linewidth]{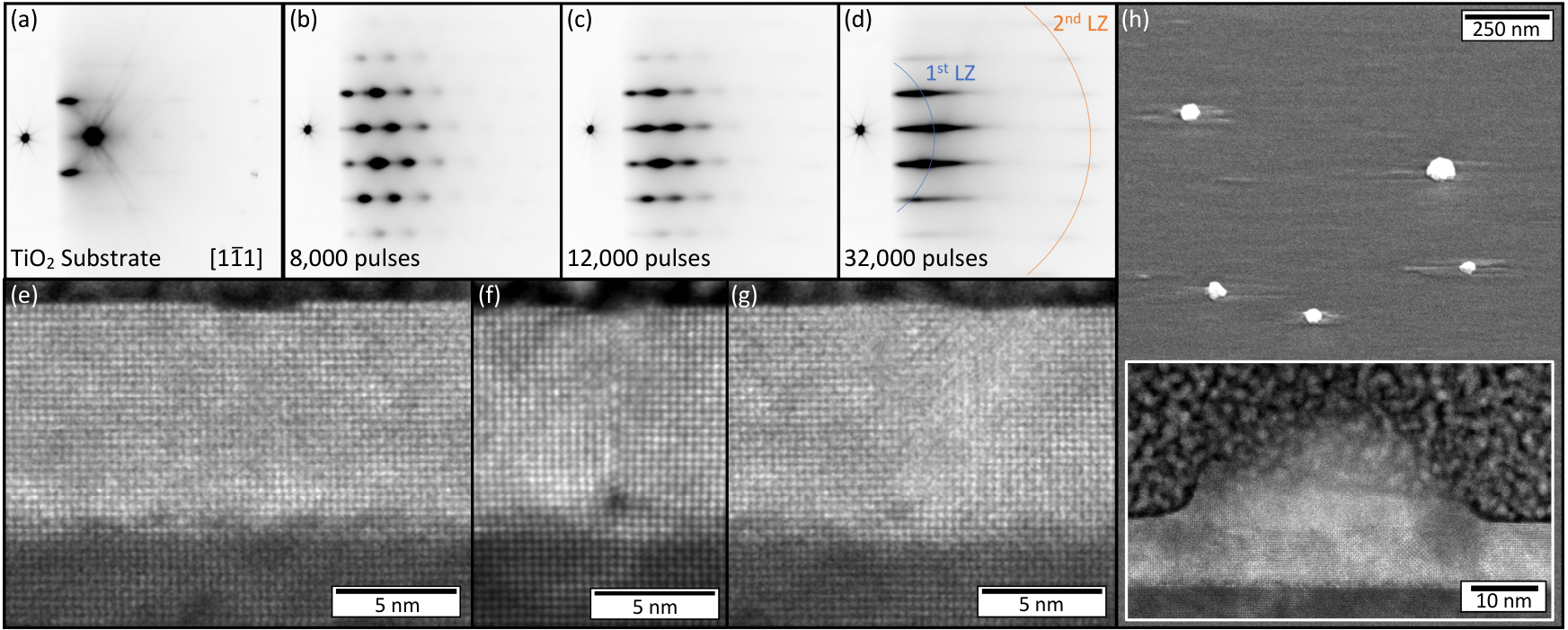}
\caption{Characterization of the island merging mechanism in RuO$_2$. RHEED images of (a) the clean TiO$_2$ substrate, (b) the RuO$_2$ islands after $8,000$ pulses, (c) the RuO$_2$ islands merging after $12,000$ pulses and (d) the closed RuO$_2$ film after $32,000$ pulses. The \nth{1} and \nth{2} order Laue zones (LZ) are marked in blue and orange. STEM measurements of single-crystalline film section (e) without defects, (f) with edge dislocation that compensates the lattice mismatch and (g) with disorder in the island merging zone. (h) SEM image show a flat RuO$_2$ surface with occasional RuO$_2$ clusters on top. The inset is a STEM cross-section through on of these clusters.}\label{fig:merging}
\end{figure*}

\subsection{Material flux control}
\label{Material Flux}

Based on these targets, we then optimized the material flux through control of the laser fluency on the target. As noted in the introduction and the paragraph above, a serious technical challenge in the PLD growth of RuO$_2$ and IrO$_2$ thin films lies in the decrease of material deposition rate despite a constant laser fluency. This is mostly due to the consecutive oxygen depletion of the target upon ablation, resulting in an increase in target metallicity/reflectivity (at 248 nm from 22\% on IrO$_2$ to 70\% on Ir, and from 20\% on RuO$_2$ to 60\% on Ru),\cite{Choi2006a,Goel1981} and a concomitant decrease in energy absorption and material flux during the film growth (Fig. 2 (a)). This change in material flux can be observed visually as a considerable shrinkage of the plasma plume during ablation: Applying $2,000$ laser pulses at standard growth parameters (p(O$_2$) = \SI{1e-3}{mbar}, J = \SI{0.7}{J/cm^2}) to a RuO$_2$ target, e.g., changes the plasma plume from being extended over a few cm in Fig. \ref{fig:Target} (c) to being barely visible in Fig. \ref{fig:Target} (d).

Aiming at saturating the target reflectivity and thus achieving a more uniform material flux, we pre-ablated the fresh target by $2,000$ laser pulses at closed shutter before the actual growth was started. As we observed a significant increase in target reflectivity and subsequent decrease of the material flux even after this step, we turned to gradually increasing the laser energy density during growth to compensate for the target depletion. Reasonable material flux stability was achieved by a step-wise increase of the laser energy density from \SI{0.7}{J/cm^2} to \SI{1.4}{J/cm^2}, using steps of \SI{0.1}{J/cm^2} after every $2,000$ laser pulses. For RuO$_2$, this resulted in a total amount of $16,000$ laser pulses until the \SI{37.5}{mm^2} scanning area was fully consumed. While this pre-ablation and subsequent ramp-up protocol during growth lends itself to future optimization, we note that, in general, the maximum film thickness achievable by PLD is mostly limited by the available target surface area and the maximum laser fluency. The overall control of material flux is, however, rather limited.

\begin{figure*}[!hbpt]
\includegraphics[width = \linewidth]{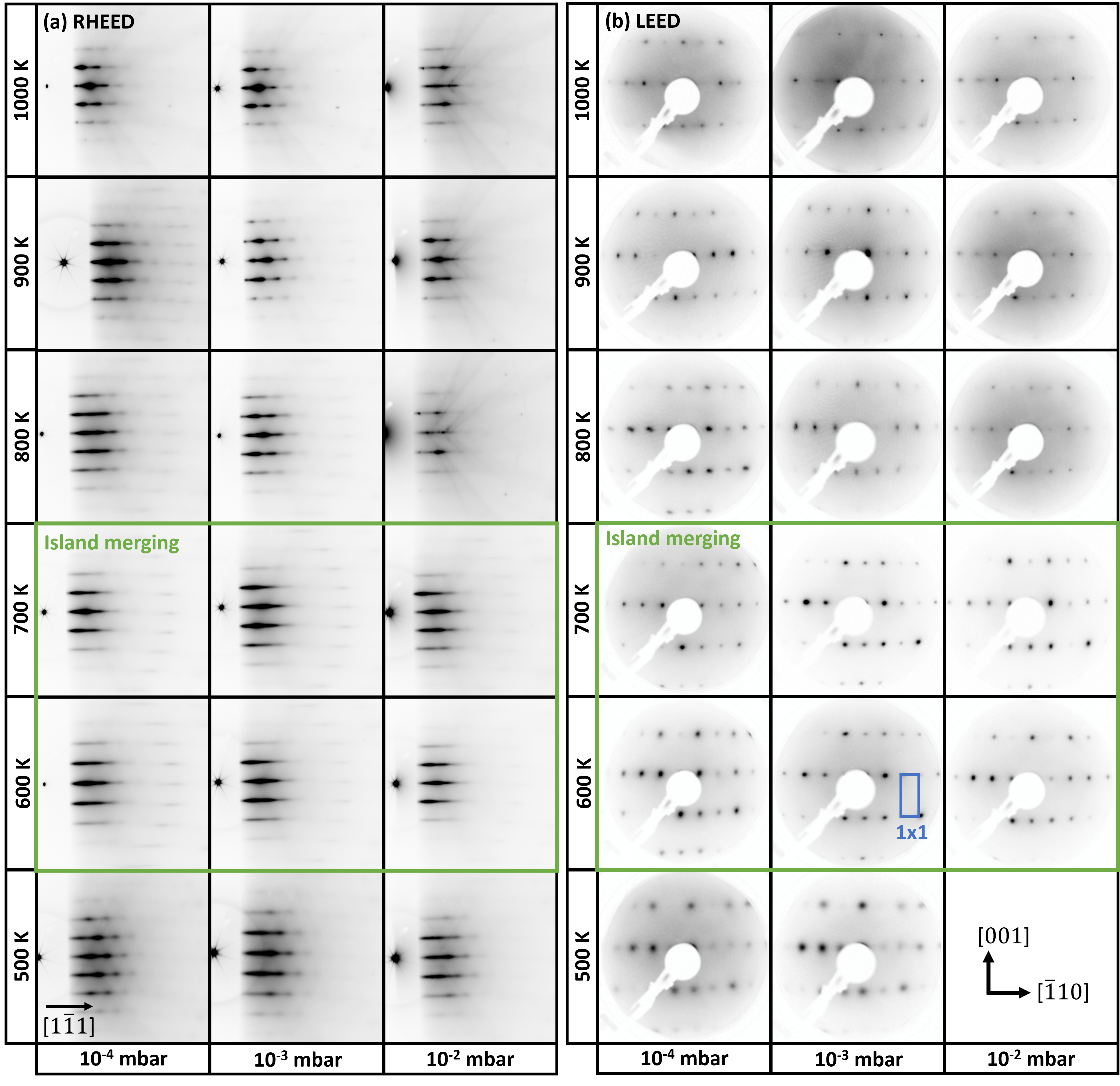}
\caption{\label{fig:RHEED_overview} (a) RHEED and (b) LEED images of as-grown RuO$_2$ films showing the influence of substrate temperature and oxygen partial pressure on surface crystallinity. The optimal temperature/pressure regime where island merging can be realized is marked in green. At \SI{500}{K} and \SI{1e-2}{mbar}, no LEED was measured due to setup maintenance.}
\end{figure*}

A second but well known reason for a decreased material flux is the deposition of ablated material onto the deposition laser entry window. The deposited material hereby acts as a low transmissivity coating that reduces the laser energy density at the target. To determine the reduction of material flux by this coating process, we initially cleaned the laser entry window and then successively grew two RuO$_2$ films at identical growth parameters of \SI{700}{K} substrate temperature and \SI{1e-3}{mbar} oxygen partial pressure, both times applying a total amount of $64,000$ laser pulses per film while scanning across four fresh \SI{37.5}{mm^2} target areas and increasing the laser energy density as described above. Although the growth parameters for both films were thus virtually identical, an analysis of XRR data yielded a film thickness of \SI{30\pm1}{nm} for the first film, and \SI{22\pm1}{nm} for the second film (Fig. \ref{fig:coating} (a)). This \SI{25}{\percent} reduction of film thickness indicates a severe loss in transparency of the laser entry window, implying a notorious difficulty to reproduce identical film thicknesses if the transparency of the laser entry window is not reset before each individual film growth. Interestingly, the reduction of laser fluency due to the window coating also induces a change in the stoichiometry of the RuO$_2$ films, as indicated by the slight high energy shift of the Ru 3d XPS core level spectra of Fig. \ref{fig:coating} (b,c). We speculate a reduced removal of oxygen from the target at lower laser energy densities to cause this slight reduction in metallicity, i.e., core level screening.


\section{Epitaxial film growth of \protect\Ru}

\subsection{Island merging in RuO$_2$}
\label{sec:Ru_merging}

As demonstrated by He et al. via PVD,\cite{He2015} one possible method to grow closed epitaxial thin films of RuO$_2$ on TiO$_2$ substrates is through island merging. Here we show that this growth mode can also be realized by PLD of RuO$_2$. The individual phases of the growth process are visualized by RHEED in Fig. \ref{fig:merging} (a)-(d). Starting from a TiO$_2$(110) substrate that was prepared according to the protocol of Sec. \ref{sec:Substrate} in Fig. \ref{fig:merging} (a) (corresponding AFM in Fig. \ref{fig:sub_prep}), initial material deposition up to about $8,000$ laser pulses results in the formation of islands, as visualized by the stripy RHEED pattern with distinct 3D transmission spots in Fig. \ref{fig:merging} (b). Further deposition of material results in the coalition of these islands, indicated by a transition of the 3D transmission spots into a modulated streak pattern at $12,000$ laser pulses (Fig. \ref{fig:merging} (c)). After $32,000$ laser pulses, the intensity modulations disappear and RHEED in Fig. \ref{fig:merging} (d) displays the characteristic streak pattern of a closed film along with higher order diffraction spots in the second Laue zone.\footnote{In case this procedure does not produce a closed film at the reported laser energy density, we suggest to initially vary this parameter by several \SI{100}{mJ/cm^2}, as optical parameters can vary strongly between PLD setups.} Indeed, exemplary STEM (Figs. \ref{fig:merging} (e)-(g)) and SEM (Figs. \ref{fig:merging} (h)) suggest a low surface corrugation of the final film on the order of only a few atoms, and a flat surface without trenches or residual islands, with occasional surface clusters (Fig. \ref{fig:merging} (h) inset) that result from the deposition process. The STEM images of Fig. \ref{fig:merging} (f) and (g), however, also reveal the drawback of this growth mode: In the merging zone of the initial islands, the compensation of the lattice mismatch leads to the formation of line defects (Fig. \ref{fig:merging} (f)) and disordered patches in the crystal lattice (Fig. \ref{fig:merging} (g)).

\begin{figure}[!htbp]
\includegraphics[width = 1\columnwidth]{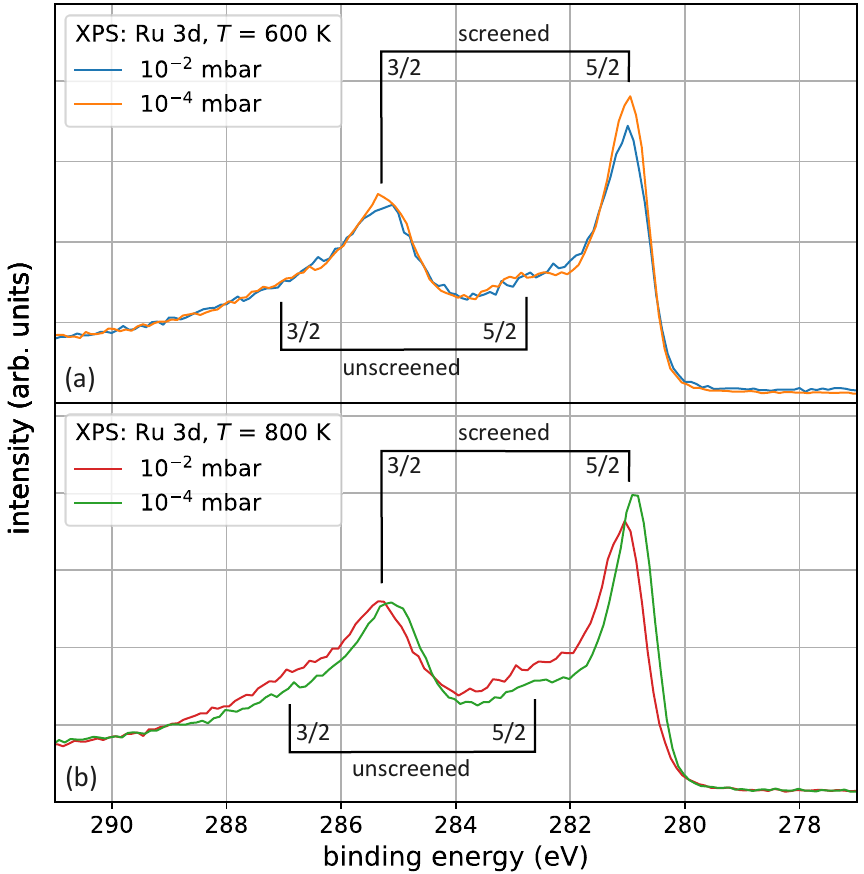}
\caption{\label{fig:XPS_pd} Ru 3d core level spectra measured by XPS for RuO$_2$ films grown with different oxygen partial pressures at (a) \SI{600}{K} and (b) \SI{800}{K}, respectively.}
\end{figure}

\begin{figure*}[!bthp]
\includegraphics[width = \linewidth]{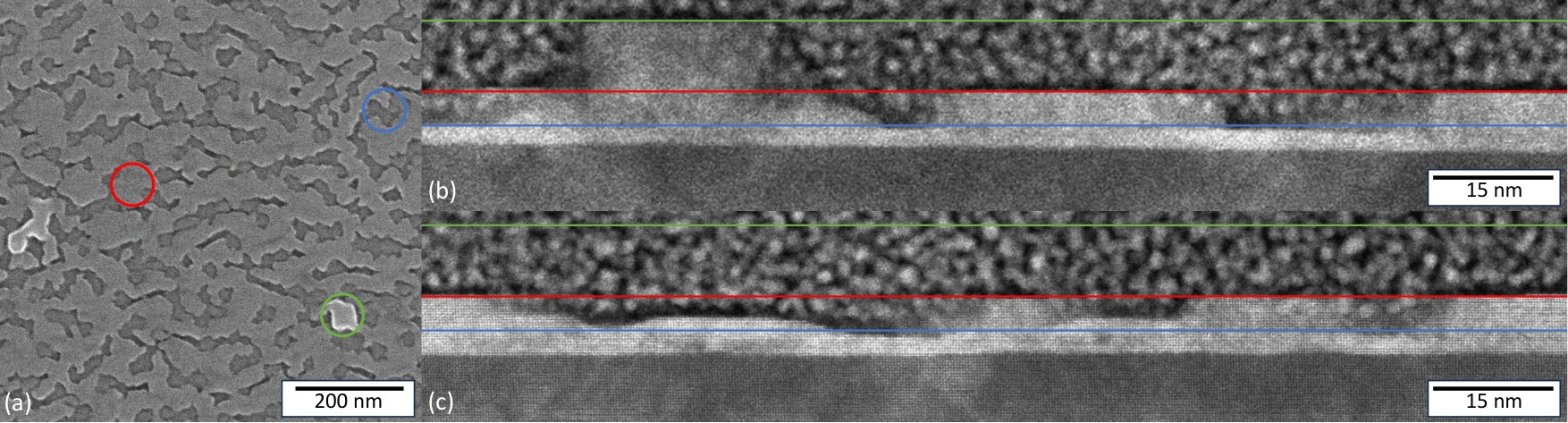}
\caption{Trenched surface morphology of RuO$_2$ films grown above \SI{700}{K}. (a) SEM image showing the trench structure and two marked regions of characteristic film thicknesses. (b,c) STEM images taken at two laterally separated parts of one sample, exhibiting three characteristic RuO$_2$ film thicknesses of \SI{6} {u.c.} (blue), \SI{20}{u.c.} (red) and \SI{43}{u.c.} (green). 
\label{fig:stable}}
\end{figure*}

\subsection{Temperature and pressure dependent RuO$_2$ nucleation}
\label{subsec: Substrate temperature and oxygen partial pressure}

Having outlined how to achieve closed epitaxial RuO$_2$ films via island merging, we now systematically study the influence of oxygen partial pressure and substrate temperature on RuO$_2$ growth in a wide parameter range from \SI{1e-4}{mbar} to \SI{1e-2}{mbar} and \SI{500}{K} to \SI{1000}{K}. All samples in this section were produced by applying a total of $16,000$ laser pulses subject to the previously defined laser energy density gradient (Sec. \ref{Material Flux}) on two fresh \SI{37.5}{mm^2} RuO$_2$ target patches. The growth order was disarrayed, i.e., samples located next to each other in the growth parameter space were not grown consecutively. In this way, we were able to distinguish intrinsic trends from systematic errors, introduced, e.g., by a proceeding metallization of the laser entry window, a changing material flux or other similar factors. The RHEED and LEED images of the as-grown films are displayed in Fig. \ref{fig:RHEED_overview} (a) and (b), respectively.

Substrate temperatures of \SI{600}{K} and \SI{700}{K} result in the island merging growth mode as discussed in Sec. \ref{sec:Ru_merging} irrespective of the oxygen partial pressure. RHEED in this temperature range (green) indicates the characteristic signature of a closed film with a streaked diffraction pattern and no transmission spots (Fig. \ref{fig:RHEED_overview} (a)), while LEED shows sharp diffraction spots and a low background intensity, indicating relatively high surface order (Fig. \ref{fig:RHEED_overview} (b)).

Increasing the substrate temperature to \SI{800}{}-\SI{1000}{K} increases the mobility of the deposited material, resulting in island formation and a trenched surface. RHEED in Fig. \ref{fig:RHEED_overview} (a) now exhibits 3D transmission spots, while LEED in Fig. \ref{fig:RHEED_overview} (b) exhibits increased background intensity indicating an increased density of surface defects. In contrast, the diffraction spots are generally still sharp, suggesting the island terraces to still order locally. 

Decreasing the substrate temperature to \SI{500}{K} lowers the kinetic energy of the deposited material and again fosters island formation, yet suppresses the capability of island merging and a subsequent step-flow growth mode. RHEED in Fig. \ref{fig:RHEED_overview} now again shows the characteristic 3D transmission spots, while LEED in Fig. \ref{fig:RHEED_overview} (b) shows high background intensity and broad diffraction peaks indicative of disordered island terraces. 

While the RHEED and LEED signatures of Figs. \ref{fig:RHEED_overview} (a) and (b) indicate the growth mode to be predominantly impacted by temperature, the oxygen partial pressure mostly influences the number and density of defects. Thus, while the augmented streakiness of RHEED patterns at higher O$_2$ pressure in Fig. \ref{fig:RHEED_overview} (a) reveals flatter films, corresponding LEED images in Fig. \ref{fig:RHEED_overview} (b) tend towards sharper spots, less overall background intensity and consequently higher surface order. We attribute this to an improved (surface) stoichiometry rationalized by the XPS in Fig. \ref{fig:XPS_pd}:  Irrespective of the growth temperature, the ratio between the unscreened (i.e. oxidized) and the screened (i.e. metallic) Ru 3d core level peaks increases for higher O$_2$ pressure, suggesting a reduced amount of oxygen deficiencies.\cite{Kim2004} However, we overall observe a lower material deposition rate at higher oxygen partial pressures, which is likely a result of Ru and RuO$_2$ over-oxidation to volatile RuO$_4$.\cite{Green1985a}

\subsection{Characteristic thicknesses of RuO$_2$}
\label{sec:stable_thickness}

As discussed above, RuO$_2$ films grown at substrate temperatures above \SI{700}{K} show a trenched surface morphology, likely related by the interplay of augmented material diffusion, surface- and volume energies as well as substrate induced strain. 
While exemplary SEM and STEM images in Fig. \ref{fig:stable} show these trenches to be randomly distributed across the sample and to not orient along preferred crystal directions, the remaining film still terminates in a flat and well defined fashion. In particular, the STEM images show the appearance of characteristic film thicknesses, namely \SI{6} {u.c.} (blue), \SI{20}{u.c.} (red) and \SI{43}{u.c.} (green) as marked by the colored horizontal lines in Figs. \ref{fig:stable} (b) and (c). As these exact thicknesses were found repeatedly irrespective of the sample or sample location, they seem to reflect a universal property of the RuO$_2$/TiO$_2$(110) interface. While we are not aware of reports of such a growth behaviour in literature, we would still like to point out certain similarities to results presented in Refs. \onlinecite{Dahal2013,Krizek2020,Yang2017a,Petit2007}.

We speculate that this tendency of RuO$_2$ to form islands of fixed thicknesses might directly compete with the tendency of RuO$_2$ to release lattice strain by forming islands that eventually coalesce and merge. The former mechanism would thus become more prevalent at higher growth temperatures, where surface diffusion is enhanced. This interpretation might be the reason why above \SI{700}{K}, island merging of RuO$_2$ could not be observed and uniformly thick closed films were unachievable.

\subsection{Layer-by-layer growth of RuO$_2$}
\begin{figure}[!htbp]
\includegraphics[width = 1\columnwidth]{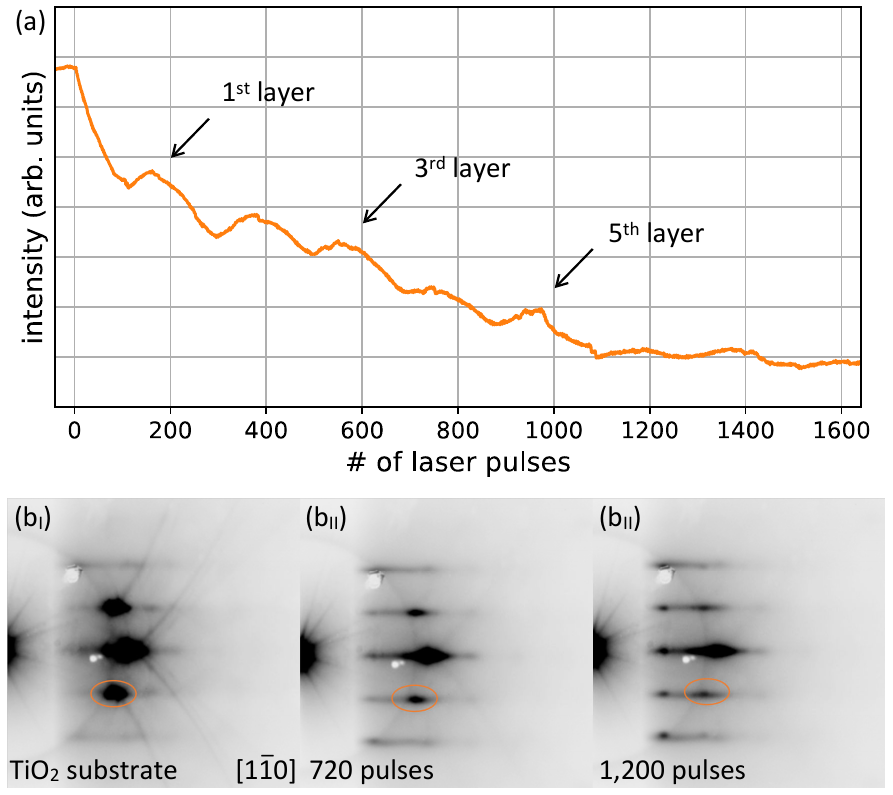}
\caption{Layer-by-layer growth of RuO$_2$. (a) RHEED oscillations can be observed up to the \nth{5} layer. (b) RHEED of the substrate (b$_{\text{I}}$), during Layer-by-layer growth (b$_{\text{II}}$) and after the formation of islands (b$_{\text{III}}$). The integration region for RHEED oscillations in (a) is marked in orange.\label{fig:layer} }
\end{figure}

Finally, we studied a growth regime where RuO$_2$(110) on TiO$_2$(110) grows layer-by-layer, allowing for the growth of closed, fully strained films up to \SI{2}{nm} thickness. This growth process is achieved if surface diffusion is greatly reduced, i.e., if the RuO$_2$ material flux is massively increased during the nucleation phase (laser energy density: \SI{2.4}{J/cm^2}, pre-ablation: $2,000$ pulses, oxygen partial pressure: \SI{1e-3}{mbar}, substrate temperature: \SI{700}{K}). As seen in Fig. \ref{fig:layer} (a), we now observe RHEED intensity oscillations and a transmission spot free diffraction pattern up to the \nth{5} layer at this laser energy density. A continuation of the growth leads to a sharp drop in the RHEED intensity, accompanied by the appearance of 3D transmission spots resulting from island growth taking over in Fig. \ref{fig:layer} (b).

\section{Epitaxial film growth of \protect\Ir}

\begin{figure*}[!htbp]
\includegraphics[width = \linewidth]{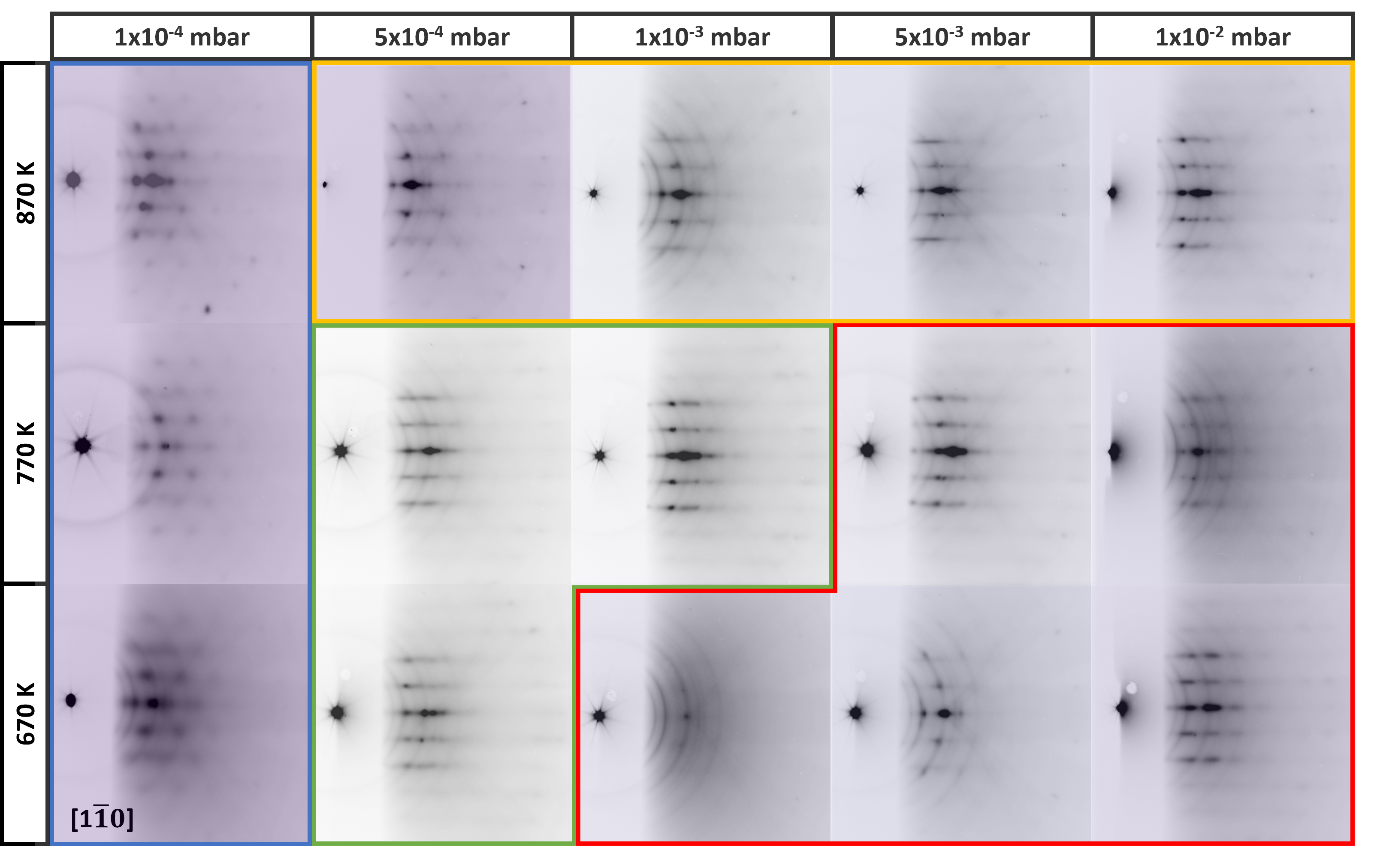}
\caption{Dependence of film quality on substrate temperature and oxygen partial pressure during growth. Shown are RHEED images of as-grown IrO$_2$ films terminated according to the stopping criterion defined in the main text. The Ir/IrO$_2$ ratio of these films as determined from the spectral decomposition of the Ir 4f peak according to Fig. \ref{fig:Ir_XPS} was used to color code the respective RHEED panels, where a dark blue represents high Ir/IrO$_2$ ratio (i.e., non-stoichiometric \ce{IrO2} films with a high \ce{Ir} concentration) and a light blue shading represents a low Ir/IrO$_2$ ratio (i.e., close to stoichiometric \ce{IrO2} films with a low \ce{Ir} concentration). The individual growth categories discussed in the text are marked by colored frames.}
\label{fig:Ir_param}
\end{figure*}

\subsection{Temperature and pressure dependent IrO$_2$ nucleation}

\begin{figure}[!htbp]
\includegraphics[width = \columnwidth]{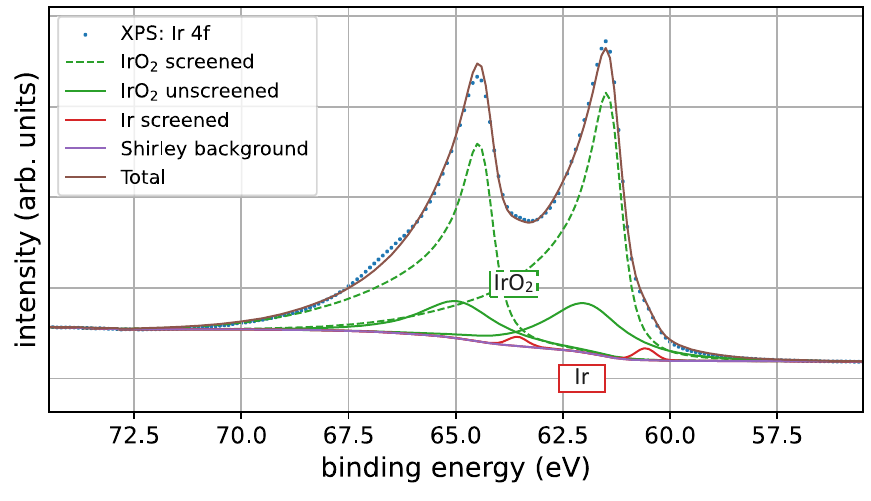}
\caption{\label{fig:Ir_XPS} Decomposition of an exemplary Ir 4f spectrum into IrO$_2$ and metallic Ir contributions. To determine the Ir/IrO$_2$ ratio, these spectra were corrected for the transmission function of the electron spectrometer and a Shirley background was subtracted.}
\end{figure}

As the \SI{6.8}{\percent} lattice mismatch between IrO$_2$(110) and TiO$_2$(110) along the [$00\bar1$] direction is considerably larger than the \SI{4.9}{\percent} between RuO$_2$(110) and TiO$_2$(110),\cite{Jain2013} the tendency of IrO$_2$ to form islands and disorder is more pronounced than for RuO$_2$. Further, while RuO$_2$ samples grown by PLD are essentially stoichiometric throughout all growth parameter regimes investigated in this study, IrO$_2$ films grown under similar conditions always contain considerable amounts of un- or underoxidized iridium, a consequence of its more ’stubborn’ oxidation behaviour as compared to ruthenium.\cite{Nair2023}

To cope with these differences between the PLD growth of RuO$_2$ and IrO$_2$ and to better understand and control the nucleation phase of IrO$_2$ on TiO$_2$(110), we employed a stopping criterion that allows for an objective comparison of \ce{IrO2} films grown at different substrate temperatures and oxygen partial pressures. This stopping criterion was reached once the intensity of a newly developed RHEED feature, such as a diffraction spot or a diffraction streak, surpassed \SI{5}{\percent} of the brightest initial substrate feature. The RHEED patterns of IrO$_2$ films grown systematically at oxygen partial pressures between \SI{1e-4}{mbar} and \SI{1e-2}{mbar}, substrate temperatures between \SI{670}{K} and \SI{870}{K}, and a laser energy density of \SI{1.1}{J/cm^2} are shown in Fig. \ref{fig:Ir_param}. Like with RuO$_2$, The growth order of IrO$_2$ films was disarrayed to avoid confusing intrinsic trends in the growth parameter space with systematic errors.

To further check for reproducibility of our results, we repeated the sample growth exemplarily at parameters \SI{5e-4}{mbar} / \SI{870}{K}, \SI{1e-4}{mbar} / \SI{770}{K} and \SI{5e-4}{mbar} / \SI{770}{K}, each yielding identical outcomes as compared to the first attempt. To benchmark the growth parameters with respect to the ’stubborn’ oxidation behaviour of \ce{Ir}, we used XPS to determine the \ce{Ir}/\ce{IrO2} ratio of every film. Hereby, the corresponding Ir 4f core-level spectra were decomposed as exemplified in Fig. \ref{fig:Ir_XPS}. After subtraction of a Shirley background and assuming a fixed intensity ratio of 4:3, the two Ir$^{4+}$ 4f 7/2 and 5/2 peaks were fitted to asymmetric Voigt-like line-shapes as provided by CasaXPS.\cite{Martin2020,Freakley2017,Fairley2021} The Ir/IrO$_2$ ratios obtained in this way were used to color code the respective RHEED images of Fig. \ref{fig:Ir_param}. 

\begin{figure*}[!htbp]
\includegraphics[width = \linewidth]{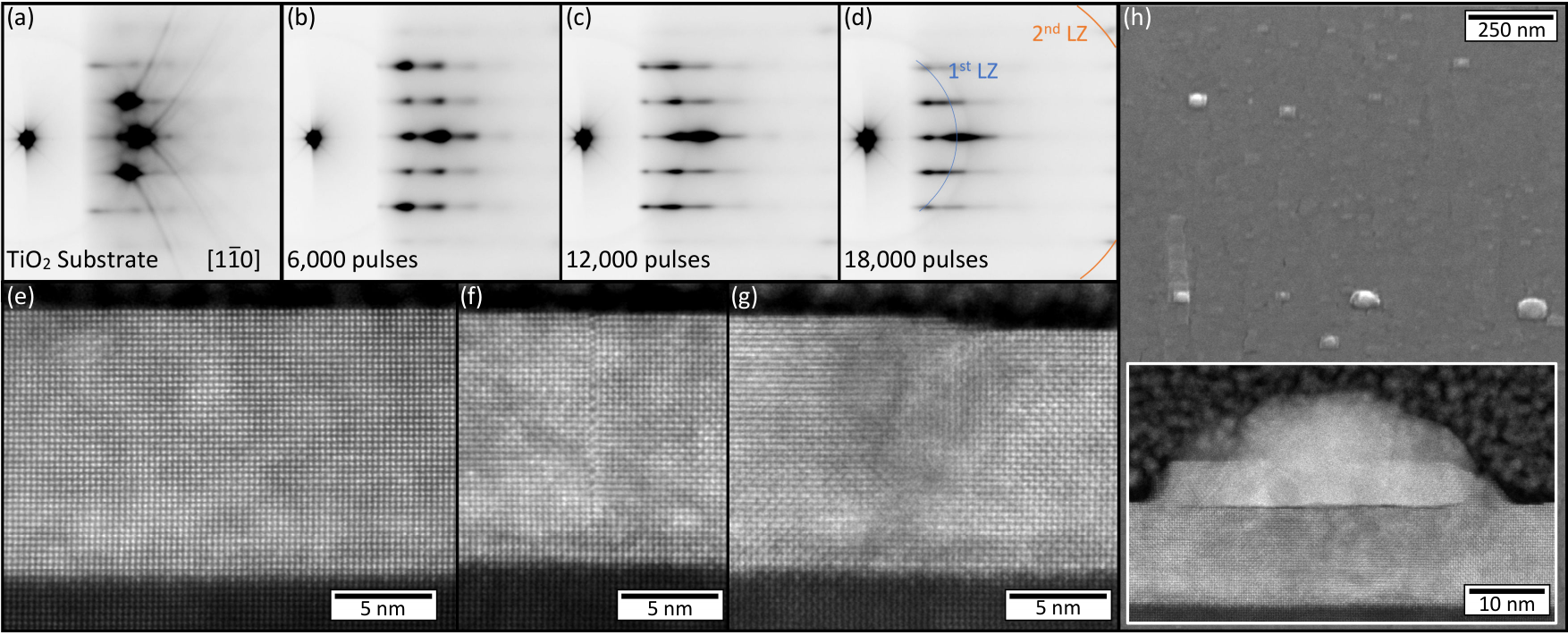}
\caption{Characterization of the island merging mechanism in IrO$_2$. RHEED images of (a) the clean TiO$_2$ substrate, (b) the IrO$_2$ islands after $6,000$ pulses, (c) the IrO$_2$ islands merging after $12,000$ pulses and (d) the closed IrO$_2$ film after $18,000$ pulses. The \nth{1} and \nth{2} order Laue zones (LZ) are marked in blue and orange, respectively. STEM measurements of a single-crystalline film section (e) without defects, (f) with edge dislocation that compensates the lattice mismatch and (g) with disorder in the island merging zone. (h) SEM image showing a flat IrO$_2$ surface with occasional Ir clusters on top. The inset is a STEM cross-section through one of these clusters.
\label{fig:Ir_merging}}
\end{figure*}


In addition to monitoring the \ce{Ir}/\ce{IrO2} ratio, we also analyzed the individual RHEED pattern of every film and identified four qualitatively different growth regimes, marked by the frame color in Fig. \ref{fig:Ir_param}:
The first category is indicated in red and represents disordered IrO$_2$ films grown at high oxygen partial pressures and low substrate temperatures. In this regime, the kinetic energy of the laser ablated particles is diminished by collisions in the oxygen gas, and diffusion on the substrate surface is limited by temperature.\cite{Ma2016,Shepelin2023}
The diffusion of the deposited material is thus insufficient to achieve overall crystalline ordering, resulting in a RHEED pattern that is a superposition of Debye rings and 3D transmission spots as a consequence of disordered growth and island formation.\cite{Nair2023}

The second category is marked in orange and represents IrO$_2$ films grown at high substrate temperatures. This growth regime is characterized by a competition between the formation of higher, volatile oxides of iridium such as IrO$_3$,\cite{Liu2004a}
and the thermal reduction of IrO$_2$ to metallic Ir,\cite{Abb2020a}
which is controlled by the oxygen partial pressure. Thus, at high oxygen partial pressures, over-oxidation of metallic Ir as well as IrO$_2$ to volatile IrO$_3$ dominates.\cite{Cordfunke1962,Chalamala1999,Liu2004a,Liu2004}
This limits the amount of deposited material and the growth rate, but keeps the Ir/IrO$_2$ ratio generally low and yields a relatively well ordered, flat surface due to augmented surface diffusion. At lower oxygen partial pressures, the desorption rate of oxygen is enhanced and reduction of IrO$_2$ takes over,\cite{Abb2020a}
leading to disorder and island growth reflected in 3D transmission spots in RHEED and the highest Ir/IrO$_2$ ratio within the explored parameters space.

This region overlaps with the third growth regime that is marked in blue and represents IrO$_2$ films grown at the lowest oxygen partial pressure. In contrast to the orange category, these samples exhibit a hexagonal RHEED pattern, i.e., a superposition of Debye rings and 3D transmission spots, indicating polycrystalline growth and island formation. Due to the lack of oxygen, oxidation of the deposited material is limited, leading to Ir/IrO$_2$ ratios well beyond 15 \% at high temperatures.

The best film quality was achieved for films within the green category, i.e., films grown at moderate substrate temperatures and oxygen partial pressures, where oxidation and surface diffusion are balanced out. RHEED displays a streaky pattern indicating relatively flat films, while XPS shows that these are essentially stoichiometric with Ir/\ce{IrO2} ratios below \SI{5}{\percent}.

\subsection{Island merging in IrO$_2$}

The last section has shown that the optimal growth parameter set to achieve flat epitaxial IrO$_2$ films of appreciable thickness requires a fine balance between diffusion of the deposited material, Ir (over-)oxidation and IrO$_2$ reduction. In particular, while the sample stoichiometry is generally better for higher oxygen partial pressures, film deposition rate and hence thickness is limited due to the formation of volatile IrO$_3$. As optimal parameter set, a temperature of \SI{770}{K} and an oxygen partial pressure of \SI{5e-4}{mbar} was found, which lends itself as an ideal starting point to initiate the growth of closed IrO$_2$ thin films.

As shown in Fig. \ref{fig:Ir_merging} and similar to what was observed for PLD and PVD\cite{He2015} of RuO$_2$, RHEED shows an initially pristine \ce{TiO2} substrate (a) followed by the formation of islands with 3D transmission spots (b), which eventually coalescence (c) to form a closed, smooth film (d). Like for RuO$_2$, high resolution STEM measurements in Fig. \ref{fig:Ir_merging} (e) reveal a well defined surface termination, with occasional line defects to cope with strain as seen in (f), and with an increased amount of disorder close to the merging zone as shown in (g). An SEM image in Fig. \ref{fig:Ir_merging} (h) further show a flat surface with tiny trenches, interrupted by occasional sprinkles and larger clusters of Ir, for which a STEM cross-section is shown in the inset.

\begin{figure}[!htbp]
\includegraphics[width = \columnwidth]{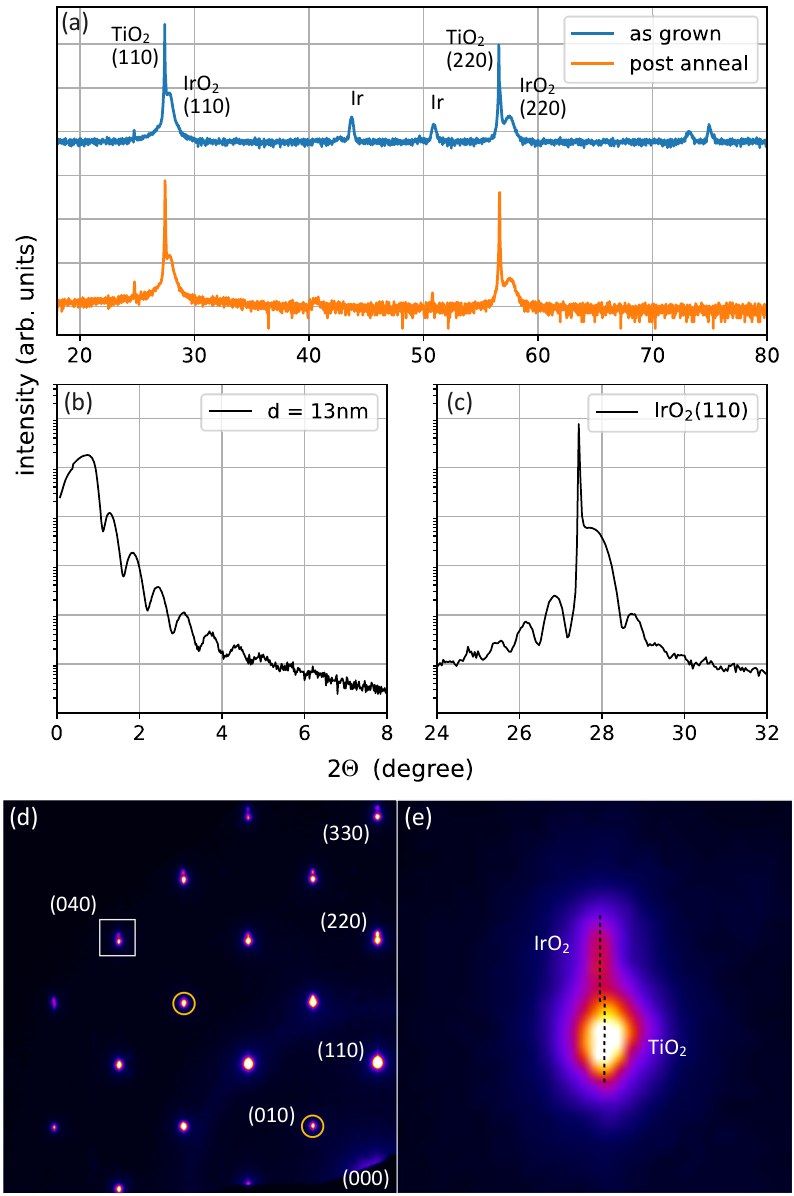}
\caption{
Characterization of an IrO$_2$ film after island merging. (a) Overview XRD scan showing the presence of an Ir phase in addition to the IrO$_2$(110) film. The Ir phase disappears after annealing the sample for 90 min at 450$^\circ$C in oxygen. (b) XRR data and (c) Laue oscillations around the IrO$_2$(110) diffraction peak yield a film thickness of \SI{13}{nm}. (d) Nano electron diffraction data obtained from STEM indicates a good crystalline order. (e) The slight lateral offset of the IrO$_2$ vs the TiO$_2$ substrate (040) peak reveals the film to be partially relaxed.
\label{fig:Ir_merging_XRD}}
\end{figure}

Consistent with the local structural information by STEM in Fig. \ref{fig:Ir_merging} (e-g), the Kiessig fringes in the XRR and the Laue-oscillations in the XRD data of Fig. \ref{fig:Ir_merging_XRD} (b) and (c) suggests an excellent crystalline quality and morphology across the entire film. Analyzing their periodicity, we find an average thickness of \SI{13}{nm} in good agreement with STEM. The XRD overview in Fig. \ref{fig:Ir_merging_XRD} (a) further reveals an unwanted partial film coverage with metallic iridium, consistent with our XPS study above and SEM and STEM images in Fig. \ref{fig:Ir_merging} (h), both suggesting that iridium clusters are mostly located at the surface. In contrast, two peaks at \SI{73.2}{\degree} and \SI{75.1}{\degree} remain unidentified. Quite notably, however, both the iridium as well as the unidentified peaks disappear after the sample is annealed for 90 min at 450$^\circ$C with an oxygen flow of 25 l/h (Fig. \ref{fig:Ir_merging_XRD} (a)).

Finally, let us turn to nano electron diffraction data obtained from STEM on the \SI{13}{nm} thick film of Fig. \ref{fig:Ir_merging_XRD} (d), showing an overall well defined film-substrate-interface. A close-up of the (040) diffraction spot in panel (e) shows a lateral offset between the IrO$_2$ and TiO$_2$ peaks, indicating a partial relaxation of the IrO$_2$ film. Note that higher order diffraction further gives rise to otherwise forbidden peaks marked by the yellow circles in Fig. \ref{fig:Ir_merging_XRD} (d).

\section{Summary and Conclusion}

In summary, we systematically investigated the PLD growth of epitaxial RuO$_2$(110) and IrO$_2$(110) films on TiO$_2$(110) substrates for a variety of substrate temperatures and oxygen partial pressures. By applying a temporal laser energy density gradient, we accounted for a decreasing material flux as a consequence of deteriorating target absorbance and an increasing coverage of the laser entry window. We find the film crystallinity and stoichiometry of RuO$_2$ to depend only mildly on the oxygen partial pressure during growth, which we varied in the range between \SI{1e-2}{mbar} to \SI{1e-4}{mbar}, with the growth rate being significantly reduced at higher pressures. The film quality, however, depends sensitively on substrate temperature. Low surface diffusion below \SI{500}{K} leads to a disordered surface and the formation of volatile RuO$_4$ above \SI{800}{K} promoting the formation of RuO$_2$ islands. The best surface quality was found for temperatures between \SI{600}{K} and \SI{700}{K}, yielding layer-by-layer growth of up to 5 unit cells and fully strained films for high deposition rates, and thicker, closed films through island merging at low deposition rates. 

In contrast to RuO$_2$, the IrO$_2$ film growth shows a pronounced dependence on both substrate temperature and oxygen partial pressure. We attribute this to the stronger resilience of Ir against oxidation and the tendency of IrO$_2$ to overoxidize to volatile IrO$_3$. Growth rates thus are generally lower than for RuO$_2$ and samples tend to exhibit considerable fractions of metallic iridium, the latter can be removed by oxygen post annealing. Minimal amounts of metallic Ir contents have been achieved at oxygen pressures of around \SI{5e-4}{mbar} and temperatures around \SI{770}{K}. In this setting, the initial island formation turns into a phase where the islands coalesce to form a closed, flat film.

\begin{acknowledgments}
We thank Hiroshi Kumigashira for supplying us with the initial RuO$_2$ target. Funding support came from the Deutsche Forschungsgemeinschaft (DFG, German Research Foundation) under Germany’s Excellence Strategy through the Würzburg-Dresden Cluster of Excellence on Complexity and Topology in Quantum Matter ct.qmat (EXC 2147, Project ID 390858490) and through the Collaborative Research Center SFB 1170 ToCoTronics (Project ID 258499086), as well as from the New Zealand Ministry of Business, Innovation and Employment (MBIE, Grant number: C05X2004). 

The following article has been submitted to APL Materials.  After it is published, it will be found at \href{https://publishing.aip.org/resources/librarians/products/journals/}{\nolinkurl{AIP Publishing}}

\end{acknowledgments}


\begin{thebibliography}{120}%
\makeatletter
\providecommand \@ifxundefined [1]{%
 \@ifx{#1\undefined}
}%
\providecommand \@ifnum [1]{%
 \ifnum #1\expandafter \@firstoftwo
 \else \expandafter \@secondoftwo
 \fi
}%
\providecommand \@ifx [1]{%
 \ifx #1\expandafter \@firstoftwo
 \else \expandafter \@secondoftwo
 \fi
}%
\providecommand \natexlab [1]{#1}%
\providecommand \enquote  [1]{``#1''}%
\providecommand \bibnamefont  [1]{#1}%
\providecommand \bibfnamefont [1]{#1}%
\providecommand \citenamefont [1]{#1}%
\providecommand \href@noop [0]{\@secondoftwo}%
\providecommand \href [0]{\begingroup \@sanitize@url \@href}%
\providecommand \@href[1]{\@@startlink{#1}\@@href}%
\providecommand \@@href[1]{\endgroup#1\@@endlink}%
\providecommand \@sanitize@url [0]{\catcode `\\12\catcode `\$12\catcode
  `\&12\catcode `\#12\catcode `\^12\catcode `\_12\catcode `\%12\relax}%
\providecommand \@@startlink[1]{}%
\providecommand \@@endlink[0]{}%
\providecommand \url  [0]{\begingroup\@sanitize@url \@url }%
\providecommand \@url [1]{\endgroup\@href {#1}{\urlprefix }}%
\providecommand \urlprefix  [0]{URL }%
\providecommand \Eprint [0]{\href }%
\providecommand \doibase [0]{http://dx.doi.org/}%
\providecommand \selectlanguage [0]{\@gobble}%
\providecommand \bibinfo  [0]{\@secondoftwo}%
\providecommand \bibfield  [0]{\@secondoftwo}%
\providecommand \translation [1]{[#1]}%
\providecommand \BibitemOpen [0]{}%
\providecommand \bibitemStop [0]{}%
\providecommand \bibitemNoStop [0]{.\EOS\space}%
\providecommand \EOS [0]{\spacefactor3000\relax}%
\providecommand \BibitemShut  [1]{\csname bibitem#1\endcsname}%
\let\auto@bib@innerbib\@empty
\bibitem [{\citenamefont {He}\ \emph {et~al.}(2015)\citenamefont {He},
  \citenamefont {Langsdorf}, \citenamefont {Li},\ and\ \citenamefont
  {Over}}]{He2015}%
  \BibitemOpen
  \bibfield  {author} {\bibinfo {author} {\bibfnamefont {Y.}~\bibnamefont
  {He}}, \bibinfo {author} {\bibfnamefont {D.}~\bibnamefont {Langsdorf}},
  \bibinfo {author} {\bibfnamefont {L.}~\bibnamefont {Li}}, \ and\ \bibinfo
  {author} {\bibfnamefont {H.}~\bibnamefont {Over}},\ }\bibfield  {title}
  {\enquote {\bibinfo {title} {{Versatile model system for studying processes
  ranging from heterogeneous to photocatalysis: Epitaxial RuO$_2$(110) on
  TiO$_2$(110)}},}\ }\href {\doibase 10.1021/jp5121405} {\bibfield  {journal}
  {\bibinfo  {journal} {Journal of Physical Chemistry C}\ }\textbf {\bibinfo
  {volume} {119}},\ \bibinfo {pages} {2692--2702} (\bibinfo {year}
  {2015})}\BibitemShut {NoStop}%
\bibitem [{\citenamefont {Over}(2012)}]{Over2012}%
  \BibitemOpen
  \bibfield  {author} {\bibinfo {author} {\bibfnamefont {H.}~\bibnamefont
  {Over}},\ }\bibfield  {title} {\enquote {\bibinfo {title} {{Surface Chemistry
  of Ruthenium Dioxide in Heterogeneous Catalysis and Electrocatalysis: From
  Fundamental to Applied Research}},}\ }\href {\doibase 10.1021/cr200247n}
  {\bibfield  {journal} {\bibinfo  {journal} {Chemical Reviews}\ }\textbf
  {\bibinfo {volume} {112}},\ \bibinfo {pages} {3356--3426} (\bibinfo {year}
  {2012})}\BibitemShut {NoStop}%
\bibitem [{\citenamefont {Weaver}(2013)}]{Weaver2013}%
  \BibitemOpen
  \bibfield  {author} {\bibinfo {author} {\bibfnamefont {J.~F.}\ \bibnamefont
  {Weaver}},\ }\bibfield  {title} {\enquote {\bibinfo {title} {{Surface
  Chemistry of Late Transition Metal Oxides}},}\ }\href {\doibase
  10.1021/cr300323w} {\bibfield  {journal} {\bibinfo  {journal} {Chemical
  Reviews}\ }\textbf {\bibinfo {volume} {113}},\ \bibinfo {pages} {4164--4215}
  (\bibinfo {year} {2013})}\BibitemShut {NoStop}%
\bibitem [{\citenamefont {Scarpelli}\ \emph {et~al.}(2022)\citenamefont
  {Scarpelli}, \citenamefont {Godbert}, \citenamefont {Crispini},\ and\
  \citenamefont {Aiello}}]{Scarpelli2022}%
  \BibitemOpen
  \bibfield  {author} {\bibinfo {author} {\bibfnamefont {F.}~\bibnamefont
  {Scarpelli}}, \bibinfo {author} {\bibfnamefont {N.}~\bibnamefont {Godbert}},
  \bibinfo {author} {\bibfnamefont {A.}~\bibnamefont {Crispini}}, \ and\
  \bibinfo {author} {\bibfnamefont {I.}~\bibnamefont {Aiello}},\ }\bibfield
  {title} {\enquote {\bibinfo {title} {{Nanostructured Iridium Oxide: State of
  the Art}},}\ }\href {\doibase 10.3390/inorganics10080115} {\bibfield
  {journal} {\bibinfo  {journal} {Inorganics}\ }\textbf {\bibinfo {volume}
  {10}},\ \bibinfo {pages} {115} (\bibinfo {year} {2022})}\BibitemShut
  {NoStop}%
\bibitem [{\citenamefont {Jang}\ and\ \citenamefont {Lee}(2020)}]{Jang2020}%
  \BibitemOpen
  \bibfield  {author} {\bibinfo {author} {\bibfnamefont {H.}~\bibnamefont
  {Jang}}\ and\ \bibinfo {author} {\bibfnamefont {J.}~\bibnamefont {Lee}},\
  }\bibfield  {title} {\enquote {\bibinfo {title} {{Iridium oxide fabrication
  and application: A review}},}\ }\href {\doibase 10.1016/j.jechem.2019.10.026}
  {\bibfield  {journal} {\bibinfo  {journal} {Journal of Energy Chemistry}\
  }\textbf {\bibinfo {volume} {46}},\ \bibinfo {pages} {152--172} (\bibinfo
  {year} {2020})}\BibitemShut {NoStop}%
\bibitem [{\citenamefont {Spöri}\ \emph {et~al.}(2017)\citenamefont {Spöri},
  \citenamefont {Kwan}, \citenamefont {Bonakdarpour}, \citenamefont
  {Wilkinson},\ and\ \citenamefont {Strasser}}]{Spoeri2017}%
  \BibitemOpen
  \bibfield  {author} {\bibinfo {author} {\bibfnamefont {C.}~\bibnamefont
  {Spöri}}, \bibinfo {author} {\bibfnamefont {J.~T.~H.}\ \bibnamefont {Kwan}},
  \bibinfo {author} {\bibfnamefont {A.}~\bibnamefont {Bonakdarpour}}, \bibinfo
  {author} {\bibfnamefont {D.~P.}\ \bibnamefont {Wilkinson}}, \ and\ \bibinfo
  {author} {\bibfnamefont {P.}~\bibnamefont {Strasser}},\ }\bibfield  {title}
  {\enquote {\bibinfo {title} {{Stabilitätsanforderungen von
  Elektrokatalysatoren für die Sauerstoffentwicklung: der Weg zu einem
  grundlegenden Verständnis und zur Minimierung der
  Katalysatordegradation}},}\ }\href {\doibase 10.1002/ange.201608601}
  {\bibfield  {journal} {\bibinfo  {journal} {Angewandte Chemie}\ }\textbf
  {\bibinfo {volume} {129}},\ \bibinfo {pages} {6088--6117} (\bibinfo {year}
  {2017})}\BibitemShut {NoStop}%
\bibitem [{\citenamefont {Weber}\ \emph {et~al.}(2022)\citenamefont {Weber},
  \citenamefont {Vonk}, \citenamefont {Abb}, \citenamefont {Evertsson},
  \citenamefont {Stierle}, \citenamefont {Lundgren},\ and\ \citenamefont
  {Over}}]{Weber2022}%
  \BibitemOpen
  \bibfield  {author} {\bibinfo {author} {\bibfnamefont {T.}~\bibnamefont
  {Weber}}, \bibinfo {author} {\bibfnamefont {V.}~\bibnamefont {Vonk}},
  \bibinfo {author} {\bibfnamefont {M.~J.}\ \bibnamefont {Abb}}, \bibinfo
  {author} {\bibfnamefont {J.}~\bibnamefont {Evertsson}}, \bibinfo {author}
  {\bibfnamefont {A.}~\bibnamefont {Stierle}}, \bibinfo {author} {\bibfnamefont
  {E.}~\bibnamefont {Lundgren}}, \ and\ \bibinfo {author} {\bibfnamefont
  {H.}~\bibnamefont {Over}},\ }\bibfield  {title} {\enquote {\bibinfo {title}
  {{In Situ Synchrotron-Based Studies of IrO$_2$(110)-TiO$_2$(110) under Harsh
  Acidic Water Splitting Conditions: Anodic Stability and Radiation
  Damages}},}\ }\href {\doibase 10.1021/acs.jpcc.2c06429} {\bibfield  {journal}
  {\bibinfo  {journal} {Journal of Physical Chemistry C}\ }\textbf {\bibinfo
  {volume} {126}},\ \bibinfo {pages} {20243--20250} (\bibinfo {year}
  {2022})}\BibitemShut {NoStop}%
\bibitem [{\citenamefont {Naito}\ \emph {et~al.}(2021)\citenamefont {Naito},
  \citenamefont {Shinagawa}, \citenamefont {Nishimoto},\ and\ \citenamefont
  {Takanabe}}]{Naito2021}%
  \BibitemOpen
  \bibfield  {author} {\bibinfo {author} {\bibfnamefont {T.}~\bibnamefont
  {Naito}}, \bibinfo {author} {\bibfnamefont {T.}~\bibnamefont {Shinagawa}},
  \bibinfo {author} {\bibfnamefont {T.}~\bibnamefont {Nishimoto}}, \ and\
  \bibinfo {author} {\bibfnamefont {K.}~\bibnamefont {Takanabe}},\ }\bibfield
  {title} {\enquote {\bibinfo {title} {{Recent advances in understanding oxygen
  evolution reaction mechanisms over iridium oxide}},}\ }\href {\doibase
  10.1039/d0qi01465f} {\bibfield  {journal} {\bibinfo  {journal} {Inorganic
  Chemistry Frontiers}\ }\textbf {\bibinfo {volume} {8}},\ \bibinfo {pages}
  {2900--2917} (\bibinfo {year} {2021})}\BibitemShut {NoStop}%
\bibitem [{\citenamefont {Hess}\ and\ \citenamefont {Over}(2023)}]{Hess2023}%
  \BibitemOpen
  \bibfield  {author} {\bibinfo {author} {\bibfnamefont {F.}~\bibnamefont
  {Hess}}\ and\ \bibinfo {author} {\bibfnamefont {H.}~\bibnamefont {Over}},\
  }\bibfield  {title} {\enquote {\bibinfo {title} {{Coordination Inversion of
  the Tetrahedrally Coordinated Ru4f Surface Complex on RuO$_2$(100) and Its
  Decisive Role in the Anodic Corrosion Process}},}\ }\href {\doibase
  10.1021/acscatal.2c06260} {\bibfield  {journal} {\bibinfo  {journal} {ACS
  Catalysis}\ }\textbf {\bibinfo {volume} {13}},\ \bibinfo {pages} {3433--3443}
  (\bibinfo {year} {2023})}\BibitemShut {NoStop}%
\bibitem [{\citenamefont {A{\ss}mann}\ \emph
  {et~al.}(2005{\natexlab{a}})\citenamefont {A{\ss}mann}, \citenamefont
  {Crihan}, \citenamefont {Knapp}, \citenamefont {Lundgren}, \citenamefont
  {L{\"{o}}ffler}, \citenamefont {Muhler}, \citenamefont {Narkhede},
  \citenamefont {Over}, \citenamefont {Schmid}, \citenamefont {Seitsonen},\
  and\ \citenamefont {Varga}}]{Assmann2005a}%
  \BibitemOpen
  \bibfield  {author} {\bibinfo {author} {\bibfnamefont {J.}~\bibnamefont
  {A{\ss}mann}}, \bibinfo {author} {\bibfnamefont {D.}~\bibnamefont {Crihan}},
  \bibinfo {author} {\bibfnamefont {M.}~\bibnamefont {Knapp}}, \bibinfo
  {author} {\bibfnamefont {E.}~\bibnamefont {Lundgren}}, \bibinfo {author}
  {\bibfnamefont {E.}~\bibnamefont {L{\"{o}}ffler}}, \bibinfo {author}
  {\bibfnamefont {M.}~\bibnamefont {Muhler}}, \bibinfo {author} {\bibfnamefont
  {V.}~\bibnamefont {Narkhede}}, \bibinfo {author} {\bibfnamefont
  {H.}~\bibnamefont {Over}}, \bibinfo {author} {\bibfnamefont {M.}~\bibnamefont
  {Schmid}}, \bibinfo {author} {\bibfnamefont {A.~P.}\ \bibnamefont
  {Seitsonen}}, \ and\ \bibinfo {author} {\bibfnamefont {P.}~\bibnamefont
  {Varga}},\ }\bibfield  {title} {\enquote {\bibinfo {title} {{Understanding
  the structural deactivation of ruthenium catalysts on an atomic scale under
  both oxidizing and reducing conditions}},}\ }\href {\doibase
  10.1002/anie.200461805} {\bibfield  {journal} {\bibinfo  {journal}
  {Angewandte Chemie - International Edition}\ }\textbf {\bibinfo {volume}
  {44}},\ \bibinfo {pages} {917--920} (\bibinfo {year}
  {2005}{\natexlab{a}})}\BibitemShut {NoStop}%
\bibitem [{\citenamefont {Stoerzinger}\ \emph {et~al.}(2014)\citenamefont
  {Stoerzinger}, \citenamefont {Qiao}, \citenamefont {Biegalski},\ and\
  \citenamefont {Shao-Horn}}]{Stoerzinger2014}%
  \BibitemOpen
  \bibfield  {author} {\bibinfo {author} {\bibfnamefont {K.~A.}\ \bibnamefont
  {Stoerzinger}}, \bibinfo {author} {\bibfnamefont {L.}~\bibnamefont {Qiao}},
  \bibinfo {author} {\bibfnamefont {M.~D.}\ \bibnamefont {Biegalski}}, \ and\
  \bibinfo {author} {\bibfnamefont {Y.}~\bibnamefont {Shao-Horn}},\ }\bibfield
  {title} {\enquote {\bibinfo {title} {{Orientation-dependent oxygen evolution
  activities of rutile IrO$_2$ and RuO$_2$}},}\ }\href {\doibase
  10.1021/jz500610u} {\bibfield  {journal} {\bibinfo  {journal} {Journal of
  Physical Chemistry Letters}\ }\textbf {\bibinfo {volume} {5}},\ \bibinfo
  {pages} {1636--1641} (\bibinfo {year} {2014})}\BibitemShut {NoStop}%
\bibitem [{\citenamefont {Moser}\ \emph {et~al.}(2021)\citenamefont {Moser},
  \citenamefont {Jovic}, \citenamefont {Consiglio}, \citenamefont {Smith},
  \citenamefont {Jozwiak}, \citenamefont {Bostwick}, \citenamefont
  {Rotenberg},\ and\ \citenamefont {{Di Sante}}}]{Moser2021}%
  \BibitemOpen
  \bibfield  {author} {\bibinfo {author} {\bibfnamefont {S.}~\bibnamefont
  {Moser}}, \bibinfo {author} {\bibfnamefont {V.}~\bibnamefont {Jovic}},
  \bibinfo {author} {\bibfnamefont {A.}~\bibnamefont {Consiglio}}, \bibinfo
  {author} {\bibfnamefont {K.~E.}\ \bibnamefont {Smith}}, \bibinfo {author}
  {\bibfnamefont {C.}~\bibnamefont {Jozwiak}}, \bibinfo {author} {\bibfnamefont
  {A.}~\bibnamefont {Bostwick}}, \bibinfo {author} {\bibfnamefont
  {E.}~\bibnamefont {Rotenberg}}, \ and\ \bibinfo {author} {\bibfnamefont
  {D.}~\bibnamefont {{Di Sante}}},\ }\bibfield  {title} {\enquote {\bibinfo
  {title} {{Momentum for catalysis: How surface reactions shape the RuO$_2$
  flat surface state}},}\ }\href {\doibase 10.1021/acscatal.0c04871} {\bibfield
   {journal} {\bibinfo  {journal} {ACS Catalysis}\ ,\ \bibinfo {pages}
  {1749--1757}} (\bibinfo {year} {2021})}\BibitemShut {NoStop}%
\bibitem [{\citenamefont {Reiser}\ \emph {et~al.}(2023)\citenamefont {Reiser},
  \citenamefont {Ke{\ss}ler}, \citenamefont {Kamp}, \citenamefont {Jovic},\
  and\ \citenamefont {Moser}}]{Reiser2023}%
  \BibitemOpen
  \bibfield  {author} {\bibinfo {author} {\bibfnamefont {C.}~\bibnamefont
  {Reiser}}, \bibinfo {author} {\bibfnamefont {P.}~\bibnamefont {Ke{\ss}ler}},
  \bibinfo {author} {\bibfnamefont {M.}~\bibnamefont {Kamp}}, \bibinfo {author}
  {\bibfnamefont {V.}~\bibnamefont {Jovic}}, \ and\ \bibinfo {author}
  {\bibfnamefont {S.}~\bibnamefont {Moser}},\ }\bibfield  {title} {\enquote
  {\bibinfo {title} {{Specific Capacitance of RuO$_2$ (110) Depends Sensitively
  on Surface Order}},}\ }\href {\doibase 10.1021/acs.jpcc.2c07217} {\bibfield
  {journal} {\bibinfo  {journal} {The Journal of Physical Chemistry C}\
  }\textbf {\bibinfo {volume} {2}} (\bibinfo {year} {2023}),\
  10.1021/acs.jpcc.2c07217}\BibitemShut {NoStop}%
\bibitem [{\citenamefont {Seki}(2010)}]{Seki2010}%
  \BibitemOpen
  \bibfield  {author} {\bibinfo {author} {\bibfnamefont {K.}~\bibnamefont
  {Seki}},\ }\bibfield  {title} {\enquote {\bibinfo {title} {{Development of
  RuO$_2$/Rutile-TiO$_2$ Catalyst for Industrial HCl Oxidation Process}},}\
  }\href {\doibase 10.1007/s10563-010-9091-7} {\bibfield  {journal} {\bibinfo
  {journal} {Catalysis Surveys from Asia}\ }\textbf {\bibinfo {volume} {14}},\
  \bibinfo {pages} {168--175} (\bibinfo {year} {2010})}\BibitemShut {NoStop}%
\bibitem [{\citenamefont {Jain}\ \emph {et~al.}(2013)\citenamefont {Jain},
  \citenamefont {Ong}, \citenamefont {Hautier}, \citenamefont {Chen},
  \citenamefont {Richards}, \citenamefont {Dacek}, \citenamefont {Cholia},
  \citenamefont {Gunter}, \citenamefont {Skinner}, \citenamefont {Ceder},\ and\
  \citenamefont {Persson}}]{Jain2013}%
  \BibitemOpen
  \bibfield  {author} {\bibinfo {author} {\bibfnamefont {A.}~\bibnamefont
  {Jain}}, \bibinfo {author} {\bibfnamefont {S.~P.}\ \bibnamefont {Ong}},
  \bibinfo {author} {\bibfnamefont {G.}~\bibnamefont {Hautier}}, \bibinfo
  {author} {\bibfnamefont {W.}~\bibnamefont {Chen}}, \bibinfo {author}
  {\bibfnamefont {W.~D.}\ \bibnamefont {Richards}}, \bibinfo {author}
  {\bibfnamefont {S.}~\bibnamefont {Dacek}}, \bibinfo {author} {\bibfnamefont
  {S.}~\bibnamefont {Cholia}}, \bibinfo {author} {\bibfnamefont
  {D.}~\bibnamefont {Gunter}}, \bibinfo {author} {\bibfnamefont
  {D.}~\bibnamefont {Skinner}}, \bibinfo {author} {\bibfnamefont
  {G.}~\bibnamefont {Ceder}}, \ and\ \bibinfo {author} {\bibfnamefont {K.~A.}\
  \bibnamefont {Persson}},\ }\bibfield  {title} {\enquote {\bibinfo {title}
  {{Commentary: The materials project: A materials genome approach to
  accelerating materials innovation}},}\ }\href {\doibase 10.1063/1.4812323}
  {\bibfield  {journal} {\bibinfo  {journal} {APL Materials}\ }\textbf
  {\bibinfo {volume} {1}} (\bibinfo {year} {2013}),\
  10.1063/1.4812323}\BibitemShut {NoStop}%
\bibitem [{\citenamefont {Uchida}\ \emph {et~al.}(2020)\citenamefont {Uchida},
  \citenamefont {Nomoto}, \citenamefont {Musashi}, \citenamefont {Arita},\ and\
  \citenamefont {Kawasaki}}]{Uchida2020}%
  \BibitemOpen
  \bibfield  {author} {\bibinfo {author} {\bibfnamefont {M.}~\bibnamefont
  {Uchida}}, \bibinfo {author} {\bibfnamefont {T.}~\bibnamefont {Nomoto}},
  \bibinfo {author} {\bibfnamefont {M.}~\bibnamefont {Musashi}}, \bibinfo
  {author} {\bibfnamefont {R.}~\bibnamefont {Arita}}, \ and\ \bibinfo {author}
  {\bibfnamefont {M.}~\bibnamefont {Kawasaki}},\ }\bibfield  {title} {\enquote
  {\bibinfo {title} {{Superconductivity in Uniquely Strained RuO$_2$ Films}},}\
  }\href {\doibase 10.1103/PhysRevLett.125.147001} {\bibfield  {journal}
  {\bibinfo  {journal} {Physical Review Letters}\ }\textbf {\bibinfo {volume}
  {125}} (\bibinfo {year} {2020}),\ 10.1103/PhysRevLett.125.147001}\BibitemShut
  {NoStop}%
\bibitem [{\citenamefont {Ruf}\ \emph {et~al.}(2021)\citenamefont {Ruf},
  \citenamefont {Paik}, \citenamefont {Schreiber}, \citenamefont {Nair},
  \citenamefont {Miao}, \citenamefont {Kawasaki}, \citenamefont {Nelson},
  \citenamefont {Faeth}, \citenamefont {Lee}, \citenamefont {Goodge},
  \citenamefont {Pamuk}, \citenamefont {Fennie}, \citenamefont {Kourkoutis},
  \citenamefont {Schlom},\ and\ \citenamefont {Shen}}]{Ruf2021}%
  \BibitemOpen
  \bibfield  {author} {\bibinfo {author} {\bibfnamefont {J.~P.}\ \bibnamefont
  {Ruf}}, \bibinfo {author} {\bibfnamefont {H.}~\bibnamefont {Paik}}, \bibinfo
  {author} {\bibfnamefont {N.~J.}\ \bibnamefont {Schreiber}}, \bibinfo {author}
  {\bibfnamefont {H.~P.}\ \bibnamefont {Nair}}, \bibinfo {author}
  {\bibfnamefont {L.}~\bibnamefont {Miao}}, \bibinfo {author} {\bibfnamefont
  {J.~K.}\ \bibnamefont {Kawasaki}}, \bibinfo {author} {\bibfnamefont {J.~N.}\
  \bibnamefont {Nelson}}, \bibinfo {author} {\bibfnamefont {B.~D.}\
  \bibnamefont {Faeth}}, \bibinfo {author} {\bibfnamefont {Y.}~\bibnamefont
  {Lee}}, \bibinfo {author} {\bibfnamefont {B.~H.}\ \bibnamefont {Goodge}},
  \bibinfo {author} {\bibfnamefont {B.}~\bibnamefont {Pamuk}}, \bibinfo
  {author} {\bibfnamefont {C.~J.}\ \bibnamefont {Fennie}}, \bibinfo {author}
  {\bibfnamefont {L.~F.}\ \bibnamefont {Kourkoutis}}, \bibinfo {author}
  {\bibfnamefont {D.~G.}\ \bibnamefont {Schlom}}, \ and\ \bibinfo {author}
  {\bibfnamefont {K.~M.}\ \bibnamefont {Shen}},\ }\bibfield  {title} {\enquote
  {\bibinfo {title} {{Strain-stabilized superconductivity}},}\ }\href {\doibase
  10.1038/s41467-020-20252-7} {\bibfield  {journal} {\bibinfo  {journal}
  {Nature Communications}\ }\textbf {\bibinfo {volume} {12}},\ \bibinfo {pages}
  {1--8} (\bibinfo {year} {2021})}\BibitemShut {NoStop}%
\bibitem [{\citenamefont {Bai}\ \emph {et~al.}(1998)\citenamefont {Bai},
  \citenamefont {Tsu}, \citenamefont {Wang},\ and\ \citenamefont
  {Foster}}]{Bai1998}%
  \BibitemOpen
  \bibfield  {author} {\bibinfo {author} {\bibfnamefont {G.}~\bibnamefont
  {Bai}}, \bibinfo {author} {\bibfnamefont {I.-f.}\ \bibnamefont {Tsu}},
  \bibinfo {author} {\bibfnamefont {A.}~\bibnamefont {Wang}}, \ and\ \bibinfo
  {author} {\bibfnamefont {C.~M.}\ \bibnamefont {Foster}},\ }\bibfield  {title}
  {\enquote {\bibinfo {title} {{By Low-Temperature Metal – Organic Chemical
  Vapor Deposition}},}\ }\href@noop {} {\bibfield  {journal} {\bibinfo
  {journal} {Applied Physics Letters}\ }\textbf {\bibinfo {volume} {72}},\
  \bibinfo {pages} {1572--1574} (\bibinfo {year} {1998})}\BibitemShut {NoStop}%
\bibitem [{\citenamefont {Mattheiss}(1976)}]{Mattheiss1976}%
  \BibitemOpen
  \bibfield  {author} {\bibinfo {author} {\bibfnamefont {L.}~\bibnamefont
  {Mattheiss}},\ }\bibfield  {title} {\enquote {\bibinfo {title} {{Electronic
  structure of RuO$_2$, OsO$_2$, and IrO$_2$}},}\ }\href {\doibase
  10.1103/PhysRevB.13.2433} {\bibfield  {journal} {\bibinfo  {journal}
  {Physical Review B}\ }\textbf {\bibinfo {volume} {13}},\ \bibinfo {pages}
  {2433--2450} (\bibinfo {year} {1976})}\BibitemShut {NoStop}%
\bibitem [{\citenamefont {Jovic}\ \emph {et~al.}(2018)\citenamefont {Jovic},
  \citenamefont {Koch}, \citenamefont {Panda}, \citenamefont {Berger},
  \citenamefont {Bugnon}, \citenamefont {Magrez}, \citenamefont {Smith},
  \citenamefont {Biermann}, \citenamefont {Jozwiak}, \citenamefont {Bostwick},
  \citenamefont {Rotenberg},\ and\ \citenamefont {Moser}}]{Jovic2018}%
  \BibitemOpen
  \bibfield  {author} {\bibinfo {author} {\bibfnamefont {V.}~\bibnamefont
  {Jovic}}, \bibinfo {author} {\bibfnamefont {R.~J.}\ \bibnamefont {Koch}},
  \bibinfo {author} {\bibfnamefont {S.~K.}\ \bibnamefont {Panda}}, \bibinfo
  {author} {\bibfnamefont {H.}~\bibnamefont {Berger}}, \bibinfo {author}
  {\bibfnamefont {P.}~\bibnamefont {Bugnon}}, \bibinfo {author} {\bibfnamefont
  {A.}~\bibnamefont {Magrez}}, \bibinfo {author} {\bibfnamefont {K.~E.}\
  \bibnamefont {Smith}}, \bibinfo {author} {\bibfnamefont {S.}~\bibnamefont
  {Biermann}}, \bibinfo {author} {\bibfnamefont {C.}~\bibnamefont {Jozwiak}},
  \bibinfo {author} {\bibfnamefont {A.}~\bibnamefont {Bostwick}}, \bibinfo
  {author} {\bibfnamefont {E.}~\bibnamefont {Rotenberg}}, \ and\ \bibinfo
  {author} {\bibfnamefont {S.}~\bibnamefont {Moser}},\ }\bibfield  {title}
  {\enquote {\bibinfo {title} {{Dirac nodal lines and flat-band surface state
  in the functional oxide RuO$_2$}},}\ }\href {\doibase
  10.1103/PhysRevB.98.241101} {\bibfield  {journal} {\bibinfo  {journal}
  {Physical Review B}\ }\textbf {\bibinfo {volume} {98}},\ \bibinfo {pages}
  {241101} (\bibinfo {year} {2018})}\BibitemShut {NoStop}%
\bibitem [{\citenamefont {Jovic}\ \emph {et~al.}(2019)\citenamefont {Jovic},
  \citenamefont {Koch}, \citenamefont {Panda}, \citenamefont {Berger},
  \citenamefont {Bugnon}, \citenamefont {Magrez}, \citenamefont {Thomale},
  \citenamefont {Smith}, \citenamefont {Biermann}, \citenamefont {Jozwiak},
  \citenamefont {Bostwick}, \citenamefont {Rotenberg}, \citenamefont {{Di
  Sante}},\ and\ \citenamefont {Moser}}]{Jovic2019}%
  \BibitemOpen
  \bibfield  {author} {\bibinfo {author} {\bibfnamefont {V.}~\bibnamefont
  {Jovic}}, \bibinfo {author} {\bibfnamefont {R.~J.}\ \bibnamefont {Koch}},
  \bibinfo {author} {\bibfnamefont {S.~K.}\ \bibnamefont {Panda}}, \bibinfo
  {author} {\bibfnamefont {H.}~\bibnamefont {Berger}}, \bibinfo {author}
  {\bibfnamefont {P.}~\bibnamefont {Bugnon}}, \bibinfo {author} {\bibfnamefont
  {A.}~\bibnamefont {Magrez}}, \bibinfo {author} {\bibfnamefont
  {R.}~\bibnamefont {Thomale}}, \bibinfo {author} {\bibfnamefont {K.~E.}\
  \bibnamefont {Smith}}, \bibinfo {author} {\bibfnamefont {S.}~\bibnamefont
  {Biermann}}, \bibinfo {author} {\bibfnamefont {C.}~\bibnamefont {Jozwiak}},
  \bibinfo {author} {\bibfnamefont {A.}~\bibnamefont {Bostwick}}, \bibinfo
  {author} {\bibfnamefont {E.}~\bibnamefont {Rotenberg}}, \bibinfo {author}
  {\bibfnamefont {D.}~\bibnamefont {{Di Sante}}}, \ and\ \bibinfo {author}
  {\bibfnamefont {S.}~\bibnamefont {Moser}},\ }\bibfield  {title} {\enquote
  {\bibinfo {title} {{The Dirac nodal line network in non-symmorphic rutile
  semimetal RuO$_2$}},}\ }\href {http://arxiv.org/abs/1908.02621} {\bibfield
  {journal} {\bibinfo  {journal} {cond-mat.mes-hall}\ } (\bibinfo {year}
  {2019})},\ \Eprint {http://arxiv.org/abs/1908.02621} {arXiv:1908.02621}
  \BibitemShut {NoStop}%
\bibitem [{\citenamefont {Sun}\ \emph {et~al.}(2017)\citenamefont {Sun},
  \citenamefont {Zhang}, \citenamefont {Liu}, \citenamefont {Felser},\ and\
  \citenamefont {Yan}}]{Sun2017}%
  \BibitemOpen
  \bibfield  {author} {\bibinfo {author} {\bibfnamefont {Y.}~\bibnamefont
  {Sun}}, \bibinfo {author} {\bibfnamefont {Y.}~\bibnamefont {Zhang}}, \bibinfo
  {author} {\bibfnamefont {C.-X.}\ \bibnamefont {Liu}}, \bibinfo {author}
  {\bibfnamefont {C.}~\bibnamefont {Felser}}, \ and\ \bibinfo {author}
  {\bibfnamefont {B.}~\bibnamefont {Yan}},\ }\bibfield  {title} {\enquote
  {\bibinfo {title} {{Dirac nodal lines and induced spin Hall effect in
  metallic rutile oxides}},}\ }\href {\doibase 10.1103/PhysRevB.95.235104}
  {\bibfield  {journal} {\bibinfo  {journal} {Physical Review B}\ }\textbf
  {\bibinfo {volume} {95}},\ \bibinfo {pages} {235104} (\bibinfo {year}
  {2017})},\ \Eprint {http://arxiv.org/abs/1701.09089} {1701.09089}
  \BibitemShut {NoStop}%
\bibitem [{\citenamefont {Ahn}\ \emph {et~al.}(2019)\citenamefont {Ahn},
  \citenamefont {Hariki}, \citenamefont {Lee},\ and\ \citenamefont
  {Kuneš}}]{Ahn2019}%
  \BibitemOpen
  \bibfield  {author} {\bibinfo {author} {\bibfnamefont {K.-H.}\ \bibnamefont
  {Ahn}}, \bibinfo {author} {\bibfnamefont {A.}~\bibnamefont {Hariki}},
  \bibinfo {author} {\bibfnamefont {K.-W.}\ \bibnamefont {Lee}}, \ and\
  \bibinfo {author} {\bibfnamefont {J.}~\bibnamefont {Kuneš}},\ }\bibfield
  {title} {\enquote {\bibinfo {title} {{Antiferromagnetism in RuO$_2$ as d
  -wave Pomeranchuk instability}},}\ }\href {\doibase
  10.1103/PhysRevB.99.184432} {\bibfield  {journal} {\bibinfo  {journal}
  {Physical Review B}\ }\textbf {\bibinfo {volume} {99}},\ \bibinfo {pages}
  {184432} (\bibinfo {year} {2019})}\BibitemShut {NoStop}%
\bibitem [{\citenamefont {Berlijn}\ \emph {et~al.}(2017)\citenamefont
  {Berlijn}, \citenamefont {Snijders}, \citenamefont {Delaire}, \citenamefont
  {Zhou}, \citenamefont {Maier}, \citenamefont {Cao}, \citenamefont {Chi},
  \citenamefont {Matsuda}, \citenamefont {Wang}, \citenamefont {Koehler},
  \citenamefont {Kent},\ and\ \citenamefont {Weitering}}]{Berlijn2017}%
  \BibitemOpen
  \bibfield  {author} {\bibinfo {author} {\bibfnamefont {T.}~\bibnamefont
  {Berlijn}}, \bibinfo {author} {\bibfnamefont {P.~C.}\ \bibnamefont
  {Snijders}}, \bibinfo {author} {\bibfnamefont {O.}~\bibnamefont {Delaire}},
  \bibinfo {author} {\bibfnamefont {H.-D.}\ \bibnamefont {Zhou}}, \bibinfo
  {author} {\bibfnamefont {T.~A.}\ \bibnamefont {Maier}}, \bibinfo {author}
  {\bibfnamefont {H.-B.}\ \bibnamefont {Cao}}, \bibinfo {author} {\bibfnamefont
  {S.-X.}\ \bibnamefont {Chi}}, \bibinfo {author} {\bibfnamefont
  {M.}~\bibnamefont {Matsuda}}, \bibinfo {author} {\bibfnamefont
  {Y.}~\bibnamefont {Wang}}, \bibinfo {author} {\bibfnamefont {M.~R.}\
  \bibnamefont {Koehler}}, \bibinfo {author} {\bibfnamefont {P.~R.~C.}\
  \bibnamefont {Kent}}, \ and\ \bibinfo {author} {\bibfnamefont {H.~H.}\
  \bibnamefont {Weitering}},\ }\bibfield  {title} {\enquote {\bibinfo {title}
  {{Itinerant Antiferromagnetism in RuO$_2$}},}\ }\href {\doibase
  10.1103/PhysRevLett.118.077201} {\bibfield  {journal} {\bibinfo  {journal}
  {Physical Review Letters}\ }\textbf {\bibinfo {volume} {118}},\ \bibinfo
  {pages} {077201} (\bibinfo {year} {2017})}\BibitemShut {NoStop}%
\bibitem [{\citenamefont {Zhu}\ \emph {et~al.}(2019)\citenamefont {Zhu},
  \citenamefont {Strempfer}, \citenamefont {Rao}, \citenamefont {Occhialini},
  \citenamefont {Pelliciari}, \citenamefont {Choi}, \citenamefont {Kawaguchi},
  \citenamefont {You}, \citenamefont {Mitchell}, \citenamefont {Shao-Horn},\
  and\ \citenamefont {Comin}}]{Zhu2019}%
  \BibitemOpen
  \bibfield  {author} {\bibinfo {author} {\bibfnamefont {Z.~H.}\ \bibnamefont
  {Zhu}}, \bibinfo {author} {\bibfnamefont {J.}~\bibnamefont {Strempfer}},
  \bibinfo {author} {\bibfnamefont {R.~R.}\ \bibnamefont {Rao}}, \bibinfo
  {author} {\bibfnamefont {C.~A.}\ \bibnamefont {Occhialini}}, \bibinfo
  {author} {\bibfnamefont {J.}~\bibnamefont {Pelliciari}}, \bibinfo {author}
  {\bibfnamefont {Y.}~\bibnamefont {Choi}}, \bibinfo {author} {\bibfnamefont
  {T.}~\bibnamefont {Kawaguchi}}, \bibinfo {author} {\bibfnamefont
  {H.}~\bibnamefont {You}}, \bibinfo {author} {\bibfnamefont {J.~F.}\
  \bibnamefont {Mitchell}}, \bibinfo {author} {\bibfnamefont {Y.}~\bibnamefont
  {Shao-Horn}}, \ and\ \bibinfo {author} {\bibfnamefont {R.}~\bibnamefont
  {Comin}},\ }\bibfield  {title} {\enquote {\bibinfo {title} {{Anomalous
  Antiferromagnetism in Metallic RuO$_2$ Determined by Resonant X-ray
  Scattering}},}\ }\href {\doibase 10.1103/PhysRevLett.122.017202} {\bibfield
  {journal} {\bibinfo  {journal} {Physical Review Letters}\ }\textbf {\bibinfo
  {volume} {122}} (\bibinfo {year} {2019}),\
  10.1103/PhysRevLett.122.017202}\BibitemShut {NoStop}%
\bibitem [{\citenamefont {Lovesey}, \citenamefont {Khalyavin},\ and\
  \citenamefont {van~der Laan}(2021)}]{Lovesey2021}%
  \BibitemOpen
  \bibfield  {author} {\bibinfo {author} {\bibfnamefont {S.~W.}\ \bibnamefont
  {Lovesey}}, \bibinfo {author} {\bibfnamefont {D.~D.}\ \bibnamefont
  {Khalyavin}}, \ and\ \bibinfo {author} {\bibfnamefont {G.}~\bibnamefont
  {van~der Laan}},\ }\bibfield  {title} {\enquote {\bibinfo {title} {{Magnetic
  properties of ruthenium dioxide (RuO$_2$) and charge-magnetic interference in
  Bragg diffraction of circularly polarized x-rays}},}\ }\href@noop {}
  {\bibfield  {journal} {\bibinfo  {journal} {arxiv}\ } (\bibinfo {year}
  {2021})}\BibitemShut {NoStop}%
\bibitem [{\citenamefont {Hiraishi}\ \emph {et~al.}(2024)\citenamefont
  {Hiraishi}, \citenamefont {Okabe}, \citenamefont {Koda}, \citenamefont
  {Kadono}, \citenamefont {Muroi}, \citenamefont {Hirai},\ and\ \citenamefont
  {Hiroi}}]{Hiraishi2024}%
  \BibitemOpen
  \bibfield  {author} {\bibinfo {author} {\bibfnamefont {M.}~\bibnamefont
  {Hiraishi}}, \bibinfo {author} {\bibfnamefont {H.}~\bibnamefont {Okabe}},
  \bibinfo {author} {\bibfnamefont {A.}~\bibnamefont {Koda}}, \bibinfo {author}
  {\bibfnamefont {R.}~\bibnamefont {Kadono}}, \bibinfo {author} {\bibfnamefont
  {T.}~\bibnamefont {Muroi}}, \bibinfo {author} {\bibfnamefont
  {D.}~\bibnamefont {Hirai}}, \ and\ \bibinfo {author} {\bibfnamefont
  {Z.}~\bibnamefont {Hiroi}},\ }\bibfield  {title} {\enquote {\bibinfo {title}
  {{Nonmagnetic Ground State in RuO$_2$ Revealed by Muon Spin Rotation}},}\
  }\href {\doibase 10.1103/PhysRevLett.132.166702} {\bibfield  {journal}
  {\bibinfo  {journal} {Physical Review Letters}\ }\textbf {\bibinfo {volume}
  {132}},\ \bibinfo {pages} {166702} (\bibinfo {year} {2024})}\BibitemShut
  {NoStop}%
  \bibitem [{\citenamefont {Ke{\ss}ler}\ \emph {et~al.}(2024)\citenamefont
  {Ke{\ss}ler}, \citenamefont {Garcia-Gassull}, \citenamefont {Suter},
  \citenamefont {Prokscha}, \citenamefont {Salman}, \citenamefont {Khalyavin},
  \citenamefont {Manuel}, \citenamefont {Orlandi}, \citenamefont {Mazin},
  \citenamefont {Valentı},\ and\ \citenamefont {Moser}}]{Kessler2024}%
  \BibitemOpen
  \bibfield  {author} {\bibinfo {author} {\bibfnamefont {P.}~\bibnamefont
  {Ke{\ss}ler}}, \bibinfo {author} {\bibfnamefont {L.}~\bibnamefont
  {Garcia-Gassull}}, \bibinfo {author} {\bibfnamefont {A.}~\bibnamefont
  {Suter}}, \bibinfo {author} {\bibfnamefont {T.}~\bibnamefont {Prokscha}},
  \bibinfo {author} {\bibfnamefont {Z.}~\bibnamefont {Salman}}, \bibinfo
  {author} {\bibfnamefont {D.}~\bibnamefont {Khalyavin}}, \bibinfo {author}
  {\bibfnamefont {P.}~\bibnamefont {Manuel}}, \bibinfo {author} {\bibfnamefont
  {F.}~\bibnamefont {Orlandi}}, \bibinfo {author} {\bibfnamefont {I.~I.}\
  \bibnamefont {Mazin}}, \bibinfo {author} {\bibfnamefont {R.}~\bibnamefont
  {Valentı}}, \ and\ \bibinfo {author} {\bibfnamefont {S.}~\bibnamefont
  {Moser}},\ }\bibfield  {title} {\enquote {\bibinfo {title} {{Absence of
  magnetic order in RuO$_2$: insights from $\mu$SR
  spectroscopy and neutron diffraction}},}\ }\href
  {http://arxiv.org/abs/2405.10820} {\ ,\ \bibinfo {pages} {1--12} (\bibinfo
  {year} {2024})},\ \Eprint {http://arxiv.org/abs/2405.10820}
  {arXiv:2405.10820} \BibitemShut {NoStop}%
\bibitem [{\citenamefont {Over}\ \emph {et~al.}(2004)\citenamefont {Over},
  \citenamefont {Knapp}, \citenamefont {Lundgren}, \citenamefont {Seitsonen},
  \citenamefont {Schmid},\ and\ \citenamefont {Varga}}]{Over2004a}%
  \BibitemOpen
  \bibfield  {author} {\bibinfo {author} {\bibfnamefont {H.}~\bibnamefont
  {Over}}, \bibinfo {author} {\bibfnamefont {M.}~\bibnamefont {Knapp}},
  \bibinfo {author} {\bibfnamefont {E.}~\bibnamefont {Lundgren}}, \bibinfo
  {author} {\bibfnamefont {A.~P.}\ \bibnamefont {Seitsonen}}, \bibinfo {author}
  {\bibfnamefont {M.}~\bibnamefont {Schmid}}, \ and\ \bibinfo {author}
  {\bibfnamefont {P.}~\bibnamefont {Varga}},\ }\bibfield  {title} {\enquote
  {\bibinfo {title} {{Visualization of atomic processes on ruthenium dioxide
  using scanning tunneling microscopy}},}\ }\href {\doibase
  10.1002/cphc.200300833} {\bibfield  {journal} {\bibinfo  {journal}
  {ChemPhysChem}\ }\textbf {\bibinfo {volume} {5}},\ \bibinfo {pages}
  {167--174} (\bibinfo {year} {2004})}\BibitemShut {NoStop}%
\bibitem [{\citenamefont {Fang}, \citenamefont {Tachiki},\ and\ \citenamefont
  {Kobayashi}(1997)}]{Fang1997}%
  \BibitemOpen
  \bibfield  {author} {\bibinfo {author} {\bibfnamefont {X.}~\bibnamefont
  {Fang}}, \bibinfo {author} {\bibfnamefont {M.}~\bibnamefont {Tachiki}}, \
  and\ \bibinfo {author} {\bibfnamefont {T.}~\bibnamefont {Kobayashi}},\
  }\bibfield  {title} {\enquote {\bibinfo {title} {{Growth of RuO$_2$ Thin
  Films on a MgO Substrate by Pulsed Laser Deposition Method}},}\ }\href
  {\doibase 10.1143/JJAP.36.L511} {\bibfield  {journal} {\bibinfo  {journal}
  {Japanese Journal of Applied Physics}\ }\textbf {\bibinfo {volume} {36}},\
  \bibinfo {pages} {L511--L514} (\bibinfo {year} {1997})}\BibitemShut {NoStop}%
\bibitem [{\citenamefont {Wang}\ \emph {et~al.}(2006)\citenamefont {Wang},
  \citenamefont {Pun}, \citenamefont {Xin},\ and\ \citenamefont
  {Zheng}}]{Wang2006}%
  \BibitemOpen
  \bibfield  {author} {\bibinfo {author} {\bibfnamefont {X.}~\bibnamefont
  {Wang}}, \bibinfo {author} {\bibfnamefont {A.~F.}\ \bibnamefont {Pun}},
  \bibinfo {author} {\bibfnamefont {Y.}~\bibnamefont {Xin}}, \ and\ \bibinfo
  {author} {\bibfnamefont {J.~P.}\ \bibnamefont {Zheng}},\ }\bibfield  {title}
  {\enquote {\bibinfo {title} {{Investigation of the growth dynamics of pulsed
  laser-deposited RuO$_2$ films using in situ resistance measurement and atomic
  force microscopy}},}\ }\href {\doibase 10.1016/j.tsf.2005.12.246} {\bibfield
  {journal} {\bibinfo  {journal} {Thin Solid Films}\ }\textbf {\bibinfo
  {volume} {510}},\ \bibinfo {pages} {82--87} (\bibinfo {year}
  {2006})}\BibitemShut {NoStop}%
\bibitem [{\citenamefont {Kim}\ \emph {et~al.}(2019)\citenamefont {Kim},
  \citenamefont {Charipar}, \citenamefont {Figueroa}, \citenamefont {Bingham},\
  and\ \citenamefont {Piqu{\'{e}}}}]{Kim2019a}%
  \BibitemOpen
  \bibfield  {author} {\bibinfo {author} {\bibfnamefont {H.}~\bibnamefont
  {Kim}}, \bibinfo {author} {\bibfnamefont {N.~A.}\ \bibnamefont {Charipar}},
  \bibinfo {author} {\bibfnamefont {J.}~\bibnamefont {Figueroa}}, \bibinfo
  {author} {\bibfnamefont {N.~S.}\ \bibnamefont {Bingham}}, \ and\ \bibinfo
  {author} {\bibfnamefont {A.}~\bibnamefont {Piqu{\'{e}}}},\ }\bibfield
  {title} {\enquote {\bibinfo {title} {{Control of metal-insulator transition
  temperature in VO 2 thin films grown on RuO$_2$ /TiO 2 templates by strain
  modification}},}\ }\href {\doibase 10.1063/1.5083848} {\bibfield  {journal}
  {\bibinfo  {journal} {AIP Advances}\ }\textbf {\bibinfo {volume} {9}}
  (\bibinfo {year} {2019}),\ 10.1063/1.5083848}\BibitemShut {NoStop}%
\bibitem [{\citenamefont {Feng}\ \emph {et~al.}(2022)\citenamefont {Feng},
  \citenamefont {Zhou}, \citenamefont {{\v{S}}mejkal}, \citenamefont {Wu},
  \citenamefont {Zhu}, \citenamefont {Guo}, \citenamefont
  {Gonz{\'{a}}lez-Hern{\'{a}}ndez}, \citenamefont {Wang}, \citenamefont {Yan},
  \citenamefont {Qin}, \citenamefont {Zhang}, \citenamefont {Wu}, \citenamefont
  {Chen}, \citenamefont {Meng}, \citenamefont {Liu}, \citenamefont {Xia},
  \citenamefont {Sinova}, \citenamefont {Jungwirth},\ and\ \citenamefont
  {Liu}}]{Feng2022}%
  \BibitemOpen
  \bibfield  {author} {\bibinfo {author} {\bibfnamefont {Z.}~\bibnamefont
  {Feng}}, \bibinfo {author} {\bibfnamefont {X.}~\bibnamefont {Zhou}}, \bibinfo
  {author} {\bibfnamefont {L.}~\bibnamefont {{\v{S}}mejkal}}, \bibinfo {author}
  {\bibfnamefont {L.}~\bibnamefont {Wu}}, \bibinfo {author} {\bibfnamefont
  {Z.}~\bibnamefont {Zhu}}, \bibinfo {author} {\bibfnamefont {H.}~\bibnamefont
  {Guo}}, \bibinfo {author} {\bibfnamefont {R.}~\bibnamefont
  {Gonz{\'{a}}lez-Hern{\'{a}}ndez}}, \bibinfo {author} {\bibfnamefont
  {X.}~\bibnamefont {Wang}}, \bibinfo {author} {\bibfnamefont {H.}~\bibnamefont
  {Yan}}, \bibinfo {author} {\bibfnamefont {P.}~\bibnamefont {Qin}}, \bibinfo
  {author} {\bibfnamefont {X.}~\bibnamefont {Zhang}}, \bibinfo {author}
  {\bibfnamefont {H.}~\bibnamefont {Wu}}, \bibinfo {author} {\bibfnamefont
  {H.}~\bibnamefont {Chen}}, \bibinfo {author} {\bibfnamefont {Z.}~\bibnamefont
  {Meng}}, \bibinfo {author} {\bibfnamefont {L.}~\bibnamefont {Liu}}, \bibinfo
  {author} {\bibfnamefont {Z.}~\bibnamefont {Xia}}, \bibinfo {author}
  {\bibfnamefont {J.}~\bibnamefont {Sinova}}, \bibinfo {author} {\bibfnamefont
  {T.}~\bibnamefont {Jungwirth}}, \ and\ \bibinfo {author} {\bibfnamefont
  {Z.}~\bibnamefont {Liu}},\ }\bibfield  {title} {\enquote {\bibinfo {title}
  {{An anomalous Hall effect in altermagnetic ruthenium dioxide}},}\ }\href
  {\doibase 10.1038/s41928-022-00866-z} {\bibfield  {journal} {\bibinfo
  {journal} {Nature Electronics}\ }\textbf {\bibinfo {volume} {5}},\ \bibinfo
  {pages} {735--743} (\bibinfo {year} {2022})}\BibitemShut {NoStop}%
\bibitem [{\citenamefont {Tschirner}\ \emph {et~al.}(2023)\citenamefont
  {Tschirner}, \citenamefont {Ke\ss ler}, \citenamefont {Betancourt},
  \citenamefont {Kotte}, \citenamefont {Kriegner}, \citenamefont {Büchner},
  \citenamefont {Dufouleur}, \citenamefont {Kamp}, \citenamefont {Jovic},
  \citenamefont {Smejkal}, \citenamefont {Sinova}, \citenamefont {Claessen},
  \citenamefont {Jungwirth}, \citenamefont {Moser}, \citenamefont {Reichlova},\
  and\ \citenamefont {Veyrat}}]{Tschirner2023}%
  \BibitemOpen
  \bibfield  {author} {\bibinfo {author} {\bibfnamefont {T.}~\bibnamefont
  {Tschirner}}, \bibinfo {author} {\bibfnamefont {P.}~\bibnamefont {Ke\ss ler}},
  \bibinfo {author} {\bibfnamefont {R.~D.~G.}\ \bibnamefont {Betancourt}},
  \bibinfo {author} {\bibfnamefont {T.}~\bibnamefont {Kotte}}, \bibinfo
  {author} {\bibfnamefont {D.}~\bibnamefont {Kriegner}}, \bibinfo {author}
  {\bibfnamefont {B.}~\bibnamefont {Büchner}}, \bibinfo {author}
  {\bibfnamefont {J.}~\bibnamefont {Dufouleur}}, \bibinfo {author}
  {\bibfnamefont {M.}~\bibnamefont {Kamp}}, \bibinfo {author} {\bibfnamefont
  {V.}~\bibnamefont {Jovic}}, \bibinfo {author} {\bibfnamefont
  {L.}~\bibnamefont {Smejkal}}, \bibinfo {author} {\bibfnamefont
  {J.}~\bibnamefont {Sinova}}, \bibinfo {author} {\bibfnamefont
  {R.}~\bibnamefont {Claessen}}, \bibinfo {author} {\bibfnamefont
  {T.}~\bibnamefont {Jungwirth}}, \bibinfo {author} {\bibfnamefont
  {S.}~\bibnamefont {Moser}}, \bibinfo {author} {\bibfnamefont
  {H.}~\bibnamefont {Reichlova}}, \ and\ \bibinfo {author} {\bibfnamefont
  {L.}~\bibnamefont {Veyrat}},\ }\bibfield  {title} {\enquote {\bibinfo {title}
  {{Saturation of the anomalous Hall effect at high magnetic fields in
  altermagnetic RuO$_2$}},}\ }\href {\doibase 10.1063/5.0160335} {\bibfield
  {journal} {\bibinfo  {journal} {APL Materials}\ }\textbf {\bibinfo {volume}
  {11}} (\bibinfo {year} {2023}),\ 10.1063/5.0160335}\BibitemShut {NoStop}%
\bibitem [{\citenamefont {Bose}\ \emph {et~al.}(2022)\citenamefont {Bose},
  \citenamefont {Schreiber}, \citenamefont {Jain}, \citenamefont {Shao},
  \citenamefont {Nair}, \citenamefont {Sun}, \citenamefont {Zhang},
  \citenamefont {Muller}, \citenamefont {Tsymbal}, \citenamefont {Schlom},\
  and\ \citenamefont {Ralph}}]{Bose2022a}%
  \BibitemOpen
  \bibfield  {author} {\bibinfo {author} {\bibfnamefont {A.}~\bibnamefont
  {Bose}}, \bibinfo {author} {\bibfnamefont {N.~J.}\ \bibnamefont {Schreiber}},
  \bibinfo {author} {\bibfnamefont {R.}~\bibnamefont {Jain}}, \bibinfo {author}
  {\bibfnamefont {D.~F.}\ \bibnamefont {Shao}}, \bibinfo {author}
  {\bibfnamefont {H.~P.}\ \bibnamefont {Nair}}, \bibinfo {author}
  {\bibfnamefont {J.}~\bibnamefont {Sun}}, \bibinfo {author} {\bibfnamefont
  {X.~S.}\ \bibnamefont {Zhang}}, \bibinfo {author} {\bibfnamefont {D.~A.}\
  \bibnamefont {Muller}}, \bibinfo {author} {\bibfnamefont {E.~Y.}\
  \bibnamefont {Tsymbal}}, \bibinfo {author} {\bibfnamefont {D.~G.}\
  \bibnamefont {Schlom}}, \ and\ \bibinfo {author} {\bibfnamefont {D.~C.}\
  \bibnamefont {Ralph}},\ }\bibfield  {title} {\enquote {\bibinfo {title}
  {{Tilted spin current generated by the collinear antiferromagnet ruthenium
  dioxide}},}\ }\href {\doibase 10.1038/s41928-022-00744-8} {\bibfield
  {journal} {\bibinfo  {journal} {Nature Electronics}\ }\textbf {\bibinfo
  {volume} {5}},\ \bibinfo {pages} {267--274} (\bibinfo {year}
  {2022})}\BibitemShut {NoStop}%
\bibitem [{\citenamefont {Shao}\ \emph {et~al.}(2021)\citenamefont {Shao},
  \citenamefont {Zhang}, \citenamefont {Li}, \citenamefont {Eom},\ and\
  \citenamefont {Tsymbal}}]{Shao2021}%
  \BibitemOpen
  \bibfield  {author} {\bibinfo {author} {\bibfnamefont {D.-F.}\ \bibnamefont
  {Shao}}, \bibinfo {author} {\bibfnamefont {S.-H.}\ \bibnamefont {Zhang}},
  \bibinfo {author} {\bibfnamefont {M.}~\bibnamefont {Li}}, \bibinfo {author}
  {\bibfnamefont {C.-B.}\ \bibnamefont {Eom}}, \ and\ \bibinfo {author}
  {\bibfnamefont {E.~Y.}\ \bibnamefont {Tsymbal}},\ }\bibfield  {title}
  {\enquote {\bibinfo {title} {{Spin-neutral currents for spintronics}},}\
  }\href {\doibase 10.1038/s41467-021-26915-3} {\bibfield  {journal} {\bibinfo
  {journal} {Nature Communications}\ }\textbf {\bibinfo {volume} {12}},\
  \bibinfo {pages} {7061} (\bibinfo {year} {2021})},\ \Eprint
  {http://arxiv.org/abs/2103.09219} {2103.09219} \BibitemShut {NoStop}%
\bibitem [{\citenamefont {{\v{S}}mejkal}, \citenamefont {Sinova},\ and\
  \citenamefont {Jungwirth}(2022{\natexlab{a}})}]{Smejkal2022}%
  \BibitemOpen
  \bibfield  {author} {\bibinfo {author} {\bibfnamefont {L.}~\bibnamefont
  {{\v{S}}mejkal}}, \bibinfo {author} {\bibfnamefont {J.}~\bibnamefont
  {Sinova}}, \ and\ \bibinfo {author} {\bibfnamefont {T.}~\bibnamefont
  {Jungwirth}},\ }\bibfield  {title} {\enquote {\bibinfo {title} {{Beyond
  Conventional Ferromagnetism and Antiferromagnetism: A Phase with
  Nonrelativistic Spin and Crystal Rotation Symmetry}},}\ }\href {\doibase
  10.1103/PhysRevX.12.031042} {\bibfield  {journal} {\bibinfo  {journal}
  {Physical Review X}\ }\textbf {\bibinfo {volume} {12}},\ \bibinfo {pages}
  {031042} (\bibinfo {year} {2022}{\natexlab{a}})}\BibitemShut {NoStop}%
\bibitem [{\citenamefont {{\v{S}}mejkal}, \citenamefont {Sinova},\ and\
  \citenamefont {Jungwirth}(2022{\natexlab{b}})}]{Smejkal2022a}%
  \BibitemOpen
  \bibfield  {author} {\bibinfo {author} {\bibfnamefont {L.}~\bibnamefont
  {{\v{S}}mejkal}}, \bibinfo {author} {\bibfnamefont {J.}~\bibnamefont
  {Sinova}}, \ and\ \bibinfo {author} {\bibfnamefont {T.}~\bibnamefont
  {Jungwirth}},\ }\bibfield  {title} {\enquote {\bibinfo {title} {{Emerging
  Research Landscape of Altermagnetism}},}\ }\href {\doibase
  10.1103/PhysRevX.12.040501} {\bibfield  {journal} {\bibinfo  {journal}
  {Physical Review X}\ }\textbf {\bibinfo {volume} {12}},\ \bibinfo {pages}
  {1--27} (\bibinfo {year} {2022}{\natexlab{b}})},\ \Eprint
  {http://arxiv.org/abs/2204.10844} {2204.10844} \BibitemShut {NoStop}%
\bibitem [{\citenamefont {Nelson}\ \emph {et~al.}(2019)\citenamefont {Nelson},
  \citenamefont {Ruf}, \citenamefont {Lee}, \citenamefont {Zeledon},
  \citenamefont {Kawasaki}, \citenamefont {Moser}, \citenamefont {Jozwiak},
  \citenamefont {Rotenberg}, \citenamefont {Bostwick}, \citenamefont {Schlom},
  \citenamefont {Shen},\ and\ \citenamefont {Moreschini}}]{Nelson2019}%
  \BibitemOpen
  \bibfield  {author} {\bibinfo {author} {\bibfnamefont {J.~N.}\ \bibnamefont
  {Nelson}}, \bibinfo {author} {\bibfnamefont {J.~P.}\ \bibnamefont {Ruf}},
  \bibinfo {author} {\bibfnamefont {Y.}~\bibnamefont {Lee}}, \bibinfo {author}
  {\bibfnamefont {C.}~\bibnamefont {Zeledon}}, \bibinfo {author} {\bibfnamefont
  {J.~K.}\ \bibnamefont {Kawasaki}}, \bibinfo {author} {\bibfnamefont
  {S.}~\bibnamefont {Moser}}, \bibinfo {author} {\bibfnamefont
  {C.}~\bibnamefont {Jozwiak}}, \bibinfo {author} {\bibfnamefont
  {E.}~\bibnamefont {Rotenberg}}, \bibinfo {author} {\bibfnamefont
  {A.}~\bibnamefont {Bostwick}}, \bibinfo {author} {\bibfnamefont {D.~G.}\
  \bibnamefont {Schlom}}, \bibinfo {author} {\bibfnamefont {K.~M.}\
  \bibnamefont {Shen}}, \ and\ \bibinfo {author} {\bibfnamefont
  {L.}~\bibnamefont {Moreschini}},\ }\bibfield  {title} {\enquote {\bibinfo
  {title} {{Dirac nodal lines protected against spin-orbit interaction in
  IrO$_2$}},}\ }\href {\doibase 10.1103/PhysRevMaterials.3.064205} {\bibfield
  {journal} {\bibinfo  {journal} {Physical Review Materials}\ }\textbf
  {\bibinfo {volume} {3}},\ \bibinfo {pages} {64205} (\bibinfo {year}
  {2019})}\BibitemShut {NoStop}%
\bibitem [{\citenamefont {Bose}\ \emph {et~al.}(2020)\citenamefont {Bose},
  \citenamefont {Nelson}, \citenamefont {Zhang}, \citenamefont {Jadaun},
  \citenamefont {Jain}, \citenamefont {Schlom}, \citenamefont {Ralph},
  \citenamefont {Muller}, \citenamefont {Shen},\ and\ \citenamefont
  {Buhrman}}]{Bose2020}%
  \BibitemOpen
  \bibfield  {author} {\bibinfo {author} {\bibfnamefont {A.}~\bibnamefont
  {Bose}}, \bibinfo {author} {\bibfnamefont {J.~N.}\ \bibnamefont {Nelson}},
  \bibinfo {author} {\bibfnamefont {X.~S.}\ \bibnamefont {Zhang}}, \bibinfo
  {author} {\bibfnamefont {P.}~\bibnamefont {Jadaun}}, \bibinfo {author}
  {\bibfnamefont {R.}~\bibnamefont {Jain}}, \bibinfo {author} {\bibfnamefont
  {D.~G.}\ \bibnamefont {Schlom}}, \bibinfo {author} {\bibfnamefont {D.~C.}\
  \bibnamefont {Ralph}}, \bibinfo {author} {\bibfnamefont {D.~A.}\ \bibnamefont
  {Muller}}, \bibinfo {author} {\bibfnamefont {K.~M.}\ \bibnamefont {Shen}}, \
  and\ \bibinfo {author} {\bibfnamefont {R.~A.}\ \bibnamefont {Buhrman}},\
  }\bibfield  {title} {\enquote {\bibinfo {title} {{Effects of Anisotropic
  Strain on Spin-Orbit Torque Produced by the Dirac Nodal Line Semimetal
  IrO$_2$}},}\ }\href {\doibase 10.1021/acsami.0c16485} {\bibfield  {journal}
  {\bibinfo  {journal} {ACS Applied Materials and Interfaces}\ }\textbf
  {\bibinfo {volume} {12}},\ \bibinfo {pages} {55411--55416} (\bibinfo {year}
  {2020})}\BibitemShut {NoStop}%
\bibitem [{\citenamefont {Patton}\ \emph {et~al.}(2023)\citenamefont {Patton},
  \citenamefont {Gurung}, \citenamefont {Shao}, \citenamefont {Noh},
  \citenamefont {Mittelstaedt}, \citenamefont {Mazur}, \citenamefont {Kim},
  \citenamefont {Ryan}, \citenamefont {Tsymbal}, \citenamefont {Choi},
  \citenamefont {Ralph}, \citenamefont {Rzchowski}, \citenamefont {Nan},\ and\
  \citenamefont {Eom}}]{Patton2023}%
  \BibitemOpen
  \bibfield  {author} {\bibinfo {author} {\bibfnamefont {M.}~\bibnamefont
  {Patton}}, \bibinfo {author} {\bibfnamefont {G.}~\bibnamefont {Gurung}},
  \bibinfo {author} {\bibfnamefont {D.~F.}\ \bibnamefont {Shao}}, \bibinfo
  {author} {\bibfnamefont {G.}~\bibnamefont {Noh}}, \bibinfo {author}
  {\bibfnamefont {J.~A.}\ \bibnamefont {Mittelstaedt}}, \bibinfo {author}
  {\bibfnamefont {M.}~\bibnamefont {Mazur}}, \bibinfo {author} {\bibfnamefont
  {J.~W.}\ \bibnamefont {Kim}}, \bibinfo {author} {\bibfnamefont {P.~J.}\
  \bibnamefont {Ryan}}, \bibinfo {author} {\bibfnamefont {E.~Y.}\ \bibnamefont
  {Tsymbal}}, \bibinfo {author} {\bibfnamefont {S.~Y.}\ \bibnamefont {Choi}},
  \bibinfo {author} {\bibfnamefont {D.~C.}\ \bibnamefont {Ralph}}, \bibinfo
  {author} {\bibfnamefont {M.~S.}\ \bibnamefont {Rzchowski}}, \bibinfo {author}
  {\bibfnamefont {T.}~\bibnamefont {Nan}}, \ and\ \bibinfo {author}
  {\bibfnamefont {C.~B.}\ \bibnamefont {Eom}},\ }\bibfield  {title} {\enquote
  {\bibinfo {title} {{Symmetry Control of Unconventional Spin–Orbit Torques
  in IrO$_2$}},}\ }\href {\doibase 10.1002/adma.202301608} {\bibfield
  {journal} {\bibinfo  {journal} {Advanced Materials}\ }\textbf {\bibinfo
  {volume} {35}},\ \bibinfo {pages} {1--8} (\bibinfo {year}
  {2023})}\BibitemShut {NoStop}%
\bibitem [{\citenamefont {Gottesfeld}\ \emph {et~al.}(1978)\citenamefont
  {Gottesfeld}, \citenamefont {McIntyre}, \citenamefont {Beni},\ and\
  \citenamefont {Shay}}]{Gottesfeld1978}%
  \BibitemOpen
  \bibfield  {author} {\bibinfo {author} {\bibfnamefont {S.}~\bibnamefont
  {Gottesfeld}}, \bibinfo {author} {\bibfnamefont {J.~D.}\ \bibnamefont
  {McIntyre}}, \bibinfo {author} {\bibfnamefont {G.}~\bibnamefont {Beni}}, \
  and\ \bibinfo {author} {\bibfnamefont {J.~L.}\ \bibnamefont {Shay}},\
  }\bibfield  {title} {\enquote {\bibinfo {title} {{Electrochromism in anodic
  iridium oxide films}},}\ }\href {\doibase 10.1063/1.90277} {\bibfield
  {journal} {\bibinfo  {journal} {Applied Physics Letters}\ }\textbf {\bibinfo
  {volume} {33}},\ \bibinfo {pages} {208--210} (\bibinfo {year}
  {1978})}\BibitemShut {NoStop}%
\bibitem [{\citenamefont {Patil}, \citenamefont {Kawar},\ and\ \citenamefont
  {Sadale}(2005)}]{Patil2005}%
  \BibitemOpen
  \bibfield  {author} {\bibinfo {author} {\bibfnamefont {P.~S.}\ \bibnamefont
  {Patil}}, \bibinfo {author} {\bibfnamefont {R.~K.}\ \bibnamefont {Kawar}}, \
  and\ \bibinfo {author} {\bibfnamefont {S.~B.}\ \bibnamefont {Sadale}},\
  }\bibfield  {title} {\enquote {\bibinfo {title} {{Electrochromism in spray
  deposited iridium oxide thin films}},}\ }\href {\doibase
  10.1016/j.electacta.2004.10.081} {\bibfield  {journal} {\bibinfo  {journal}
  {Electrochimica Acta}\ }\textbf {\bibinfo {volume} {50}},\ \bibinfo {pages}
  {2527--2532} (\bibinfo {year} {2005})}\BibitemShut {NoStop}%
\bibitem [{\citenamefont {Fujiwara}\ \emph {et~al.}(2013)\citenamefont
  {Fujiwara}, \citenamefont {Fukuma}, \citenamefont {Matsuno}, \citenamefont
  {Idzuchi}, \citenamefont {Niimi}, \citenamefont {Otani},\ and\ \citenamefont
  {Takagi}}]{Fujiwara2013}%
  \BibitemOpen
  \bibfield  {author} {\bibinfo {author} {\bibfnamefont {K.}~\bibnamefont
  {Fujiwara}}, \bibinfo {author} {\bibfnamefont {Y.}~\bibnamefont {Fukuma}},
  \bibinfo {author} {\bibfnamefont {J.}~\bibnamefont {Matsuno}}, \bibinfo
  {author} {\bibfnamefont {H.}~\bibnamefont {Idzuchi}}, \bibinfo {author}
  {\bibfnamefont {Y.}~\bibnamefont {Niimi}}, \bibinfo {author} {\bibfnamefont
  {Y.}~\bibnamefont {Otani}}, \ and\ \bibinfo {author} {\bibfnamefont
  {H.}~\bibnamefont {Takagi}},\ }\bibfield  {title} {\enquote {\bibinfo {title}
  {5d iridium oxide as a material for spin-current detection},}\ }\href
  {\doibase 10.1038/ncomms3893} {\bibfield  {journal} {\bibinfo  {journal}
  {Nature Communications}\ }\textbf {\bibinfo {volume} {4}},\ \bibinfo {pages}
  {2893} (\bibinfo {year} {2013})}\BibitemShut {NoStop}%
\bibitem [{\citenamefont {Uchida}\ \emph {et~al.}(2015)\citenamefont {Uchida},
  \citenamefont {Sano}, \citenamefont {Takahashi}, \citenamefont {Koretsune},
  \citenamefont {Kozuka}, \citenamefont {Arita}, \citenamefont {Tokura},\ and\
  \citenamefont {Kawasaki}}]{Uchida2015}%
  \BibitemOpen
  \bibfield  {author} {\bibinfo {author} {\bibfnamefont {M.}~\bibnamefont
  {Uchida}}, \bibinfo {author} {\bibfnamefont {W.}~\bibnamefont {Sano}},
  \bibinfo {author} {\bibfnamefont {K.~S.}\ \bibnamefont {Takahashi}}, \bibinfo
  {author} {\bibfnamefont {T.}~\bibnamefont {Koretsune}}, \bibinfo {author}
  {\bibfnamefont {Y.}~\bibnamefont {Kozuka}}, \bibinfo {author} {\bibfnamefont
  {R.}~\bibnamefont {Arita}}, \bibinfo {author} {\bibfnamefont
  {Y.}~\bibnamefont {Tokura}}, \ and\ \bibinfo {author} {\bibfnamefont
  {M.}~\bibnamefont {Kawasaki}},\ }\bibfield  {title} {\enquote {\bibinfo
  {title} {{Field-direction control of the type of charge carriers in
  nonsymmorphic IrO$_2$ }},}\ }\href {\doibase 10.1103/PhysRevB.91.241119}
  {\bibfield  {journal} {\bibinfo  {journal} {Physical Review B}\ }\textbf
  {\bibinfo {volume} {91}},\ \bibinfo {pages} {241119} (\bibinfo {year}
  {2015})}\BibitemShut {NoStop}%
\bibitem [{\citenamefont {Ping}, \citenamefont {Galli},\ and\ \citenamefont
  {Goddard}(2015)}]{Ping2015}%
  \BibitemOpen
  \bibfield  {author} {\bibinfo {author} {\bibfnamefont {Y.}~\bibnamefont
  {Ping}}, \bibinfo {author} {\bibfnamefont {G.}~\bibnamefont {Galli}}, \ and\
  \bibinfo {author} {\bibfnamefont {W.~A.}\ \bibnamefont {Goddard}},\
  }\bibfield  {title} {\enquote {\bibinfo {title} {{Electronic structure of
  IrO$_2$: The role of the metal d orbitals}},}\ }\href {\doibase
  10.1021/acs.jpcc.5b00861} {\bibfield  {journal} {\bibinfo  {journal} {Journal
  of Physical Chemistry C}\ }\textbf {\bibinfo {volume} {119}},\ \bibinfo
  {pages} {11570--11577} (\bibinfo {year} {2015})}\BibitemShut {NoStop}%
\bibitem [{\citenamefont {Martin}, \citenamefont {Maria},\ and\ \citenamefont
  {Schlom}(2024)}]{Martin2024}%
  \BibitemOpen
  \bibfield  {author} {\bibinfo {author} {\bibfnamefont {L.~W.}\ \bibnamefont
  {Martin}}, \bibinfo {author} {\bibfnamefont {J.-P.}\ \bibnamefont {Maria}}, \
  and\ \bibinfo {author} {\bibfnamefont {D.~G.}\ \bibnamefont {Schlom}},\
  }\bibfield  {title} {\enquote {\bibinfo {title} {{Lifting the fog in
  ferroelectric thin-film synthesis}},}\ }\href {\doibase
  10.1038/s41563-023-01732-9} {\bibfield  {journal} {\bibinfo  {journal}
  {Nature Materials}\ }\textbf {\bibinfo {volume} {23}},\ \bibinfo {pages}
  {9--10} (\bibinfo {year} {2024})}\BibitemShut {NoStop}%
\bibitem [{\citenamefont {He}\ \emph {et~al.}(2008)\citenamefont {He},
  \citenamefont {Stierle}, \citenamefont {Li}, \citenamefont {Farkas},
  \citenamefont {Kasper},\ and\ \citenamefont {Over}}]{He2008}%
  \BibitemOpen
  \bibfield  {author} {\bibinfo {author} {\bibfnamefont {Y.~B.}\ \bibnamefont
  {He}}, \bibinfo {author} {\bibfnamefont {A.}~\bibnamefont {Stierle}},
  \bibinfo {author} {\bibfnamefont {W.~X.}\ \bibnamefont {Li}}, \bibinfo
  {author} {\bibfnamefont {A.}~\bibnamefont {Farkas}}, \bibinfo {author}
  {\bibfnamefont {N.}~\bibnamefont {Kasper}}, \ and\ \bibinfo {author}
  {\bibfnamefont {H.}~\bibnamefont {Over}},\ }\bibfield  {title} {\enquote
  {\bibinfo {title} {{Oxidation of Ir(111): From O-Ir-O trilayer to bulk oxide
  formation}},}\ }\href {\doibase 10.1021/jp803607y} {\bibfield  {journal}
  {\bibinfo  {journal} {Journal of Physical Chemistry C}\ }\textbf {\bibinfo
  {volume} {112}},\ \bibinfo {pages} {11946--11953} (\bibinfo {year}
  {2008})}\BibitemShut {NoStop}%
\bibitem [{\citenamefont {Chung}\ \emph {et~al.}(2012)\citenamefont {Chung},
  \citenamefont {Tsai}, \citenamefont {Fan}, \citenamefont {Yang},\ and\
  \citenamefont {Huang}}]{Chung2012}%
  \BibitemOpen
  \bibfield  {author} {\bibinfo {author} {\bibfnamefont {W.-H.}\ \bibnamefont
  {Chung}}, \bibinfo {author} {\bibfnamefont {D.-S.}\ \bibnamefont {Tsai}},
  \bibinfo {author} {\bibfnamefont {L.-J.}\ \bibnamefont {Fan}}, \bibinfo
  {author} {\bibfnamefont {Y.-W.}\ \bibnamefont {Yang}}, \ and\ \bibinfo
  {author} {\bibfnamefont {Y.-S.}\ \bibnamefont {Huang}},\ }\bibfield  {title}
  {\enquote {\bibinfo {title} {{Surface oxides of Ir(111) prepared by gas-phase
  oxygen atoms}},}\ }\href {\doibase 10.1016/j.susc.2012.08.020} {\bibfield
  {journal} {\bibinfo  {journal} {Surface Science}\ }\textbf {\bibinfo {volume}
  {606}},\ \bibinfo {pages} {1965--1971} (\bibinfo {year} {2012})}\BibitemShut
  {NoStop}%
\bibitem [{\citenamefont {Nair}\ \emph {et~al.}(2023)\citenamefont {Nair},
  \citenamefont {Yang}, \citenamefont {Lee}, \citenamefont {Guo}, \citenamefont
  {Sadowski}, \citenamefont {Johnson}, \citenamefont {Saboor}, \citenamefont
  {Li}, \citenamefont {Zhou}, \citenamefont {Comes}, \citenamefont {Jin},
  \citenamefont {Mkhoyan}, \citenamefont {Janotti},\ and\ \citenamefont
  {Jalan}}]{Nair2023}%
  \BibitemOpen
  \bibfield  {author} {\bibinfo {author} {\bibfnamefont {S.}~\bibnamefont
  {Nair}}, \bibinfo {author} {\bibfnamefont {Z.}~\bibnamefont {Yang}}, \bibinfo
  {author} {\bibfnamefont {D.}~\bibnamefont {Lee}}, \bibinfo {author}
  {\bibfnamefont {S.}~\bibnamefont {Guo}}, \bibinfo {author} {\bibfnamefont
  {J.~T.}\ \bibnamefont {Sadowski}}, \bibinfo {author} {\bibfnamefont
  {S.}~\bibnamefont {Johnson}}, \bibinfo {author} {\bibfnamefont
  {A.}~\bibnamefont {Saboor}}, \bibinfo {author} {\bibfnamefont
  {Y.}~\bibnamefont {Li}}, \bibinfo {author} {\bibfnamefont {H.}~\bibnamefont
  {Zhou}}, \bibinfo {author} {\bibfnamefont {R.~B.}\ \bibnamefont {Comes}},
  \bibinfo {author} {\bibfnamefont {W.}~\bibnamefont {Jin}}, \bibinfo {author}
  {\bibfnamefont {K.~A.}\ \bibnamefont {Mkhoyan}}, \bibinfo {author}
  {\bibfnamefont {A.}~\bibnamefont {Janotti}}, \ and\ \bibinfo {author}
  {\bibfnamefont {B.}~\bibnamefont {Jalan}},\ }\bibfield  {title} {\enquote
  {\bibinfo {title} {{Engineering metal oxidation using epitaxial strain}},}\
  }\href {\doibase 10.1038/s41565-023-01397-0} {\bibfield  {journal} {\bibinfo
  {journal} {Nature Nanotechnology}\ } (\bibinfo {year} {2023}),\
  10.1038/s41565-023-01397-0}\BibitemShut {NoStop}%
\bibitem [{\citenamefont {Bhat}\ \emph {et~al.}(2017)\citenamefont {Bhat},
  \citenamefont {Koshy}, \citenamefont {Pittala},\ and\ \citenamefont
  {Kumar}}]{Bhat2017}%
  \BibitemOpen
  \bibfield  {author} {\bibinfo {author} {\bibfnamefont {S.~G.}\ \bibnamefont
  {Bhat}}, \bibinfo {author} {\bibfnamefont {A.~M.}\ \bibnamefont {Koshy}},
  \bibinfo {author} {\bibfnamefont {S.}~\bibnamefont {Pittala}}, \ and\
  \bibinfo {author} {\bibfnamefont {P.~S.}\ \bibnamefont {Kumar}},\ }\bibfield
  {title} {\enquote {\bibinfo {title} {{Tuning the growth of IrO$_2$ on SrTiO3
  (100) for spin-hall effect based oxide devices}},}\ }\href {\doibase
  10.1063/1.4990160} {\bibfield  {journal} {\bibinfo  {journal} {AIP Conference
  Proceedings}\ }\textbf {\bibinfo {volume} {1859}} (\bibinfo {year} {2017}),\
  10.1063/1.4990160}\BibitemShut {NoStop}%
\bibitem [{\citenamefont {Arias-Egido}\ \emph {et~al.}(2021)\citenamefont
  {Arias-Egido}, \citenamefont {Laguna-Marco}, \citenamefont {Piquer},
  \citenamefont {Jim{\'{e}}nez-Cavero}, \citenamefont {Lucas}, \citenamefont
  {Morell{\'{o}}n}, \citenamefont {Gallego}, \citenamefont {Rivera-Calzada},
  \citenamefont {Cabero-Piris}, \citenamefont {Santamaria}, \citenamefont
  {Fabbris}, \citenamefont {Haskel}, \citenamefont {Boada},\ and\ \citenamefont
  {D{\'{i}}az-Moreno}}]{Arias-Egido2021}%
  \BibitemOpen
  \bibfield  {author} {\bibinfo {author} {\bibfnamefont {E.}~\bibnamefont
  {Arias-Egido}}, \bibinfo {author} {\bibfnamefont {M.~A.}\ \bibnamefont
  {Laguna-Marco}}, \bibinfo {author} {\bibfnamefont {C.}~\bibnamefont
  {Piquer}}, \bibinfo {author} {\bibfnamefont {P.}~\bibnamefont
  {Jim{\'{e}}nez-Cavero}}, \bibinfo {author} {\bibfnamefont {I.}~\bibnamefont
  {Lucas}}, \bibinfo {author} {\bibfnamefont {L.}~\bibnamefont
  {Morell{\'{o}}n}}, \bibinfo {author} {\bibfnamefont {F.}~\bibnamefont
  {Gallego}}, \bibinfo {author} {\bibfnamefont {A.}~\bibnamefont
  {Rivera-Calzada}}, \bibinfo {author} {\bibfnamefont {M.}~\bibnamefont
  {Cabero-Piris}}, \bibinfo {author} {\bibfnamefont {J.}~\bibnamefont
  {Santamaria}}, \bibinfo {author} {\bibfnamefont {G.}~\bibnamefont {Fabbris}},
  \bibinfo {author} {\bibfnamefont {D.}~\bibnamefont {Haskel}}, \bibinfo
  {author} {\bibfnamefont {R.}~\bibnamefont {Boada}}, \ and\ \bibinfo {author}
  {\bibfnamefont {S.}~\bibnamefont {D{\'{i}}az-Moreno}},\ }\bibfield  {title}
  {\enquote {\bibinfo {title} {{Dimensionality-driven metal-insulator
  transition in spin-orbit-coupled IrO$_2$}},}\ }\href {\doibase
  10.1039/d1nr04207f} {\bibfield  {journal} {\bibinfo  {journal} {Nanoscale}\
  }\textbf {\bibinfo {volume} {13}},\ \bibinfo {pages} {17125--17135} (\bibinfo
  {year} {2021})}\BibitemShut {NoStop}%
\bibitem [{\citenamefont {Abb}, \citenamefont {Herd},\ and\ \citenamefont
  {Over}(2018)}]{Abb2018}%
  \BibitemOpen
  \bibfield  {author} {\bibinfo {author} {\bibfnamefont {M.~J.}\ \bibnamefont
  {Abb}}, \bibinfo {author} {\bibfnamefont {B.}~\bibnamefont {Herd}}, \ and\
  \bibinfo {author} {\bibfnamefont {H.}~\bibnamefont {Over}},\ }\bibfield
  {title} {\enquote {\bibinfo {title} {{Template-Assisted Growth of Ultrathin
  Single-Crystalline IrO$_2$(110) Films on RuO$_2$(110)/Ru(0001) and Its
  Thermal Stability}},}\ }\href {\doibase 10.1021/acs.jpcc.8b04375} {\bibfield
  {journal} {\bibinfo  {journal} {Journal of Physical Chemistry C}\ }\textbf
  {\bibinfo {volume} {122}},\ \bibinfo {pages} {14725--14732} (\bibinfo {year}
  {2018})}\BibitemShut {NoStop}%
\bibitem [{\citenamefont {Herd}, \citenamefont {Knapp},\ and\ \citenamefont
  {Over}(2012)}]{Herd2012}%
  \BibitemOpen
  \bibfield  {author} {\bibinfo {author} {\bibfnamefont {B.}~\bibnamefont
  {Herd}}, \bibinfo {author} {\bibfnamefont {M.}~\bibnamefont {Knapp}}, \ and\
  \bibinfo {author} {\bibfnamefont {H.}~\bibnamefont {Over}},\ }\bibfield
  {title} {\enquote {\bibinfo {title} {{Atomic scale insights into the initial
  oxidation of Ru(0001) using molecular oxygen: A scanning tunneling microscopy
  study}},}\ }\href {\doibase 10.1021/jp3085155} {\bibfield  {journal}
  {\bibinfo  {journal} {Journal of Physical Chemistry C}\ }\textbf {\bibinfo
  {volume} {116}},\ \bibinfo {pages} {24649--24660} (\bibinfo {year}
  {2012})}\BibitemShut {NoStop}%
\bibitem [{\citenamefont {Weber}\ \emph {et~al.}(2019)\citenamefont {Weber},
  \citenamefont {Pfrommer}, \citenamefont {Abb}, \citenamefont {Herd},
  \citenamefont {Khalid}, \citenamefont {Rohnke}, \citenamefont {Lakner},
  \citenamefont {Evertsson}, \citenamefont {Volkov}, \citenamefont {Bertram},
  \citenamefont {Znaiguia}, \citenamefont {Carla}, \citenamefont {Vonk},
  \citenamefont {Lundgren}, \citenamefont {Stierle},\ and\ \citenamefont
  {Over}}]{Weber2019}%
  \BibitemOpen
  \bibfield  {author} {\bibinfo {author} {\bibfnamefont {T.}~\bibnamefont
  {Weber}}, \bibinfo {author} {\bibfnamefont {J.}~\bibnamefont {Pfrommer}},
  \bibinfo {author} {\bibfnamefont {M.~J.}\ \bibnamefont {Abb}}, \bibinfo
  {author} {\bibfnamefont {B.}~\bibnamefont {Herd}}, \bibinfo {author}
  {\bibfnamefont {O.}~\bibnamefont {Khalid}}, \bibinfo {author} {\bibfnamefont
  {M.}~\bibnamefont {Rohnke}}, \bibinfo {author} {\bibfnamefont {P.~H.}\
  \bibnamefont {Lakner}}, \bibinfo {author} {\bibfnamefont {J.}~\bibnamefont
  {Evertsson}}, \bibinfo {author} {\bibfnamefont {S.}~\bibnamefont {Volkov}},
  \bibinfo {author} {\bibfnamefont {F.}~\bibnamefont {Bertram}}, \bibinfo
  {author} {\bibfnamefont {R.}~\bibnamefont {Znaiguia}}, \bibinfo {author}
  {\bibfnamefont {F.}~\bibnamefont {Carla}}, \bibinfo {author} {\bibfnamefont
  {V.}~\bibnamefont {Vonk}}, \bibinfo {author} {\bibfnamefont {E.}~\bibnamefont
  {Lundgren}}, \bibinfo {author} {\bibfnamefont {A.}~\bibnamefont {Stierle}}, \
  and\ \bibinfo {author} {\bibfnamefont {H.}~\bibnamefont {Over}},\ }\bibfield
  {title} {\enquote {\bibinfo {title} {{Potential-Induced Pitting Corrosion of
  an IrO$_2$(110)-RuO$_2$(110)/Ru(0001) Model Electrode under Oxygen Evolution
  Reaction Conditions}},}\ }\href {\doibase 10.1021/acscatal.9b01402}
  {\bibfield  {journal} {\bibinfo  {journal} {ACS Catalysis}\ }\textbf
  {\bibinfo {volume} {9}},\ \bibinfo {pages} {6530--6539} (\bibinfo {year}
  {2019})}\BibitemShut {NoStop}%
\bibitem [{\citenamefont {Rai}\ \emph {et~al.}(2016)\citenamefont {Rai},
  \citenamefont {Li}, \citenamefont {Liang}, \citenamefont {Kim}, \citenamefont
  {Asthagiri},\ and\ \citenamefont {Weaver}}]{Rai2016}%
  \BibitemOpen
  \bibfield  {author} {\bibinfo {author} {\bibfnamefont {R.}~\bibnamefont
  {Rai}}, \bibinfo {author} {\bibfnamefont {T.}~\bibnamefont {Li}}, \bibinfo
  {author} {\bibfnamefont {Z.}~\bibnamefont {Liang}}, \bibinfo {author}
  {\bibfnamefont {M.}~\bibnamefont {Kim}}, \bibinfo {author} {\bibfnamefont
  {A.}~\bibnamefont {Asthagiri}}, \ and\ \bibinfo {author} {\bibfnamefont
  {J.~F.}\ \bibnamefont {Weaver}},\ }\bibfield  {title} {\enquote {\bibinfo
  {title} {{Growth and termination of a rutile IrO$_2$(100) layer on
  Ir(111)}},}\ }\href {\doibase 10.1016/j.susc.2016.01.018} {\bibfield
  {journal} {\bibinfo  {journal} {Surface Science}\ }\textbf {\bibinfo {volume}
  {652}},\ \bibinfo {pages} {213--221} (\bibinfo {year} {2016})}\BibitemShut
  {NoStop}%
\bibitem [{\citenamefont {Liang}\ \emph {et~al.}(2017)\citenamefont {Liang},
  \citenamefont {Li}, \citenamefont {Kim}, \citenamefont {Asthagiri},\ and\
  \citenamefont {Weaver}}]{Liang2017a}%
  \BibitemOpen
  \bibfield  {author} {\bibinfo {author} {\bibfnamefont {Z.}~\bibnamefont
  {Liang}}, \bibinfo {author} {\bibfnamefont {T.}~\bibnamefont {Li}}, \bibinfo
  {author} {\bibfnamefont {M.}~\bibnamefont {Kim}}, \bibinfo {author}
  {\bibfnamefont {A.}~\bibnamefont {Asthagiri}}, \ and\ \bibinfo {author}
  {\bibfnamefont {J.~F.}\ \bibnamefont {Weaver}},\ }\bibfield  {title}
  {\enquote {\bibinfo {title} {{Low-temperature activation of methane on the
  IrO$_2$(110) surface}},}\ }\href {\doibase 10.1126/science.aam9147}
  {\bibfield  {journal} {\bibinfo  {journal} {Science}\ }\textbf {\bibinfo
  {volume} {356}},\ \bibinfo {pages} {299--303} (\bibinfo {year}
  {2017})}\BibitemShut {NoStop}%
\bibitem [{\citenamefont {Kim}, \citenamefont {Gao},\ and\ \citenamefont
  {Chambers}(1997)}]{Kim1997}%
  \BibitemOpen
  \bibfield  {author} {\bibinfo {author} {\bibfnamefont {Y.}~\bibnamefont
  {Kim}}, \bibinfo {author} {\bibfnamefont {Y.}~\bibnamefont {Gao}}, \ and\
  \bibinfo {author} {\bibfnamefont {S.}~\bibnamefont {Chambers}},\ }\bibfield
  {title} {\enquote {\bibinfo {title} {{Core-level X-ray photoelectron spectra
  and X-ray photoelectron diffraction of RuO$_2$(110) grown by molecular beam
  epitaxy on TiO$_2$(110)}},}\ }\href {\doibase 10.1016/S0169-4332(97)00233-X}
  {\bibfield  {journal} {\bibinfo  {journal} {Applied Surface Science}\
  }\textbf {\bibinfo {volume} {120}},\ \bibinfo {pages} {250--260} (\bibinfo
  {year} {1997})}\BibitemShut {NoStop}%
\bibitem [{\citenamefont {Nunn}\ \emph {et~al.}(2021)\citenamefont {Nunn},
  \citenamefont {Nair}, \citenamefont {Yun}, \citenamefont {Manjeshwar},
  \citenamefont {Rajapitamahuni}, \citenamefont {Lee}, \citenamefont
  {Mkhoyan},\ and\ \citenamefont {Jalan}}]{Nunn2021}%
  \BibitemOpen
  \bibfield  {author} {\bibinfo {author} {\bibfnamefont {W.}~\bibnamefont
  {Nunn}}, \bibinfo {author} {\bibfnamefont {S.}~\bibnamefont {Nair}}, \bibinfo
  {author} {\bibfnamefont {H.}~\bibnamefont {Yun}}, \bibinfo {author}
  {\bibfnamefont {A.~K.}\ \bibnamefont {Manjeshwar}}, \bibinfo {author}
  {\bibfnamefont {A.}~\bibnamefont {Rajapitamahuni}}, \bibinfo {author}
  {\bibfnamefont {D.}~\bibnamefont {Lee}}, \bibinfo {author} {\bibfnamefont
  {K.~A.}\ \bibnamefont {Mkhoyan}}, \ and\ \bibinfo {author} {\bibfnamefont
  {B.}~\bibnamefont {Jalan}},\ }\bibfield  {title} {\enquote {\bibinfo {title}
  {{Solid-source metal-organic molecular beam epitaxy of epitaxial RuO$_2$}},}\
  }\href {\doibase 10.1063/5.0062726} {\bibfield  {journal} {\bibinfo
  {journal} {APL Materials}\ }\textbf {\bibinfo {volume} {9}} (\bibinfo {year}
  {2021}),\ 10.1063/5.0062726}\BibitemShut {NoStop}%
\bibitem [{\citenamefont {Kawasaki}\ \emph
  {et~al.}(2018{\natexlab{a}})\citenamefont {Kawasaki}, \citenamefont {Baek},
  \citenamefont {Paik}, \citenamefont {Nair}, \citenamefont {Kourkoutis},
  \citenamefont {Schlom},\ and\ \citenamefont {Shen}}]{Kawasaki2018b}%
  \BibitemOpen
  \bibfield  {author} {\bibinfo {author} {\bibfnamefont {J.~K.}\ \bibnamefont
  {Kawasaki}}, \bibinfo {author} {\bibfnamefont {D.}~\bibnamefont {Baek}},
  \bibinfo {author} {\bibfnamefont {H.}~\bibnamefont {Paik}}, \bibinfo {author}
  {\bibfnamefont {H.~P.}\ \bibnamefont {Nair}}, \bibinfo {author}
  {\bibfnamefont {L.~F.}\ \bibnamefont {Kourkoutis}}, \bibinfo {author}
  {\bibfnamefont {D.~G.}\ \bibnamefont {Schlom}}, \ and\ \bibinfo {author}
  {\bibfnamefont {K.~M.}\ \bibnamefont {Shen}},\ }\bibfield  {title} {\enquote
  {\bibinfo {title} {{Rutile IrO$_2$/TiO$_2$ superlattices: A hyperconnected
  analog to the Ruddelsden-Popper structure}},}\ }\href {\doibase
  10.1103/PhysRevMaterials.2.054206} {\bibfield  {journal} {\bibinfo  {journal}
  {Physical Review Materials}\ }\textbf {\bibinfo {volume} {2}} (\bibinfo
  {year} {2018}{\natexlab{a}}),\ 10.1103/PhysRevMaterials.2.054206}\BibitemShut
  {NoStop}%
\bibitem [{\citenamefont {Kawasaki}\ \emph
  {et~al.}(2018{\natexlab{b}})\citenamefont {Kawasaki}, \citenamefont {Kim},
  \citenamefont {Nelson}, \citenamefont {Crisp}, \citenamefont {Zollner},
  \citenamefont {Biegenwald}, \citenamefont {Heron}, \citenamefont {Fennie},
  \citenamefont {Schlom},\ and\ \citenamefont {Shen}}]{Kawasaki2018a}%
  \BibitemOpen
  \bibfield  {author} {\bibinfo {author} {\bibfnamefont {J.~K.}\ \bibnamefont
  {Kawasaki}}, \bibinfo {author} {\bibfnamefont {C.~H.}\ \bibnamefont {Kim}},
  \bibinfo {author} {\bibfnamefont {J.~N.}\ \bibnamefont {Nelson}}, \bibinfo
  {author} {\bibfnamefont {S.}~\bibnamefont {Crisp}}, \bibinfo {author}
  {\bibfnamefont {C.~J.}\ \bibnamefont {Zollner}}, \bibinfo {author}
  {\bibfnamefont {E.}~\bibnamefont {Biegenwald}}, \bibinfo {author}
  {\bibfnamefont {J.~T.}\ \bibnamefont {Heron}}, \bibinfo {author}
  {\bibfnamefont {C.~J.}\ \bibnamefont {Fennie}}, \bibinfo {author}
  {\bibfnamefont {D.~G.}\ \bibnamefont {Schlom}}, \ and\ \bibinfo {author}
  {\bibfnamefont {K.~M.}\ \bibnamefont {Shen}},\ }\bibfield  {title} {\enquote
  {\bibinfo {title} {{Engineering Carrier Effective Masses in Ultrathin Quantum
  Wells of IrO$_2$}},}\ }\href {\doibase 10.1103/PhysRevLett.121.176802}
  {\bibfield  {journal} {\bibinfo  {journal} {Physical Review Letters}\
  }\textbf {\bibinfo {volume} {121}} (\bibinfo {year} {2018}{\natexlab{b}}),\
  10.1103/PhysRevLett.121.176802}\BibitemShut {NoStop}%
\bibitem [{\citenamefont {Kawasaki}\ \emph {et~al.}(2016)\citenamefont
  {Kawasaki}, \citenamefont {Uchida}, \citenamefont {Paik}, \citenamefont
  {Schlom},\ and\ \citenamefont {Shen}}]{Kawasaki2016}%
  \BibitemOpen
  \bibfield  {author} {\bibinfo {author} {\bibfnamefont {J.~K.}\ \bibnamefont
  {Kawasaki}}, \bibinfo {author} {\bibfnamefont {M.}~\bibnamefont {Uchida}},
  \bibinfo {author} {\bibfnamefont {H.}~\bibnamefont {Paik}}, \bibinfo {author}
  {\bibfnamefont {D.~G.}\ \bibnamefont {Schlom}}, \ and\ \bibinfo {author}
  {\bibfnamefont {K.~M.}\ \bibnamefont {Shen}},\ }\bibfield  {title} {\enquote
  {\bibinfo {title} {{Evolution of electronic correlations across the rutile,
  perovskite, and Ruddelsden-Popper iridates with octahedral connectivity}},}\
  }\href {\doibase 10.1103/PhysRevB.94.121104} {\bibfield  {journal} {\bibinfo
  {journal} {Physical Review B}\ }\textbf {\bibinfo {volume} {94}},\ \bibinfo
  {pages} {121104} (\bibinfo {year} {2016})}\BibitemShut {NoStop}%
\bibitem [{\citenamefont {Kuo}\ \emph {et~al.}(2019)\citenamefont {Kuo},
  \citenamefont {Paik}, \citenamefont {Nelson}, \citenamefont {Shen},
  \citenamefont {Schlom},\ and\ \citenamefont {Suntivich}}]{Kuo2019}%
  \BibitemOpen
  \bibfield  {author} {\bibinfo {author} {\bibfnamefont {D.~Y.}\ \bibnamefont
  {Kuo}}, \bibinfo {author} {\bibfnamefont {H.}~\bibnamefont {Paik}}, \bibinfo
  {author} {\bibfnamefont {J.~N.}\ \bibnamefont {Nelson}}, \bibinfo {author}
  {\bibfnamefont {K.~M.}\ \bibnamefont {Shen}}, \bibinfo {author}
  {\bibfnamefont {D.~G.}\ \bibnamefont {Schlom}}, \ and\ \bibinfo {author}
  {\bibfnamefont {J.}~\bibnamefont {Suntivich}},\ }\bibfield  {title} {\enquote
  {\bibinfo {title} {{Chlorine evolution reaction electrocatalysis on RuO$_2$
  (110) and IrO$_2$ (110) grown using molecular-beam epitaxy}},}\ }\href
  {\doibase 10.1063/1.5051429} {\bibfield  {journal} {\bibinfo  {journal}
  {Journal of Chemical Physics}\ }\textbf {\bibinfo {volume} {150}} (\bibinfo
  {year} {2019}),\ 10.1063/1.5051429}\BibitemShut {NoStop}%
\bibitem [{\citenamefont {Iembo}\ \emph {et~al.}(1997)\citenamefont {Iembo},
  \citenamefont {Fuso}, \citenamefont {Arimondo}, \citenamefont {Ciofi},
  \citenamefont {Pennelli}, \citenamefont {Curr{\`{o}}}, \citenamefont {Neri},\
  and\ \citenamefont {Allegrini}}]{Iembo1997a}%
  \BibitemOpen
  \bibfield  {author} {\bibinfo {author} {\bibfnamefont {A.}~\bibnamefont
  {Iembo}}, \bibinfo {author} {\bibfnamefont {F.}~\bibnamefont {Fuso}},
  \bibinfo {author} {\bibfnamefont {E.}~\bibnamefont {Arimondo}}, \bibinfo
  {author} {\bibfnamefont {C.}~\bibnamefont {Ciofi}}, \bibinfo {author}
  {\bibfnamefont {G.}~\bibnamefont {Pennelli}}, \bibinfo {author}
  {\bibfnamefont {G.~M.}\ \bibnamefont {Curr{\`{o}}}}, \bibinfo {author}
  {\bibfnamefont {F.}~\bibnamefont {Neri}}, \ and\ \bibinfo {author}
  {\bibfnamefont {M.}~\bibnamefont {Allegrini}},\ }\bibfield  {title} {\enquote
  {\bibinfo {title} {{Pulsed laser deposition and characterization of
  conductive RuO$_2$ thin films}},}\ }\href {\doibase 10.1557/JMR.1997.0195}
  {\bibfield  {journal} {\bibinfo  {journal} {Journal of Materials Research}\
  }\textbf {\bibinfo {volume} {12}},\ \bibinfo {pages} {1433--1436} (\bibinfo
  {year} {1997})}\BibitemShut {NoStop}%
\bibitem [{\citenamefont {Zhang}\ \emph {et~al.}(2007)\citenamefont {Zhang},
  \citenamefont {Gong}, \citenamefont {Wang},\ and\ \citenamefont
  {Shen}}]{Zhang2007}%
  \BibitemOpen
  \bibfield  {author} {\bibinfo {author} {\bibfnamefont {L.~M.}\ \bibnamefont
  {Zhang}}, \bibinfo {author} {\bibfnamefont {Y.~S.}\ \bibnamefont {Gong}},
  \bibinfo {author} {\bibfnamefont {C.~B.}\ \bibnamefont {Wang}}, \ and\
  \bibinfo {author} {\bibfnamefont {Q.}~\bibnamefont {Shen}},\ }\bibfield
  {title} {\enquote {\bibinfo {title} {{Microstructure and Resistivity of
  Iridium Oxide Thin Films by Pulsed Laser Deposition Technique}},}\ }\href
  {\doibase 10.4028/www.scientific.net/KEM.336-338.2215} {\bibfield  {journal}
  {\bibinfo  {journal} {Key Engineering Materials}\ }\textbf {\bibinfo {volume}
  {336-338}},\ \bibinfo {pages} {2215--2217} (\bibinfo {year}
  {2007})}\BibitemShut {NoStop}%
\bibitem [{\citenamefont {{El Khakani}}\ and\ \citenamefont
  {Chaker}(1997)}]{ElKhakani1997}%
  \BibitemOpen
  \bibfield  {author} {\bibinfo {author} {\bibfnamefont {M.}~\bibnamefont {{El
  Khakani}}}\ and\ \bibinfo {author} {\bibfnamefont {M.}~\bibnamefont
  {Chaker}},\ }\bibfield  {title} {\enquote {\bibinfo {title} {{Highly
  Conductive and Optically Transparent Polycrystalline Iridium Oxide Thin Films
  Grown by Reactive Pulsed Laser Deposition}},}\ }\href {\doibase
  10.1557/PROC-472-373} {\bibfield  {journal} {\bibinfo  {journal} {MRS
  Proceedings}\ }\textbf {\bibinfo {volume} {472}},\ \bibinfo {pages} {373}
  (\bibinfo {year} {1997})}\BibitemShut {NoStop}%
\bibitem [{\citenamefont {{El Khakani}}\ and\ \citenamefont
  {Chaker}(1998)}]{ElKhakani1998}%
  \BibitemOpen
  \bibfield  {author} {\bibinfo {author} {\bibfnamefont {M.}~\bibnamefont {{El
  Khakani}}}\ and\ \bibinfo {author} {\bibfnamefont {M.}~\bibnamefont
  {Chaker}},\ }\bibfield  {title} {\enquote {\bibinfo {title} {{Reactive pulsed
  laser deposition of iridium oxide thin films}},}\ }\href {\doibase
  10.1016/S0040-6090(98)00862-1} {\bibfield  {journal} {\bibinfo  {journal}
  {Thin Solid Films}\ }\textbf {\bibinfo {volume} {335}},\ \bibinfo {pages}
  {6--12} (\bibinfo {year} {1998})}\BibitemShut {NoStop}%
\bibitem [{\citenamefont {{El Khakani}}, \citenamefont {Chaker},\ and\
  \citenamefont {Gat}(1996)}]{ElKhakani1996}%
  \BibitemOpen
  \bibfield  {author} {\bibinfo {author} {\bibfnamefont {M.~A.}\ \bibnamefont
  {{El Khakani}}}, \bibinfo {author} {\bibfnamefont {M.}~\bibnamefont
  {Chaker}}, \ and\ \bibinfo {author} {\bibfnamefont {E.}~\bibnamefont {Gat}},\
  }\bibfield  {title} {\enquote {\bibinfo {title} {{Pulsed laser deposition of
  highly conductive iridium oxide thin films}},}\ }\href {\doibase
  10.1063/1.116868} {\bibfield  {journal} {\bibinfo  {journal} {Applied Physics
  Letters}\ }\textbf {\bibinfo {volume} {69}},\ \bibinfo {pages} {2027--2029}
  (\bibinfo {year} {1996})}\BibitemShut {NoStop}%
\bibitem [{\citenamefont {Kim}\ \emph {et~al.}(2016)\citenamefont {Kim},
  \citenamefont {Kim}, \citenamefont {Kim}, \citenamefont {Sohn}, \citenamefont
  {Korneta}, \citenamefont {Chae},\ and\ \citenamefont {Noh}}]{Kim2016b}%
  \BibitemOpen
  \bibfield  {author} {\bibinfo {author} {\bibfnamefont {W.~J.}\ \bibnamefont
  {Kim}}, \bibinfo {author} {\bibfnamefont {S.~Y.}\ \bibnamefont {Kim}},
  \bibinfo {author} {\bibfnamefont {C.~H.}\ \bibnamefont {Kim}}, \bibinfo
  {author} {\bibfnamefont {C.~H.}\ \bibnamefont {Sohn}}, \bibinfo {author}
  {\bibfnamefont {O.~B.}\ \bibnamefont {Korneta}}, \bibinfo {author}
  {\bibfnamefont {S.~C.}\ \bibnamefont {Chae}}, \ and\ \bibinfo {author}
  {\bibfnamefont {T.~W.}\ \bibnamefont {Noh}},\ }\bibfield  {title} {\enquote
  {\bibinfo {title} {{Spin-orbit coupling induced band structure change and
  orbital character of epitaxial IrO$_2$ films}},}\ }\href {\doibase
  10.1103/PhysRevB.93.045104} {\bibfield  {journal} {\bibinfo  {journal}
  {Physical Review B}\ }\textbf {\bibinfo {volume} {93}},\ \bibinfo {pages}
  {045104} (\bibinfo {year} {2016})}\BibitemShut {NoStop}%
\bibitem [{\citenamefont {Hou}\ \emph {et~al.}(2017)\citenamefont {Hou},
  \citenamefont {Takahashi}, \citenamefont {Yamamoto},\ and\ \citenamefont
  {Lippmaa}}]{Hou2017a}%
  \BibitemOpen
  \bibfield  {author} {\bibinfo {author} {\bibfnamefont {X.}~\bibnamefont
  {Hou}}, \bibinfo {author} {\bibfnamefont {R.}~\bibnamefont {Takahashi}},
  \bibinfo {author} {\bibfnamefont {T.}~\bibnamefont {Yamamoto}}, \ and\
  \bibinfo {author} {\bibfnamefont {M.}~\bibnamefont {Lippmaa}},\ }\bibfield
  {title} {\enquote {\bibinfo {title} {{Microstructure analysis of IrO$_2$ thin
  films}},}\ }\href {\doibase 10.1016/j.jcrysgro.2016.12.104} {\bibfield
  {journal} {\bibinfo  {journal} {Journal of Crystal Growth}\ }\textbf
  {\bibinfo {volume} {462}},\ \bibinfo {pages} {24--28} (\bibinfo {year}
  {2017})}\BibitemShut {NoStop}%
\bibitem [{\citenamefont {Wang}\ \emph {et~al.}(1996)\citenamefont {Wang},
  \citenamefont {Gladfelter}, \citenamefont {{Fennell Evans}}, \citenamefont
  {Fan},\ and\ \citenamefont {Franciosi}}]{Wang1996}%
  \BibitemOpen
  \bibfield  {author} {\bibinfo {author} {\bibfnamefont {Q.}~\bibnamefont
  {Wang}}, \bibinfo {author} {\bibfnamefont {W.~L.}\ \bibnamefont
  {Gladfelter}}, \bibinfo {author} {\bibfnamefont {D.}~\bibnamefont {{Fennell
  Evans}}}, \bibinfo {author} {\bibfnamefont {Y.}~\bibnamefont {Fan}}, \ and\
  \bibinfo {author} {\bibfnamefont {A.}~\bibnamefont {Franciosi}},\ }\bibfield
  {title} {\enquote {\bibinfo {title} {{Reactive-sputter deposition and
  structure of RuO$_2$ films on sapphire and strontium titanate}},}\ }\href
  {\doibase 10.1116/1.580382} {\bibfield  {journal} {\bibinfo  {journal}
  {Journal of Vacuum Science {\&} Technology A: Vacuum, Surfaces, and Films}\
  }\textbf {\bibinfo {volume} {14}},\ \bibinfo {pages} {747--752} (\bibinfo
  {year} {1996})}\BibitemShut {NoStop}%
\bibitem [{\citenamefont {Huang}(2001)}]{Huang2001}%
  \BibitemOpen
  \bibfield  {author} {\bibinfo {author} {\bibfnamefont {J.}~\bibnamefont
  {Huang}},\ }\bibfield  {title} {\enquote {\bibinfo {title} {{Material
  characteristics and electrical property of reactively sputtered RuO$_2$ thin
  films}},}\ }\href {\doibase 10.1016/S0040-6090(00)01777-6} {\bibfield
  {journal} {\bibinfo  {journal} {Thin Solid Films}\ }\textbf {\bibinfo
  {volume} {382}},\ \bibinfo {pages} {139--145} (\bibinfo {year}
  {2001})}\BibitemShut {NoStop}%
\bibitem [{\citenamefont {Meng}\ and\ \citenamefont {{Dos
  Santos}}(1999)}]{Meng1999}%
  \BibitemOpen
  \bibfield  {author} {\bibinfo {author} {\bibfnamefont {L.~J.}\ \bibnamefont
  {Meng}}\ and\ \bibinfo {author} {\bibfnamefont {M.~P.}\ \bibnamefont {{Dos
  Santos}}},\ }\bibfield  {title} {\enquote {\bibinfo {title}
  {{Characterization of RuO$_2$ films prepared by rf reactive magnetron
  sputtering}},}\ }\href {\doibase 10.1016/S0169-4332(99)00089-6} {\bibfield
  {journal} {\bibinfo  {journal} {Applied Surface Science}\ }\textbf {\bibinfo
  {volume} {147}},\ \bibinfo {pages} {94--100} (\bibinfo {year}
  {1999})}\BibitemShut {NoStop}%
\bibitem [{\citenamefont {Paoli}\ \emph {et~al.}(2015)\citenamefont {Paoli},
  \citenamefont {Masini}, \citenamefont {Frydendal}, \citenamefont {Deiana},
  \citenamefont {Schlaup}, \citenamefont {Malizia}, \citenamefont {Hansen},
  \citenamefont {Horch}, \citenamefont {Stephens},\ and\ \citenamefont
  {Chorkendorff}}]{Paoli2015}%
  \BibitemOpen
  \bibfield  {author} {\bibinfo {author} {\bibfnamefont {E.~A.}\ \bibnamefont
  {Paoli}}, \bibinfo {author} {\bibfnamefont {F.}~\bibnamefont {Masini}},
  \bibinfo {author} {\bibfnamefont {R.}~\bibnamefont {Frydendal}}, \bibinfo
  {author} {\bibfnamefont {D.}~\bibnamefont {Deiana}}, \bibinfo {author}
  {\bibfnamefont {C.}~\bibnamefont {Schlaup}}, \bibinfo {author} {\bibfnamefont
  {M.}~\bibnamefont {Malizia}}, \bibinfo {author} {\bibfnamefont {T.~W.}\
  \bibnamefont {Hansen}}, \bibinfo {author} {\bibfnamefont {S.}~\bibnamefont
  {Horch}}, \bibinfo {author} {\bibfnamefont {I.~E.~L.}\ \bibnamefont
  {Stephens}}, \ and\ \bibinfo {author} {\bibfnamefont {I.}~\bibnamefont
  {Chorkendorff}},\ }\bibfield  {title} {\enquote {\bibinfo {title} {{Oxygen
  evolution on well-characterized mass-selected Ru and RuO$_2$
  nanoparticles}},}\ }\href {\doibase 10.1039/C4SC02685C} {\bibfield  {journal}
  {\bibinfo  {journal} {Chemical Science}\ }\textbf {\bibinfo {volume} {6}},\
  \bibinfo {pages} {190--196} (\bibinfo {year} {2015})}\BibitemShut {NoStop}%
\bibitem [{\citenamefont {Liu}\ \emph {et~al.}(2023)\citenamefont {Liu},
  \citenamefont {Bai}, \citenamefont {Song}, \citenamefont {Ji}, \citenamefont
  {Lou}, \citenamefont {Zhang}, \citenamefont {Song},\ and\ \citenamefont
  {Jin}}]{Liu2023}%
  \BibitemOpen
  \bibfield  {author} {\bibinfo {author} {\bibfnamefont {Y.}~\bibnamefont
  {Liu}}, \bibinfo {author} {\bibfnamefont {H.}~\bibnamefont {Bai}}, \bibinfo
  {author} {\bibfnamefont {Y.}~\bibnamefont {Song}}, \bibinfo {author}
  {\bibfnamefont {Z.}~\bibnamefont {Ji}}, \bibinfo {author} {\bibfnamefont
  {S.}~\bibnamefont {Lou}}, \bibinfo {author} {\bibfnamefont {Z.}~\bibnamefont
  {Zhang}}, \bibinfo {author} {\bibfnamefont {C.}~\bibnamefont {Song}}, \ and\
  \bibinfo {author} {\bibfnamefont {Q.}~\bibnamefont {Jin}},\ }\bibfield
  {title} {\enquote {\bibinfo {title} {{Inverse Altermagnetic Spin Splitting
  Effect‐Induced Terahertz Emission in RuO$_2$}},}\ }\href {\doibase
  10.1002/adom.202300177} {\bibfield  {journal} {\bibinfo  {journal} {Advanced
  Optical Materials}\ }\textbf {\bibinfo {volume} {11}},\ \bibinfo {pages}
  {1--7} (\bibinfo {year} {2023})}\BibitemShut {NoStop}%
\bibitem [{\citenamefont {Klein}, \citenamefont {Clauson},\ and\ \citenamefont
  {Cogan}(1995{\natexlab{a}})}]{Klein1995}%
  \BibitemOpen
  \bibfield  {author} {\bibinfo {author} {\bibfnamefont {J.}~\bibnamefont
  {Klein}}, \bibinfo {author} {\bibfnamefont {S.}~\bibnamefont {Clauson}}, \
  and\ \bibinfo {author} {\bibfnamefont {S.}~\bibnamefont {Cogan}},\ }\bibfield
   {title} {\enquote {\bibinfo {title} {{Reactive IrO$_2$ sputtering in
  reducing/oxidizing atmospheres}},}\ }\href {\doibase 10.1557/JMR.1995.0328}
  {\bibfield  {journal} {\bibinfo  {journal} {Journal of Materials Research}\
  }\textbf {\bibinfo {volume} {10}},\ \bibinfo {pages} {328--333} (\bibinfo
  {year} {1995}{\natexlab{a}})}\BibitemShut {NoStop}%
\bibitem [{\citenamefont {Klein}, \citenamefont {Clauson},\ and\ \citenamefont
  {Cogan}(1995{\natexlab{b}})}]{Klein1995a}%
  \BibitemOpen
  \bibfield  {author} {\bibinfo {author} {\bibfnamefont {J.}~\bibnamefont
  {Klein}}, \bibinfo {author} {\bibfnamefont {S.}~\bibnamefont {Clauson}}, \
  and\ \bibinfo {author} {\bibfnamefont {S.}~\bibnamefont {Cogan}},\ }\bibfield
   {title} {\enquote {\bibinfo {title} {{Reactive IrO$_2$ sputtering in
  reducing/oxidizing atmospheres}},}\ }\href {\doibase 10.1557/JMR.1995.0328}
  {\bibfield  {journal} {\bibinfo  {journal} {Journal of Materials Research}\
  }\textbf {\bibinfo {volume} {10}},\ \bibinfo {pages} {328--333} (\bibinfo
  {year} {1995}{\natexlab{b}})}\BibitemShut {NoStop}%
\bibitem [{\citenamefont {Balu}\ \emph {et~al.}(1996)\citenamefont {Balu},
  \citenamefont {Chen}, \citenamefont {Jiang}, \citenamefont {Kuah},
  \citenamefont {Lee}, \citenamefont {Chu}, \citenamefont {Jones},
  \citenamefont {Zurcher}, \citenamefont {Taylor},\ and\ \citenamefont
  {Gillespie}}]{Balu1996}%
  \BibitemOpen
  \bibfield  {author} {\bibinfo {author} {\bibfnamefont {V.}~\bibnamefont
  {Balu}}, \bibinfo {author} {\bibfnamefont {T.-S.}\ \bibnamefont {Chen}},
  \bibinfo {author} {\bibfnamefont {B.}~\bibnamefont {Jiang}}, \bibinfo
  {author} {\bibfnamefont {S.-H.}\ \bibnamefont {Kuah}}, \bibinfo {author}
  {\bibfnamefont {J.~C.}\ \bibnamefont {Lee}}, \bibinfo {author} {\bibfnamefont
  {P.}~\bibnamefont {Chu}}, \bibinfo {author} {\bibfnamefont {R.~E.}\
  \bibnamefont {Jones}}, \bibinfo {author} {\bibfnamefont {P.}~\bibnamefont
  {Zurcher}}, \bibinfo {author} {\bibfnamefont {D.~J.}\ \bibnamefont {Taylor}},
  \ and\ \bibinfo {author} {\bibfnamefont {S.}~\bibnamefont {Gillespie}},\
  }\bibfield  {title} {\enquote {\bibinfo {title} {{Electrode Materials for
  Ferroelectric Capacitors: Properties of Reactive DC Sputtered IrO$_2$ Thin
  Films}},}\ }\href {\doibase 10.1557/PROC-433-139} {\bibfield  {journal}
  {\bibinfo  {journal} {MRS Proceedings}\ }\textbf {\bibinfo {volume} {433}},\
  \bibinfo {pages} {139} (\bibinfo {year} {1996})}\BibitemShut {NoStop}%
\bibitem [{\citenamefont {Shimizu}\ \emph {et~al.}(1997)\citenamefont
  {Shimizu}, \citenamefont {Fujisawa}, \citenamefont {Hyodo}, \citenamefont
  {Nakashima}, \citenamefont {Niu}, \citenamefont {Okino},\ and\ \citenamefont
  {Shiosaki}}]{Shimizu1997}%
  \BibitemOpen
  \bibfield  {author} {\bibinfo {author} {\bibfnamefont {M.}~\bibnamefont
  {Shimizu}}, \bibinfo {author} {\bibfnamefont {H.}~\bibnamefont {Fujisawa}},
  \bibinfo {author} {\bibfnamefont {S.}~\bibnamefont {Hyodo}}, \bibinfo
  {author} {\bibfnamefont {S.}~\bibnamefont {Nakashima}}, \bibinfo {author}
  {\bibfnamefont {H.}~\bibnamefont {Niu}}, \bibinfo {author} {\bibfnamefont
  {H.}~\bibnamefont {Okino}}, \ and\ \bibinfo {author} {\bibfnamefont
  {T.}~\bibnamefont {Shiosaki}},\ }\bibfield  {title} {\enquote {\bibinfo
  {title} {{Effects of Sputtered Ir and IrO$_2$ Electrodes on the Properties of
  PZT Thin Films Deposited By MOCVD}},}\ }\href {\doibase 10.1557/PROC-493-159}
  {\bibfield  {journal} {\bibinfo  {journal} {MRS Proceedings}\ }\textbf
  {\bibinfo {volume} {493}},\ \bibinfo {pages} {159} (\bibinfo {year}
  {1997})}\BibitemShut {NoStop}%
\bibitem [{\citenamefont {Liu}\ \emph {et~al.}(2008)\citenamefont {Liu},
  \citenamefont {Yu}, \citenamefont {Son},\ and\ \citenamefont
  {Joo}}]{Liu2008}%
  \BibitemOpen
  \bibfield  {author} {\bibinfo {author} {\bibfnamefont {D.-Q.}\ \bibnamefont
  {Liu}}, \bibinfo {author} {\bibfnamefont {S.-H.}\ \bibnamefont {Yu}},
  \bibinfo {author} {\bibfnamefont {S.-W.}\ \bibnamefont {Son}}, \ and\
  \bibinfo {author} {\bibfnamefont {S.-K.}\ \bibnamefont {Joo}},\ }\bibfield
  {title} {\enquote {\bibinfo {title} {{Supercapacitive Studies on IrO$_2$ Thin
  Film Electrodes Prepared by Radio Frequency Magnetron Sputtering}},}\ }\href
  {\doibase 10.1149/1.2977787} {\bibfield  {journal} {\bibinfo  {journal}
  {Electrochemical and Solid-State Letters}\ }\textbf {\bibinfo {volume}
  {11}},\ \bibinfo {pages} {A206} (\bibinfo {year} {2008})}\BibitemShut
  {NoStop}%
\bibitem [{\citenamefont {van Ooyen}\ \emph {et~al.}(2009)\citenamefont {van
  Ooyen}, \citenamefont {Topalov}, \citenamefont {Ganske}, \citenamefont
  {Mokwa},\ and\ \citenamefont {Schnakenberg}}]{VanOoyen2009}%
  \BibitemOpen
  \bibfield  {author} {\bibinfo {author} {\bibfnamefont {A.}~\bibnamefont {van
  Ooyen}}, \bibinfo {author} {\bibfnamefont {G.}~\bibnamefont {Topalov}},
  \bibinfo {author} {\bibfnamefont {G.}~\bibnamefont {Ganske}}, \bibinfo
  {author} {\bibfnamefont {W.}~\bibnamefont {Mokwa}}, \ and\ \bibinfo {author}
  {\bibfnamefont {U.}~\bibnamefont {Schnakenberg}},\ }\bibfield  {title}
  {\enquote {\bibinfo {title} {{Iridium oxide deposited by pulsed dc-sputtering
  for stimulation electrodes}},}\ }\href {\doibase
  10.1088/0960-1317/19/7/074009} {\bibfield  {journal} {\bibinfo  {journal}
  {Journal of Micromechanics and Microengineering}\ }\textbf {\bibinfo {volume}
  {19}},\ \bibinfo {pages} {074009} (\bibinfo {year} {2009})}\BibitemShut
  {NoStop}%
\bibitem [{\citenamefont {Cogan}\ \emph {et~al.}(2009)\citenamefont {Cogan},
  \citenamefont {Ehrlich}, \citenamefont {Plante}, \citenamefont {Smirnov},
  \citenamefont {Shire}, \citenamefont {Gingerich},\ and\ \citenamefont
  {Rizzo}}]{Cogan2009}%
  \BibitemOpen
  \bibfield  {author} {\bibinfo {author} {\bibfnamefont {S.~F.}\ \bibnamefont
  {Cogan}}, \bibinfo {author} {\bibfnamefont {J.}~\bibnamefont {Ehrlich}},
  \bibinfo {author} {\bibfnamefont {T.~D.}\ \bibnamefont {Plante}}, \bibinfo
  {author} {\bibfnamefont {A.}~\bibnamefont {Smirnov}}, \bibinfo {author}
  {\bibfnamefont {D.~B.}\ \bibnamefont {Shire}}, \bibinfo {author}
  {\bibfnamefont {M.}~\bibnamefont {Gingerich}}, \ and\ \bibinfo {author}
  {\bibfnamefont {J.~F.}\ \bibnamefont {Rizzo}},\ }\bibfield  {title} {\enquote
  {\bibinfo {title} {{Sputtered iridium oxide films for neural stimulation
  electrodes}},}\ }\href {\doibase 10.1002/jbm.b.31223} {\bibfield  {journal}
  {\bibinfo  {journal} {Journal of Biomedical Materials Research Part B:
  Applied Biomaterials}\ }\textbf {\bibinfo {volume} {89B}},\ \bibinfo {pages}
  {353--361} (\bibinfo {year} {2009})}\BibitemShut {NoStop}%
\bibitem [{\citenamefont {Kim}\ \emph {et~al.}(2007)\citenamefont {Kim},
  \citenamefont {Kil}, \citenamefont {Yeom}, \citenamefont {Roh}, \citenamefont
  {Kwak},\ and\ \citenamefont {Kim}}]{Kim2007}%
  \BibitemOpen
  \bibfield  {author} {\bibinfo {author} {\bibfnamefont {J.-H.}\ \bibnamefont
  {Kim}}, \bibinfo {author} {\bibfnamefont {D.-S.}\ \bibnamefont {Kil}},
  \bibinfo {author} {\bibfnamefont {S.-J.}\ \bibnamefont {Yeom}}, \bibinfo
  {author} {\bibfnamefont {J.-S.}\ \bibnamefont {Roh}}, \bibinfo {author}
  {\bibfnamefont {N.-J.}\ \bibnamefont {Kwak}}, \ and\ \bibinfo {author}
  {\bibfnamefont {J.-W.}\ \bibnamefont {Kim}},\ }\bibfield  {title} {\enquote
  {\bibinfo {title} {{Modified atomic layer deposition of RuO$_2$ thin films
  for capacitor electrodes}},}\ }\href {\doibase 10.1063/1.2767769} {\bibfield
  {journal} {\bibinfo  {journal} {Applied Physics Letters}\ }\textbf {\bibinfo
  {volume} {91}},\ \bibinfo {pages} {1--4} (\bibinfo {year}
  {2007})}\BibitemShut {NoStop}%
\bibitem [{\citenamefont {Kwon}\ \emph {et~al.}(2007)\citenamefont {Kwon},
  \citenamefont {Kwon}, \citenamefont {Kim}, \citenamefont {Jeong},
  \citenamefont {Kim},\ and\ \citenamefont {Kang}}]{Kwon2007}%
  \BibitemOpen
  \bibfield  {author} {\bibinfo {author} {\bibfnamefont {S.-H.}\ \bibnamefont
  {Kwon}}, \bibinfo {author} {\bibfnamefont {O.-K.}\ \bibnamefont {Kwon}},
  \bibinfo {author} {\bibfnamefont {J.-H.}\ \bibnamefont {Kim}}, \bibinfo
  {author} {\bibfnamefont {S.-J.}\ \bibnamefont {Jeong}}, \bibinfo {author}
  {\bibfnamefont {S.-W.}\ \bibnamefont {Kim}}, \ and\ \bibinfo {author}
  {\bibfnamefont {S.-W.}\ \bibnamefont {Kang}},\ }\bibfield  {title} {\enquote
  {\bibinfo {title} {{Improvement of the Morphological Stability by Stacking
  RuO$_2$ on Ru Thin Films with Atomic Layer Deposition}},}\ }\href {\doibase
  10.1149/1.2750448} {\bibfield  {journal} {\bibinfo  {journal} {Journal of The
  Electrochemical Society}\ }\textbf {\bibinfo {volume} {154}},\ \bibinfo
  {pages} {H773} (\bibinfo {year} {2007})}\BibitemShut {NoStop}%
\bibitem [{\citenamefont {Park}\ \emph {et~al.}(2014)\citenamefont {Park},
  \citenamefont {Yeo}, \citenamefont {Cheon}, \citenamefont {Kim},
  \citenamefont {Kim}, \citenamefont {Kim}, \citenamefont {Hong},\ and\
  \citenamefont {Lee}}]{Park2014}%
  \BibitemOpen
  \bibfield  {author} {\bibinfo {author} {\bibfnamefont {J.-Y.}\ \bibnamefont
  {Park}}, \bibinfo {author} {\bibfnamefont {S.}~\bibnamefont {Yeo}}, \bibinfo
  {author} {\bibfnamefont {T.}~\bibnamefont {Cheon}}, \bibinfo {author}
  {\bibfnamefont {S.-H.}\ \bibnamefont {Kim}}, \bibinfo {author} {\bibfnamefont
  {M.-K.}\ \bibnamefont {Kim}}, \bibinfo {author} {\bibfnamefont
  {H.}~\bibnamefont {Kim}}, \bibinfo {author} {\bibfnamefont {T.~E.}\
  \bibnamefont {Hong}}, \ and\ \bibinfo {author} {\bibfnamefont {D.-J.}\
  \bibnamefont {Lee}},\ }\bibfield  {title} {\enquote {\bibinfo {title}
  {{Growth of highly conformal ruthenium-oxide thin films with enhanced
  nucleation by atomic layer deposition}},}\ }\href {\doibase
  10.1016/j.jallcom.2014.04.186} {\bibfield  {journal} {\bibinfo  {journal}
  {Journal of Alloys and Compounds}\ }\textbf {\bibinfo {volume} {610}},\
  \bibinfo {pages} {529--539} (\bibinfo {year} {2014})}\BibitemShut {NoStop}%
\bibitem [{\citenamefont {Jung}\ \emph {et~al.}(2014)\citenamefont {Jung},
  \citenamefont {Han}, \citenamefont {Jung}, \citenamefont {Park},
  \citenamefont {Hwang}, \citenamefont {Son}, \citenamefont {Kim},
  \citenamefont {Chung},\ and\ \citenamefont {An}}]{Jung2014}%
  \BibitemOpen
  \bibfield  {author} {\bibinfo {author} {\bibfnamefont {H.~J.}\ \bibnamefont
  {Jung}}, \bibinfo {author} {\bibfnamefont {J.~H.}\ \bibnamefont {Han}},
  \bibinfo {author} {\bibfnamefont {E.~A.}\ \bibnamefont {Jung}}, \bibinfo
  {author} {\bibfnamefont {B.~K.}\ \bibnamefont {Park}}, \bibinfo {author}
  {\bibfnamefont {J.-H.}\ \bibnamefont {Hwang}}, \bibinfo {author}
  {\bibfnamefont {S.~U.}\ \bibnamefont {Son}}, \bibinfo {author} {\bibfnamefont
  {C.~G.}\ \bibnamefont {Kim}}, \bibinfo {author} {\bibfnamefont {T.-m.}\
  \bibnamefont {Chung}}, \ and\ \bibinfo {author} {\bibfnamefont {K.-s.}\
  \bibnamefont {An}},\ }\bibfield  {title} {\enquote {\bibinfo {title} {{Atomic
  Layer Deposition of Ruthenium and Ruthenium Oxide Thin Films from a
  Zero-Valent (1,5-Hexadiene)(1-isopropyl-4-methylbenzene)ruthenium Complex and
  O 2}},}\ }\href {\doibase 10.1021/cm5035485} {\bibfield  {journal} {\bibinfo
  {journal} {Chemistry of Materials}\ }\textbf {\bibinfo {volume} {26}},\
  \bibinfo {pages} {7083--7090} (\bibinfo {year} {2014})},\ \Eprint
  {http://arxiv.org/abs/10.1021/cm5035485} {10.1021/cm5035485 [dx.doi.org]}
  \BibitemShut {NoStop}%
\bibitem [{\citenamefont {H{\"{a}}m{\"{a}}l{\"{a}}inen}\ \emph
  {et~al.}(2011)\citenamefont {H{\"{a}}m{\"{a}}l{\"{a}}inen}, \citenamefont
  {Hatanp{\"{a}}{\"{a}}}, \citenamefont {Puukilainen}, \citenamefont
  {Sajavaara}, \citenamefont {Ritala},\ and\ \citenamefont
  {Leskel{\"{a}}}}]{Hamalainen2011}%
  \BibitemOpen
  \bibfield  {author} {\bibinfo {author} {\bibfnamefont {J.}~\bibnamefont
  {H{\"{a}}m{\"{a}}l{\"{a}}inen}}, \bibinfo {author} {\bibfnamefont
  {T.}~\bibnamefont {Hatanp{\"{a}}{\"{a}}}}, \bibinfo {author} {\bibfnamefont
  {E.}~\bibnamefont {Puukilainen}}, \bibinfo {author} {\bibfnamefont
  {T.}~\bibnamefont {Sajavaara}}, \bibinfo {author} {\bibfnamefont
  {M.}~\bibnamefont {Ritala}}, \ and\ \bibinfo {author} {\bibfnamefont
  {M.}~\bibnamefont {Leskel{\"{a}}}},\ }\bibfield  {title} {\enquote {\bibinfo
  {title} {{Iridium metal and iridium oxide thin films grown by atomic layer
  deposition at low temperatures}},}\ }\href {\doibase 10.1039/c1jm12245b}
  {\bibfield  {journal} {\bibinfo  {journal} {Journal of Materials Chemistry}\
  }\textbf {\bibinfo {volume} {21}},\ \bibinfo {pages} {16488} (\bibinfo {year}
  {2011})}\BibitemShut {NoStop}%
\bibitem [{\citenamefont {Mattinen}\ \emph {et~al.}(2016)\citenamefont
  {Mattinen}, \citenamefont {H{\"{a}}m{\"{a}}l{\"{a}}inen}, \citenamefont
  {Gao}, \citenamefont {Jalkanen}, \citenamefont {Mizohata}, \citenamefont
  {R{\"{a}}is{\"{a}}nen}, \citenamefont {Puurunen}, \citenamefont {Ritala},\
  and\ \citenamefont {Leskel{\"{a}}}}]{Mattinen2016}%
  \BibitemOpen
  \bibfield  {author} {\bibinfo {author} {\bibfnamefont {M.}~\bibnamefont
  {Mattinen}}, \bibinfo {author} {\bibfnamefont {J.}~\bibnamefont
  {H{\"{a}}m{\"{a}}l{\"{a}}inen}}, \bibinfo {author} {\bibfnamefont
  {F.}~\bibnamefont {Gao}}, \bibinfo {author} {\bibfnamefont {P.}~\bibnamefont
  {Jalkanen}}, \bibinfo {author} {\bibfnamefont {K.}~\bibnamefont {Mizohata}},
  \bibinfo {author} {\bibfnamefont {J.}~\bibnamefont {R{\"{a}}is{\"{a}}nen}},
  \bibinfo {author} {\bibfnamefont {R.~L.}\ \bibnamefont {Puurunen}}, \bibinfo
  {author} {\bibfnamefont {M.}~\bibnamefont {Ritala}}, \ and\ \bibinfo {author}
  {\bibfnamefont {M.}~\bibnamefont {Leskel{\"{a}}}},\ }\bibfield  {title}
  {\enquote {\bibinfo {title} {{Nucleation and Conformality of Iridium and
  Iridium Oxide Thin Films Grown by Atomic Layer Deposition}},}\ }\href
  {\doibase 10.1021/acs.langmuir.6b03007} {\bibfield  {journal} {\bibinfo
  {journal} {Langmuir}\ }\textbf {\bibinfo {volume} {32}},\ \bibinfo {pages}
  {10559--10569} (\bibinfo {year} {2016})}\BibitemShut {NoStop}%
\bibitem [{\citenamefont {Kim}\ \emph {et~al.}(2008)\citenamefont {Kim},
  \citenamefont {Kwon}, \citenamefont {Kwak},\ and\ \citenamefont
  {Kang}}]{Kim2008a}%
  \BibitemOpen
  \bibfield  {author} {\bibinfo {author} {\bibfnamefont {S.-W.}\ \bibnamefont
  {Kim}}, \bibinfo {author} {\bibfnamefont {S.-H.}\ \bibnamefont {Kwon}},
  \bibinfo {author} {\bibfnamefont {D.-K.}\ \bibnamefont {Kwak}}, \ and\
  \bibinfo {author} {\bibfnamefont {S.-W.}\ \bibnamefont {Kang}},\ }\bibfield
  {title} {\enquote {\bibinfo {title} {{Phase control of iridium and iridium
  oxide thin films in atomic layer deposition}},}\ }\href {\doibase
  10.1063/1.2836965} {\bibfield  {journal} {\bibinfo  {journal} {Journal of
  Applied Physics}\ }\textbf {\bibinfo {volume} {103}},\ \bibinfo {pages}
  {1--7} (\bibinfo {year} {2008})}\BibitemShut {NoStop}%
\bibitem [{\citenamefont {H{\"{a}}m{\"{a}}l{\"{a}}inen}\ \emph
  {et~al.}(2008)\citenamefont {H{\"{a}}m{\"{a}}l{\"{a}}inen}, \citenamefont
  {Kemell}, \citenamefont {Munnik}, \citenamefont {Kreissig}, \citenamefont
  {Ritala},\ and\ \citenamefont {Leskel{\"{a}}}}]{Hamalainen2008}%
  \BibitemOpen
  \bibfield  {author} {\bibinfo {author} {\bibfnamefont {J.}~\bibnamefont
  {H{\"{a}}m{\"{a}}l{\"{a}}inen}}, \bibinfo {author} {\bibfnamefont
  {M.}~\bibnamefont {Kemell}}, \bibinfo {author} {\bibfnamefont
  {F.}~\bibnamefont {Munnik}}, \bibinfo {author} {\bibfnamefont
  {U.}~\bibnamefont {Kreissig}}, \bibinfo {author} {\bibfnamefont
  {M.}~\bibnamefont {Ritala}}, \ and\ \bibinfo {author} {\bibfnamefont
  {M.}~\bibnamefont {Leskel{\"{a}}}},\ }\bibfield  {title} {\enquote {\bibinfo
  {title} {{Atomic Layer Deposition of Iridium Oxide Thin Films from
  Ir(acac)$_3$ and Ozone}},}\ }\href {\doibase 10.1021/cm7030224} {\bibfield
  {journal} {\bibinfo  {journal} {Chemistry of Materials}\ }\textbf {\bibinfo
  {volume} {20}},\ \bibinfo {pages} {2903--2907} (\bibinfo {year}
  {2008})}\BibitemShut {NoStop}%
\bibitem [{\citenamefont {Simon}\ \emph {et~al.}(2021)\citenamefont {Simon},
  \citenamefont {Asplund}, \citenamefont {Stieglitz},\ and\ \citenamefont
  {Bucher}}]{Simon2021}%
  \BibitemOpen
  \bibfield  {author} {\bibinfo {author} {\bibfnamefont {N.}~\bibnamefont
  {Simon}}, \bibinfo {author} {\bibfnamefont {M.}~\bibnamefont {Asplund}},
  \bibinfo {author} {\bibfnamefont {T.}~\bibnamefont {Stieglitz}}, \ and\
  \bibinfo {author} {\bibfnamefont {V.}~\bibnamefont {Bucher}},\ }\bibfield
  {title} {\enquote {\bibinfo {title} {{Plasma Enhanced Atomic Layer Deposition
  of Iridium Oxide for Application in Miniaturized Neural Implants}},}\ }\href
  {\doibase 10.1515/cdbme-2021-2137} {\bibfield  {journal} {\bibinfo  {journal}
  {Current Directions in Biomedical Engineering}\ }\textbf {\bibinfo {volume}
  {7}},\ \bibinfo {pages} {539--542} (\bibinfo {year} {2021})}\BibitemShut
  {NoStop}%
\bibitem [{\citenamefont {Kim}, \citenamefont {Mannhart},\ and\ \citenamefont
  {Braun}(2021)}]{Kim2021}%
  \BibitemOpen
  \bibfield  {author} {\bibinfo {author} {\bibfnamefont {D.~Y.}\ \bibnamefont
  {Kim}}, \bibinfo {author} {\bibfnamefont {J.}~\bibnamefont {Mannhart}}, \
  and\ \bibinfo {author} {\bibfnamefont {W.}~\bibnamefont {Braun}},\ }\bibfield
   {title} {\enquote {\bibinfo {title} {{Thermal laser evaporation for the
  growth of oxide films}},}\ }\href {\doibase 10.1063/5.0055237} {\bibfield
  {journal} {\bibinfo  {journal} {APL Materials}\ }\textbf {\bibinfo {volume}
  {9}},\ \bibinfo {pages} {081105} (\bibinfo {year} {2021})}\BibitemShut
  {NoStop}%
\bibitem [{\citenamefont {Knapp}\ \emph {et~al.}(2007)\citenamefont {Knapp},
  \citenamefont {Crihan}, \citenamefont {Seitsonen}, \citenamefont {Lundgren},
  \citenamefont {Resta}, \citenamefont {Andersen},\ and\ \citenamefont
  {Over}}]{Knapp2007b}%
  \BibitemOpen
  \bibfield  {author} {\bibinfo {author} {\bibfnamefont {M.}~\bibnamefont
  {Knapp}}, \bibinfo {author} {\bibfnamefont {D.}~\bibnamefont {Crihan}},
  \bibinfo {author} {\bibfnamefont {A.~P.}\ \bibnamefont {Seitsonen}}, \bibinfo
  {author} {\bibfnamefont {E.}~\bibnamefont {Lundgren}}, \bibinfo {author}
  {\bibfnamefont {A.}~\bibnamefont {Resta}}, \bibinfo {author} {\bibfnamefont
  {J.~N.}\ \bibnamefont {Andersen}}, \ and\ \bibinfo {author} {\bibfnamefont
  {H.}~\bibnamefont {Over}},\ }\bibfield  {title} {\enquote {\bibinfo {title}
  {{Complex interaction of hydrogen with the RuO$_2$(110) surface}},}\ }\href
  {\doibase 10.1021/jp0667339} {\bibfield  {journal} {\bibinfo  {journal}
  {Journal of Physical Chemistry C}\ }\textbf {\bibinfo {volume} {111}},\
  \bibinfo {pages} {5363--5373} (\bibinfo {year} {2007})}\BibitemShut {NoStop}%
\bibitem [{\citenamefont {Kim}, \citenamefont {Seitsonen},\ and\ \citenamefont
  {Over}(2000)}]{Kim2000}%
  \BibitemOpen
  \bibfield  {author} {\bibinfo {author} {\bibfnamefont {Y.~D.}\ \bibnamefont
  {Kim}}, \bibinfo {author} {\bibfnamefont {A.~P.}\ \bibnamefont {Seitsonen}},
  \ and\ \bibinfo {author} {\bibfnamefont {H.}~\bibnamefont {Over}},\
  }\bibfield  {title} {\enquote {\bibinfo {title} {{Atomic geometry of
  oxygen-rich Ru(0001) surfaces: Coexistence of ($1\times1$)O and RuO$_2$(110)
  domains}},}\ }\href {\doibase 10.1016/S0039-6028(00)00733-0} {\bibfield
  {journal} {\bibinfo  {journal} {Surface Science}\ }\textbf {\bibinfo {volume}
  {465}},\ \bibinfo {pages} {1--8} (\bibinfo {year} {2000})}\BibitemShut
  {NoStop}%
\bibitem [{\citenamefont {A{\ss}mann}\ \emph
  {et~al.}(2005{\natexlab{b}})\citenamefont {A{\ss}mann}, \citenamefont
  {Crihan}, \citenamefont {Knapp}, \citenamefont {Lundgren}, \citenamefont
  {L{\"{o}}ffler}, \citenamefont {Muhler}, \citenamefont {Narkhede},
  \citenamefont {Over}, \citenamefont {Schmid}, \citenamefont {Seitsonen},\
  and\ \citenamefont {Varga}}]{Assmann2005}%
  \BibitemOpen
  \bibfield  {author} {\bibinfo {author} {\bibfnamefont {J.}~\bibnamefont
  {A{\ss}mann}}, \bibinfo {author} {\bibfnamefont {D.}~\bibnamefont {Crihan}},
  \bibinfo {author} {\bibfnamefont {M.}~\bibnamefont {Knapp}}, \bibinfo
  {author} {\bibfnamefont {E.}~\bibnamefont {Lundgren}}, \bibinfo {author}
  {\bibfnamefont {E.}~\bibnamefont {L{\"{o}}ffler}}, \bibinfo {author}
  {\bibfnamefont {M.}~\bibnamefont {Muhler}}, \bibinfo {author} {\bibfnamefont
  {V.}~\bibnamefont {Narkhede}}, \bibinfo {author} {\bibfnamefont
  {H.}~\bibnamefont {Over}}, \bibinfo {author} {\bibfnamefont {M.}~\bibnamefont
  {Schmid}}, \bibinfo {author} {\bibfnamefont {A.~P.}\ \bibnamefont
  {Seitsonen}}, \ and\ \bibinfo {author} {\bibfnamefont {P.}~\bibnamefont
  {Varga}},\ }\bibfield  {title} {\enquote {\bibinfo {title} {{Understanding
  the structural deactivation of ruthenium catalysts on an atomic scale under
  both oxidizing and reducing conditions}},}\ }\href {\doibase
  10.1002/anie.200461805} {\bibfield  {journal} {\bibinfo  {journal}
  {Angewandte Chemie - International Edition}\ }\textbf {\bibinfo {volume}
  {44}},\ \bibinfo {pages} {917--920} (\bibinfo {year}
  {2005}{\natexlab{b}})}\BibitemShut {NoStop}%
\bibitem [{\citenamefont {Jia}\ \emph {et~al.}(1995)\citenamefont {Jia},
  \citenamefont {Wu}, \citenamefont {Foltyn}, \citenamefont {Findikoglu},
  \citenamefont {Tiwari}, \citenamefont {Zheng},\ and\ \citenamefont
  {Jow}}]{Jia1995}%
  \BibitemOpen
  \bibfield  {author} {\bibinfo {author} {\bibfnamefont {Q.~X.}\ \bibnamefont
  {Jia}}, \bibinfo {author} {\bibfnamefont {X.~D.}\ \bibnamefont {Wu}},
  \bibinfo {author} {\bibfnamefont {S.~R.}\ \bibnamefont {Foltyn}}, \bibinfo
  {author} {\bibfnamefont {A.~T.}\ \bibnamefont {Findikoglu}}, \bibinfo
  {author} {\bibfnamefont {P.}~\bibnamefont {Tiwari}}, \bibinfo {author}
  {\bibfnamefont {J.~P.}\ \bibnamefont {Zheng}}, \ and\ \bibinfo {author}
  {\bibfnamefont {T.~R.}\ \bibnamefont {Jow}},\ }\bibfield  {title} {\enquote
  {\bibinfo {title} {{Heteroepitaxial growth of highly conductive metal oxide
  RuO$_2$ thin films by pulsed laser deposition}},}\ }\href {\doibase
  10.1063/1.115054} {\bibfield  {journal} {\bibinfo  {journal} {Applied Physics
  Letters}\ }\textbf {\bibinfo {volume} {67}},\ \bibinfo {pages} {1677}
  (\bibinfo {year} {1995})}\BibitemShut {NoStop}%
\bibitem [{\citenamefont {Abb}\ \emph {et~al.}(2019)\citenamefont {Abb},
  \citenamefont {Weber}, \citenamefont {Glatthaar},\ and\ \citenamefont
  {Over}}]{Abb2019}%
  \BibitemOpen
  \bibfield  {author} {\bibinfo {author} {\bibfnamefont {M.~J.}\ \bibnamefont
  {Abb}}, \bibinfo {author} {\bibfnamefont {T.}~\bibnamefont {Weber}}, \bibinfo
  {author} {\bibfnamefont {L.}~\bibnamefont {Glatthaar}}, \ and\ \bibinfo
  {author} {\bibfnamefont {H.}~\bibnamefont {Over}},\ }\bibfield  {title}
  {\enquote {\bibinfo {title} {{Growth of Ultrathin Single-Crystalline
  IrO$_2$(110) Films on a TiO$_2$(110) Single Crystal}},}\ }\href {\doibase
  10.1021/acs.langmuir.9b00667} {\bibfield  {journal} {\bibinfo  {journal}
  {Langmuir}\ }\textbf {\bibinfo {volume} {35}},\ \bibinfo {pages} {7720--7726}
  (\bibinfo {year} {2019})}\BibitemShut {NoStop}%
\bibitem [{\citenamefont {Braun}\ and\ \citenamefont
  {Mannhart}(2019)}]{Braun2019}%
  \BibitemOpen
  \bibfield  {author} {\bibinfo {author} {\bibfnamefont {W.}~\bibnamefont
  {Braun}}\ and\ \bibinfo {author} {\bibfnamefont {J.}~\bibnamefont
  {Mannhart}},\ }\bibfield  {title} {\enquote {\bibinfo {title} {{Film
  deposition by thermal laser evaporation}},}\ }\href {\doibase
  10.1063/1.5111678} {\bibfield  {journal} {\bibinfo  {journal} {AIP Advances}\
  }\textbf {\bibinfo {volume} {9}},\ \bibinfo {pages} {085310} (\bibinfo {year}
  {2019})}\BibitemShut {NoStop}%
\bibitem [{\citenamefont {Zheng}\ and\ \citenamefont {Kwok}(1993)}]{Zheng1993}%
  \BibitemOpen
  \bibfield  {author} {\bibinfo {author} {\bibfnamefont {J.~P.}\ \bibnamefont
  {Zheng}}\ and\ \bibinfo {author} {\bibfnamefont {H.~S.}\ \bibnamefont
  {Kwok}},\ }\bibfield  {title} {\enquote {\bibinfo {title} {{Low resistivity
  indium tin oxide films by pulsed laser deposition}},}\ }\href {\doibase
  10.1063/1.109736} {\bibfield  {journal} {\bibinfo  {journal} {Applied Physics
  Letters}\ }\textbf {\bibinfo {volume} {63}},\ \bibinfo {pages} {1--3}
  (\bibinfo {year} {1993})}\BibitemShut {NoStop}%
\bibitem [{\citenamefont {Adiga}\ \emph {et~al.}(2022)\citenamefont {Adiga},
  \citenamefont {Nunn}, \citenamefont {Wong}, \citenamefont {Manjeshwar},
  \citenamefont {Nair}, \citenamefont {Jalan},\ and\ \citenamefont
  {Stoerzinger}}]{Adiga2022}%
  \BibitemOpen
  \bibfield  {author} {\bibinfo {author} {\bibfnamefont {P.}~\bibnamefont
  {Adiga}}, \bibinfo {author} {\bibfnamefont {W.}~\bibnamefont {Nunn}},
  \bibinfo {author} {\bibfnamefont {C.}~\bibnamefont {Wong}}, \bibinfo {author}
  {\bibfnamefont {A.~K.}\ \bibnamefont {Manjeshwar}}, \bibinfo {author}
  {\bibfnamefont {S.}~\bibnamefont {Nair}}, \bibinfo {author} {\bibfnamefont
  {B.}~\bibnamefont {Jalan}}, \ and\ \bibinfo {author} {\bibfnamefont {K.~A.}\
  \bibnamefont {Stoerzinger}},\ }\bibfield  {title} {\enquote {\bibinfo {title}
  {{Breaking OER and CER scaling relations via strain and its relaxation in
  RuO$_2$ (101)}},}\ }\href {\doibase 10.1016/j.mtener.2022.101087} {\bibfield
  {journal} {\bibinfo  {journal} {Materials Today Energy}\ }\textbf {\bibinfo
  {volume} {28}},\ \bibinfo {pages} {101087} (\bibinfo {year}
  {2022})}\BibitemShut {NoStop}%
\bibitem [{\citenamefont {Serventi}\ \emph {et~al.}(2001)\citenamefont
  {Serventi}, \citenamefont {Khakani}, \citenamefont {Saint-Jacques},\ and\
  \citenamefont {Rickerby}}]{Serventi2001}%
  \BibitemOpen
  \bibfield  {author} {\bibinfo {author} {\bibfnamefont {A.~M.}\ \bibnamefont
  {Serventi}}, \bibinfo {author} {\bibfnamefont {M.~A.~E.}\ \bibnamefont
  {Khakani}}, \bibinfo {author} {\bibfnamefont {R.~G.}\ \bibnamefont
  {Saint-Jacques}}, \ and\ \bibinfo {author} {\bibfnamefont {D.~G.}\
  \bibnamefont {Rickerby}},\ }\bibfield  {title} {\enquote {\bibinfo {title}
  {{Highly textured nanostructure of pulsed laser deposited IrO$_2$ thin films
  as investigated by transmission electron microscopy}},}\ }\href {\doibase
  10.1557/JMR.2001.0320} {\bibfield  {journal} {\bibinfo  {journal} {Journal of
  Materials Research}\ }\textbf {\bibinfo {volume} {16}},\ \bibinfo {pages}
  {2336--2342} (\bibinfo {year} {2001})}\BibitemShut {NoStop}%
\bibitem [{\citenamefont {Wang}\ \emph {et~al.}(2016)\citenamefont {Wang},
  \citenamefont {Lee}, \citenamefont {Vilmercati}, \citenamefont {Lee},
  \citenamefont {Weitering},\ and\ \citenamefont {Snijders}}]{Wang2016a}%
  \BibitemOpen
  \bibfield  {author} {\bibinfo {author} {\bibfnamefont {Y.}~\bibnamefont
  {Wang}}, \bibinfo {author} {\bibfnamefont {S.}~\bibnamefont {Lee}}, \bibinfo
  {author} {\bibfnamefont {P.}~\bibnamefont {Vilmercati}}, \bibinfo {author}
  {\bibfnamefont {H.~N.}\ \bibnamefont {Lee}}, \bibinfo {author} {\bibfnamefont
  {H.~H.}\ \bibnamefont {Weitering}}, \ and\ \bibinfo {author} {\bibfnamefont
  {P.~C.}\ \bibnamefont {Snijders}},\ }\bibfield  {title} {\enquote {\bibinfo
  {title} {{Atomically flat reconstructed rutile TiO$_2$(001) surfaces for
  oxide film growth}},}\ }\href {\doibase 10.1063/1.4942967} {\bibfield
  {journal} {\bibinfo  {journal} {Applied Physics Letters}\ }\textbf {\bibinfo
  {volume} {108}},\ \bibinfo {pages} {8--10} (\bibinfo {year}
  {2016})}\BibitemShut {NoStop}%
\bibitem [{\citenamefont {Choi}\ \emph {et~al.}(2006)\citenamefont {Choi},
  \citenamefont {Seo}, \citenamefont {Kim}, \citenamefont {Noh}, \citenamefont
  {Kim},\ and\ \citenamefont {Shin}}]{Choi2006a}%
  \BibitemOpen
  \bibfield  {author} {\bibinfo {author} {\bibfnamefont {W.~S.}\ \bibnamefont
  {Choi}}, \bibinfo {author} {\bibfnamefont {S.~S.~A.}\ \bibnamefont {Seo}},
  \bibinfo {author} {\bibfnamefont {K.~W.}\ \bibnamefont {Kim}}, \bibinfo
  {author} {\bibfnamefont {T.~W.}\ \bibnamefont {Noh}}, \bibinfo {author}
  {\bibfnamefont {M.~Y.}\ \bibnamefont {Kim}}, \ and\ \bibinfo {author}
  {\bibfnamefont {S.}~\bibnamefont {Shin}},\ }\bibfield  {title} {\enquote
  {\bibinfo {title} {{Dielectric constants of Ir, Ru, Pt, and IrO$_2$ :
  Contributions from bound charges}},}\ }\href {\doibase
  10.1103/PhysRevB.74.205117} {\bibfield  {journal} {\bibinfo  {journal}
  {Physical Review B}\ }\textbf {\bibinfo {volume} {74}},\ \bibinfo {pages}
  {205117} (\bibinfo {year} {2006})}\BibitemShut {NoStop}%
\bibitem [{\citenamefont {Goel}, \citenamefont {Skorinko},\ and\ \citenamefont
  {Pollak}(1981)}]{Goel1981}%
  \BibitemOpen
  \bibfield  {author} {\bibinfo {author} {\bibfnamefont {A.~K.}\ \bibnamefont
  {Goel}}, \bibinfo {author} {\bibfnamefont {G.}~\bibnamefont {Skorinko}}, \
  and\ \bibinfo {author} {\bibfnamefont {F.~H.}\ \bibnamefont {Pollak}},\
  }\bibfield  {title} {\enquote {\bibinfo {title} {{Optical properties of
  single-crystal rutile RuO$_2$ and IrO$_2$ in the range 0.5 to 9.5 eV}},}\
  }\href {\doibase 10.1103/PhysRevB.24.7342} {\bibfield  {journal} {\bibinfo
  {journal} {Physical Review B}\ }\textbf {\bibinfo {volume} {24}},\ \bibinfo
  {pages} {7342--7350} (\bibinfo {year} {1981})}\BibitemShut {NoStop}%
\bibitem [{Note1()}]{Note1}%
  \BibitemOpen
  \bibinfo {note} {In case this procedure does not produce a closed film at the
  reported laser energy density, we suggest to initially vary this parameter by
  several \SI {100}{mJ/cm^2}, as optical parameters can vary strongly between
  PLD setups.}\BibitemShut {Stop}%
\bibitem [{\citenamefont {Kim}\ \emph {et~al.}(2004)\citenamefont {Kim},
  \citenamefont {Noh}, \citenamefont {Kim},\ and\ \citenamefont
  {Oh}}]{Kim2004}%
  \BibitemOpen
  \bibfield  {author} {\bibinfo {author} {\bibfnamefont {H.-D.}\ \bibnamefont
  {Kim}}, \bibinfo {author} {\bibfnamefont {H.-J.}\ \bibnamefont {Noh}},
  \bibinfo {author} {\bibfnamefont {K.~H.}\ \bibnamefont {Kim}}, \ and\
  \bibinfo {author} {\bibfnamefont {S.-J.}\ \bibnamefont {Oh}},\ }\bibfield
  {title} {\enquote {\bibinfo {title} {{Core-Level X-Ray Photoemission
  Satellites in Ruthenates: A New Mechanism Revealing The Mott Transition}},}\
  }\href {\doibase 10.1103/PhysRevLett.93.126404} {\bibfield  {journal}
  {\bibinfo  {journal} {Physical Review Letters}\ }\textbf {\bibinfo {volume}
  {93}},\ \bibinfo {pages} {126404} (\bibinfo {year} {2004})}\BibitemShut
  {NoStop}%
\bibitem [{\citenamefont {Green}\ \emph {et~al.}(1985)\citenamefont {Green},
  \citenamefont {Gross}, \citenamefont {Papa}, \citenamefont {Schnoes},\ and\
  \citenamefont {Brasen}}]{Green1985a}%
  \BibitemOpen
  \bibfield  {author} {\bibinfo {author} {\bibfnamefont {M.~L.}\ \bibnamefont
  {Green}}, \bibinfo {author} {\bibfnamefont {M.~E.}\ \bibnamefont {Gross}},
  \bibinfo {author} {\bibfnamefont {L.~E.}\ \bibnamefont {Papa}}, \bibinfo
  {author} {\bibfnamefont {K.~J.}\ \bibnamefont {Schnoes}}, \ and\ \bibinfo
  {author} {\bibfnamefont {D.}~\bibnamefont {Brasen}},\ }\bibfield  {title}
  {\enquote {\bibinfo {title} {{Chemical Vapor Deposition of Ruthenium and
  Ruthenium Dioxide Films}},}\ }\href {\doibase 10.1149/1.2113647} {\bibfield
  {journal} {\bibinfo  {journal} {Journal of the Electrochemical Society}\
  }\textbf {\bibinfo {volume} {132}},\ \bibinfo {pages} {2677--2685} (\bibinfo
  {year} {1985})}\BibitemShut {NoStop}%
\bibitem [{\citenamefont {Dahal}\ \emph {et~al.}(2013)\citenamefont {Dahal},
  \citenamefont {Coy-Diaz}, \citenamefont {Addou}, \citenamefont {Lallo},
  \citenamefont {Sutter},\ and\ \citenamefont {Batzill}}]{Dahal2013}%
  \BibitemOpen
  \bibfield  {author} {\bibinfo {author} {\bibfnamefont {A.}~\bibnamefont
  {Dahal}}, \bibinfo {author} {\bibfnamefont {H.}~\bibnamefont {Coy-Diaz}},
  \bibinfo {author} {\bibfnamefont {R.}~\bibnamefont {Addou}}, \bibinfo
  {author} {\bibfnamefont {J.}~\bibnamefont {Lallo}}, \bibinfo {author}
  {\bibfnamefont {E.}~\bibnamefont {Sutter}}, \ and\ \bibinfo {author}
  {\bibfnamefont {M.}~\bibnamefont {Batzill}},\ }\bibfield  {title} {\enquote
  {\bibinfo {title} {{Preparation and characterization of
  Ni(111)/graphene/Y$_2$O$_3$(111) heterostructures}},}\ }\href {\doibase
  10.1063/1.4805042} {\bibfield  {journal} {\bibinfo  {journal} {Journal of
  Applied Physics}\ }\textbf {\bibinfo {volume} {113}} (\bibinfo {year}
  {2013}),\ 10.1063/1.4805042}\BibitemShut {NoStop}%
\bibitem [{\citenamefont {Krizek}\ \emph {et~al.}(2020)\citenamefont {Krizek},
  \citenamefont {Ka{\v{s}}par}, \citenamefont {Vetushka}, \citenamefont
  {Kriegner}, \citenamefont {Fiordaliso}, \citenamefont {Michalicka},
  \citenamefont {Man}, \citenamefont {Zub{\'{a}}{\v{c}}}, \citenamefont
  {Brajer}, \citenamefont {Hills}, \citenamefont {Edmonds}, \citenamefont
  {Wadley}, \citenamefont {Campion}, \citenamefont {Olejn{\'{i}}k},
  \citenamefont {Jungwirth},\ and\ \citenamefont {Nov{\'{a}}k}}]{Krizek2020}%
  \BibitemOpen
  \bibfield  {author} {\bibinfo {author} {\bibfnamefont {F.}~\bibnamefont
  {Krizek}}, \bibinfo {author} {\bibfnamefont {Z.}~\bibnamefont
  {Ka{\v{s}}par}}, \bibinfo {author} {\bibfnamefont {A.}~\bibnamefont
  {Vetushka}}, \bibinfo {author} {\bibfnamefont {D.}~\bibnamefont {Kriegner}},
  \bibinfo {author} {\bibfnamefont {E.~M.}\ \bibnamefont {Fiordaliso}},
  \bibinfo {author} {\bibfnamefont {J.}~\bibnamefont {Michalicka}}, \bibinfo
  {author} {\bibfnamefont {O.}~\bibnamefont {Man}}, \bibinfo {author}
  {\bibfnamefont {J.}~\bibnamefont {Zub{\'{a}}{\v{c}}}}, \bibinfo {author}
  {\bibfnamefont {M.}~\bibnamefont {Brajer}}, \bibinfo {author} {\bibfnamefont
  {V.~A.}\ \bibnamefont {Hills}}, \bibinfo {author} {\bibfnamefont {K.~W.}\
  \bibnamefont {Edmonds}}, \bibinfo {author} {\bibfnamefont {P.}~\bibnamefont
  {Wadley}}, \bibinfo {author} {\bibfnamefont {R.~P.}\ \bibnamefont {Campion}},
  \bibinfo {author} {\bibfnamefont {K.}~\bibnamefont {Olejn{\'{i}}k}}, \bibinfo
  {author} {\bibfnamefont {T.}~\bibnamefont {Jungwirth}}, \ and\ \bibinfo
  {author} {\bibfnamefont {V.}~\bibnamefont {Nov{\'{a}}k}},\ }\bibfield
  {title} {\enquote {\bibinfo {title} {{Molecular beam epitaxy of CuMnAs}},}\
  }\href {\doibase 10.1103/PhysRevMaterials.4.014409} {\bibfield  {journal}
  {\bibinfo  {journal} {Physical Review Materials}\ }\textbf {\bibinfo {volume}
  {4}} (\bibinfo {year} {2020}),\ 10.1103/PhysRevMaterials.4.014409},\ \Eprint
  {http://arxiv.org/abs/1911.01794} {1911.01794} \BibitemShut {NoStop}%
\bibitem [{\citenamefont {Yang}\ \emph {et~al.}(2017)\citenamefont {Yang},
  \citenamefont {Gao}, \citenamefont {Yang}, \citenamefont {Zhu}, \citenamefont
  {Huang},\ and\ \citenamefont {Ye}}]{Yang2017a}%
  \BibitemOpen
  \bibfield  {author} {\bibinfo {author} {\bibfnamefont {X.}~\bibnamefont
  {Yang}}, \bibinfo {author} {\bibfnamefont {P.}~\bibnamefont {Gao}}, \bibinfo
  {author} {\bibfnamefont {Z.}~\bibnamefont {Yang}}, \bibinfo {author}
  {\bibfnamefont {J.}~\bibnamefont {Zhu}}, \bibinfo {author} {\bibfnamefont
  {F.}~\bibnamefont {Huang}}, \ and\ \bibinfo {author} {\bibfnamefont
  {J.}~\bibnamefont {Ye}},\ }\bibfield  {title} {\enquote {\bibinfo {title}
  {{Optimizing ultrathin Ag films for high performance oxide-metal-oxide
  flexible transparent electrodes through surface energy modulation and
  template-stripping procedures}},}\ }\href {\doibase 10.1038/srep44576}
  {\bibfield  {journal} {\bibinfo  {journal} {Scientific Reports}\ }\textbf
  {\bibinfo {volume} {7}},\ \bibinfo {pages} {1--9} (\bibinfo {year}
  {2017})}\BibitemShut {NoStop}%
\bibitem [{\citenamefont {Petit}\ \emph {et~al.}(2007)\citenamefont {Petit},
  \citenamefont {Hayakawa}, \citenamefont {Wakayama},\ and\ \citenamefont
  {Chikyow}}]{Petit2007}%
  \BibitemOpen
  \bibfield  {author} {\bibinfo {author} {\bibfnamefont {M.}~\bibnamefont
  {Petit}}, \bibinfo {author} {\bibfnamefont {R.}~\bibnamefont {Hayakawa}},
  \bibinfo {author} {\bibfnamefont {Y.}~\bibnamefont {Wakayama}}, \ and\
  \bibinfo {author} {\bibfnamefont {T.}~\bibnamefont {Chikyow}},\ }\bibfield
  {title} {\enquote {\bibinfo {title} {{Early stage of growth of a perylene
  diimide derivative thin film growth on various Si(001) substrates}},}\ }\href
  {\doibase 10.1021/jp071876w} {\bibfield  {journal} {\bibinfo  {journal}
  {Journal of Physical Chemistry C}\ }\textbf {\bibinfo {volume} {111}},\
  \bibinfo {pages} {12747--12751} (\bibinfo {year} {2007})}\BibitemShut
  {NoStop}%
\bibitem [{\citenamefont {Martin}\ \emph {et~al.}(2020)\citenamefont {Martin},
  \citenamefont {Kim}, \citenamefont {Lee}, \citenamefont {Mehar},
  \citenamefont {Albertin}, \citenamefont {Hejral}, \citenamefont {Merte},
  \citenamefont {Lundgren}, \citenamefont {Asthagiri},\ and\ \citenamefont
  {Weaver}}]{Martin2020}%
  \BibitemOpen
  \bibfield  {author} {\bibinfo {author} {\bibfnamefont {R.}~\bibnamefont
  {Martin}}, \bibinfo {author} {\bibfnamefont {M.}~\bibnamefont {Kim}},
  \bibinfo {author} {\bibfnamefont {C.~J.}\ \bibnamefont {Lee}}, \bibinfo
  {author} {\bibfnamefont {V.}~\bibnamefont {Mehar}}, \bibinfo {author}
  {\bibfnamefont {S.}~\bibnamefont {Albertin}}, \bibinfo {author}
  {\bibfnamefont {U.}~\bibnamefont {Hejral}}, \bibinfo {author} {\bibfnamefont
  {L.~R.}\ \bibnamefont {Merte}}, \bibinfo {author} {\bibfnamefont
  {E.}~\bibnamefont {Lundgren}}, \bibinfo {author} {\bibfnamefont
  {A.}~\bibnamefont {Asthagiri}}, \ and\ \bibinfo {author} {\bibfnamefont
  {J.~F.}\ \bibnamefont {Weaver}},\ }\bibfield  {title} {\enquote {\bibinfo
  {title} {{High-Resolution X-ray Photoelectron Spectroscopy of an IrO$_2$(110)
  Film on Ir(100)}},}\ }\href {\doibase 10.1021/acs.jpclett.0c01805} {\bibfield
   {journal} {\bibinfo  {journal} {The Journal of Physical Chemistry Letters}\
  }\textbf {\bibinfo {volume} {11}},\ \bibinfo {pages} {7184--7189} (\bibinfo
  {year} {2020})}\BibitemShut {NoStop}%
\bibitem [{\citenamefont {Freakley}, \citenamefont {Ruiz-Esquius},\ and\
  \citenamefont {Morgan}(2017)}]{Freakley2017}%
  \BibitemOpen
  \bibfield  {author} {\bibinfo {author} {\bibfnamefont {S.~J.}\ \bibnamefont
  {Freakley}}, \bibinfo {author} {\bibfnamefont {J.}~\bibnamefont
  {Ruiz-Esquius}}, \ and\ \bibinfo {author} {\bibfnamefont {D.~J.}\
  \bibnamefont {Morgan}},\ }\bibfield  {title} {\enquote {\bibinfo {title}
  {{The X-ray photoelectron spectra of Ir, IrO$_2$ and IrCl$_3$ revisited}},}\
  }\href {\doibase 10.1002/sia.6225} {\bibfield  {journal} {\bibinfo  {journal}
  {Surface and Interface Analysis}\ }\textbf {\bibinfo {volume} {49}},\
  \bibinfo {pages} {794--799} (\bibinfo {year} {2017})}\BibitemShut {NoStop}%
\bibitem [{\citenamefont {Fairley}\ \emph {et~al.}(2021)\citenamefont
  {Fairley}, \citenamefont {Fernandez}, \citenamefont {Richard‐Plouet},
  \citenamefont {Guillot-Deudon}, \citenamefont {Walton}, \citenamefont
  {Smith}, \citenamefont {Flahaut}, \citenamefont {Greiner}, \citenamefont
  {Biesinger}, \citenamefont {Tougaard}, \citenamefont {Morgan},\ and\
  \citenamefont {Baltrusaitis}}]{Fairley2021}%
  \BibitemOpen
  \bibfield  {author} {\bibinfo {author} {\bibfnamefont {N.}~\bibnamefont
  {Fairley}}, \bibinfo {author} {\bibfnamefont {V.}~\bibnamefont {Fernandez}},
  \bibinfo {author} {\bibfnamefont {M.}~\bibnamefont {Richard‐Plouet}},
  \bibinfo {author} {\bibfnamefont {C.}~\bibnamefont {Guillot-Deudon}},
  \bibinfo {author} {\bibfnamefont {J.}~\bibnamefont {Walton}}, \bibinfo
  {author} {\bibfnamefont {E.}~\bibnamefont {Smith}}, \bibinfo {author}
  {\bibfnamefont {D.}~\bibnamefont {Flahaut}}, \bibinfo {author} {\bibfnamefont
  {M.}~\bibnamefont {Greiner}}, \bibinfo {author} {\bibfnamefont
  {M.}~\bibnamefont {Biesinger}}, \bibinfo {author} {\bibfnamefont
  {S.}~\bibnamefont {Tougaard}}, \bibinfo {author} {\bibfnamefont
  {D.}~\bibnamefont {Morgan}}, \ and\ \bibinfo {author} {\bibfnamefont
  {J.}~\bibnamefont {Baltrusaitis}},\ }\bibfield  {title} {\enquote {\bibinfo
  {title} {{Systematic and collaborative approach to problem solving using
  X-ray photoelectron spectroscopy}},}\ }\href {\doibase
  10.1016/j.apsadv.2021.100112} {\bibfield  {journal} {\bibinfo  {journal}
  {Applied Surface Science Advances}\ }\textbf {\bibinfo {volume} {5}},\
  \bibinfo {pages} {100112} (\bibinfo {year} {2021})}\BibitemShut {NoStop}%
\bibitem [{\citenamefont {Ma}\ and\ \citenamefont {Chen}(2016)}]{Ma2016}%
  \BibitemOpen
  \bibfield  {author} {\bibinfo {author} {\bibfnamefont {C.}~\bibnamefont
  {Ma}}\ and\ \bibinfo {author} {\bibfnamefont {C.}~\bibnamefont {Chen}},\
  }\bibfield  {title} {\enquote {\bibinfo {title} {{Pulsed Laser Deposition for
  Complex Oxide Thin Film and Nanostructure}},}\ }in\ \href {\doibase
  10.1002/9783527696406.ch1} {\emph {\bibinfo {booktitle} {Advanced Nano
  Deposition Methods}}}\ (\bibinfo  {publisher} {Wiley},\ \bibinfo {year}
  {2016})\ pp.\ \bibinfo {pages} {1--31}\BibitemShut {NoStop}%
\bibitem [{\citenamefont {Shepelin}\ \emph {et~al.}(2023)\citenamefont
  {Shepelin}, \citenamefont {Tehrani}, \citenamefont {Ohannessian},
  \citenamefont {Schneider}, \citenamefont {Pergolesi},\ and\ \citenamefont
  {Lippert}}]{Shepelin2023}%
  \BibitemOpen
  \bibfield  {author} {\bibinfo {author} {\bibfnamefont {N.~A.}\ \bibnamefont
  {Shepelin}}, \bibinfo {author} {\bibfnamefont {Z.~P.}\ \bibnamefont
  {Tehrani}}, \bibinfo {author} {\bibfnamefont {N.}~\bibnamefont
  {Ohannessian}}, \bibinfo {author} {\bibfnamefont {C.~W.}\ \bibnamefont
  {Schneider}}, \bibinfo {author} {\bibfnamefont {D.}~\bibnamefont
  {Pergolesi}}, \ and\ \bibinfo {author} {\bibfnamefont {T.}~\bibnamefont
  {Lippert}},\ }\bibfield  {title} {\enquote {\bibinfo {title} {{A practical
  guide to pulsed laser deposition}},}\ }\href {\doibase 10.1039/D2CS00938B}
  {\bibfield  {journal} {\bibinfo  {journal} {Chemical Society Reviews}\
  }\textbf {\bibinfo {volume} {52}},\ \bibinfo {pages} {2294--2321} (\bibinfo
  {year} {2023})}\BibitemShut {NoStop}%
\bibitem [{\citenamefont {Liu}, \citenamefont {Masumoto},\ and\ \citenamefont
  {Goto}(2004{\natexlab{a}})}]{Liu2004a}%
  \BibitemOpen
  \bibfield  {author} {\bibinfo {author} {\bibfnamefont {Y.}~\bibnamefont
  {Liu}}, \bibinfo {author} {\bibfnamefont {H.}~\bibnamefont {Masumoto}}, \
  and\ \bibinfo {author} {\bibfnamefont {T.}~\bibnamefont {Goto}},\ }\bibfield
  {title} {\enquote {\bibinfo {title} {{Preparation of IrO$_2$ Thin Films by
  Oxidating Laser-ablated Ir}},}\ }\href {\doibase 10.2320/matertrans.45.900}
  {\bibfield  {journal} {\bibinfo  {journal} {MATERIALS TRANSACTIONS}\ }\textbf
  {\bibinfo {volume} {45}},\ \bibinfo {pages} {900--903} (\bibinfo {year}
  {2004}{\natexlab{a}})}\BibitemShut {NoStop}%
\bibitem [{\citenamefont {Abb}\ \emph {et~al.}(2020)\citenamefont {Abb},
  \citenamefont {Weber}, \citenamefont {Langsdorf}, \citenamefont {Koller},
  \citenamefont {Gericke}, \citenamefont {Pfaff}, \citenamefont {Busch},
  \citenamefont {Zetterberg}, \citenamefont {Preobrajenski}, \citenamefont
  {Gr{\"{o}}nbeck}, \citenamefont {Lundgren},\ and\ \citenamefont
  {Over}}]{Abb2020a}%
  \BibitemOpen
  \bibfield  {author} {\bibinfo {author} {\bibfnamefont {M.~J.~S.}\
  \bibnamefont {Abb}}, \bibinfo {author} {\bibfnamefont {T.}~\bibnamefont
  {Weber}}, \bibinfo {author} {\bibfnamefont {D.}~\bibnamefont {Langsdorf}},
  \bibinfo {author} {\bibfnamefont {V.}~\bibnamefont {Koller}}, \bibinfo
  {author} {\bibfnamefont {S.~M.}\ \bibnamefont {Gericke}}, \bibinfo {author}
  {\bibfnamefont {S.}~\bibnamefont {Pfaff}}, \bibinfo {author} {\bibfnamefont
  {M.}~\bibnamefont {Busch}}, \bibinfo {author} {\bibfnamefont
  {J.}~\bibnamefont {Zetterberg}}, \bibinfo {author} {\bibfnamefont
  {A.}~\bibnamefont {Preobrajenski}}, \bibinfo {author} {\bibfnamefont
  {H.}~\bibnamefont {Gr{\"{o}}nbeck}}, \bibinfo {author} {\bibfnamefont
  {E.}~\bibnamefont {Lundgren}}, \ and\ \bibinfo {author} {\bibfnamefont
  {H.}~\bibnamefont {Over}},\ }\bibfield  {title} {\enquote {\bibinfo {title}
  {{Thermal Stability of Single-Crystalline IrO$_2$ (110) Layers: Spectroscopic
  and Adsorption Studies}},}\ }\href {\doibase 10.1021/acs.jpcc.0c04373}
  {\bibfield  {journal} {\bibinfo  {journal} {The Journal of Physical Chemistry
  C}\ }\textbf {\bibinfo {volume} {124}},\ \bibinfo {pages} {15324--15336}
  (\bibinfo {year} {2020})}\BibitemShut {NoStop}%
\bibitem [{\citenamefont {Cordfunke}\ and\ \citenamefont
  {Meyer}(1962)}]{Cordfunke1962}%
  \BibitemOpen
  \bibfield  {author} {\bibinfo {author} {\bibfnamefont {E.~H.~P.}\
  \bibnamefont {Cordfunke}}\ and\ \bibinfo {author} {\bibfnamefont
  {G.}~\bibnamefont {Meyer}},\ }\bibfield  {title} {\enquote {\bibinfo {title}
  {{The system iridium ‐ oxygen I. Measurements on the volatile oxide of
  iridium}},}\ }\href {\doibase 10.1002/recl.19620810608} {\bibfield  {journal}
  {\bibinfo  {journal} {Recueil des Travaux Chimiques des Pays-Bas}\ }\textbf
  {\bibinfo {volume} {81}},\ \bibinfo {pages} {495--504} (\bibinfo {year}
  {1962})}\BibitemShut {NoStop}%
\bibitem [{\citenamefont {Chalamala}\ \emph {et~al.}(1999)\citenamefont
  {Chalamala}, \citenamefont {Wei}, \citenamefont {Reuss}, \citenamefont
  {Aggarwal}, \citenamefont {Gnade}, \citenamefont {Ramesh}, \citenamefont
  {Bernhard}, \citenamefont {Sosa},\ and\ \citenamefont
  {Golden}}]{Chalamala1999}%
  \BibitemOpen
  \bibfield  {author} {\bibinfo {author} {\bibfnamefont {B.~R.}\ \bibnamefont
  {Chalamala}}, \bibinfo {author} {\bibfnamefont {Y.}~\bibnamefont {Wei}},
  \bibinfo {author} {\bibfnamefont {R.~H.}\ \bibnamefont {Reuss}}, \bibinfo
  {author} {\bibfnamefont {S.}~\bibnamefont {Aggarwal}}, \bibinfo {author}
  {\bibfnamefont {B.~E.}\ \bibnamefont {Gnade}}, \bibinfo {author}
  {\bibfnamefont {R.}~\bibnamefont {Ramesh}}, \bibinfo {author} {\bibfnamefont
  {J.~M.}\ \bibnamefont {Bernhard}}, \bibinfo {author} {\bibfnamefont {E.~D.}\
  \bibnamefont {Sosa}}, \ and\ \bibinfo {author} {\bibfnamefont {D.~E.}\
  \bibnamefont {Golden}},\ }\bibfield  {title} {\enquote {\bibinfo {title}
  {{Effect of growth conditions on surface morphology and photoelectric work
  function characteristics of iridium oxide thin films}},}\ }\href {\doibase
  10.1063/1.123561} {\bibfield  {journal} {\bibinfo  {journal} {Applied Physics
  Letters}\ }\textbf {\bibinfo {volume} {74}},\ \bibinfo {pages} {1394--1396}
  (\bibinfo {year} {1999})}\BibitemShut {NoStop}%
\bibitem [{\citenamefont {Liu}, \citenamefont {Masumoto},\ and\ \citenamefont
  {Goto}(2004{\natexlab{b}})}]{Liu2004}%
  \BibitemOpen
  \bibfield  {author} {\bibinfo {author} {\bibfnamefont {Y.}~\bibnamefont
  {Liu}}, \bibinfo {author} {\bibfnamefont {H.}~\bibnamefont {Masumoto}}, \
  and\ \bibinfo {author} {\bibfnamefont {T.}~\bibnamefont {Goto}},\ }\bibfield
  {title} {\enquote {\bibinfo {title} {{Electrical and Optical Properties of
  IrO$_2$ Thin Films Prepared by Laser-ablation}},}\ }\href {\doibase
  10.2320/matertrans.45.3023} {\bibfield  {journal} {\bibinfo  {journal}
  {MATERIALS TRANSACTIONS}\ }\textbf {\bibinfo {volume} {45}},\ \bibinfo
  {pages} {3023--3027} (\bibinfo {year} {2004}{\natexlab{b}})}\BibitemShut
  {NoStop}%
\end{thebibliography}

\providecommand{\noopsort}[1]{}\providecommand{\singleletter}[1]{#1}%

\end{document}